\documentclass[aps,prc,twocolumn,superscriptaddress,10pt,english,preprintnumbers]{revtex4-1}
\pdfoutput=1
\usepackage[T1]{fontenc}
\usepackage[utf8]{inputenc}
\usepackage{epsfig}
\usepackage{graphicx}
\usepackage{xcolor}
\usepackage{latexsym}
\usepackage{textcomp}
\usepackage{amssymb}
\usepackage[colorlinks=true,linkcolor=blue,citecolor=blue, urlcolor=blue]{hyperref}   %
\usepackage{amsfonts,amsthm,amstext,amscd}
\usepackage{amsmath}

\def\trento{T\raisebox{-0.5ex}{R}ENTo}
\def\zetamax{(\zeta/s)_{\rm max}}
\def\zetawidth{(\zeta/s)_{\rm width}}
\def\etamin{(\eta/s)_{\rm min}}
\def\etaslope{(\eta/s)_{\rm slope}}
\def\vfs{v_{\rm fs}}
\def\vtwo{v_2\{2\}}
\def\vtwofour{v_2\{4\}}

\DeclareMathOperator{\res}{Res}
\DeclareMathOperator{\re}{Re}

\newcommand{\fig}[1]{Fig.~\ref{#1}}

\begin{document}
\title{Bayesian analysis of heavy ion collisions with the heavy ion computational framework \emph{Trajectum}}
\author{Govert Nijs}
\affiliation{Center for Theoretical Physics, Massachusetts Institute of Technology, Cambridge, MA 02139, USA}
\affiliation{Institute for Theoretical Physics and Center for Extreme Matter and Emergent Phenomena, Utrecht University, 3584 CC Utrecht, The Netherlands}
\author{Wilke van der Schee}
\affiliation{Theoretical Physics Department, CERN, CH-1211 Gen\`eve 23, Switzerland}
\author{Umut G\"ursoy}
\affiliation{Institute for Theoretical Physics and Center for Extreme Matter and Emergent Phenomena, Utrecht University, 3584 CC Utrecht, The Netherlands}
\author{Raimond Snellings}
\affiliation{Nikhef, 1098 XG Amsterdam, The Netherlands}
\affiliation{Institute for Gravitational and Subatomic Physics (GRASP), Utrecht University, 3584 CC Utrecht, The Netherlands}
\begin{abstract}
We introduce a model for heavy ion collisions named \emph{Trajectum}, which includes an expanded initial stage with a variable free streaming velocity $v_{\rm fs}$ and a hydrodynamic stage with three varying second order transport coefficients. We describe how to obtain a Gaussian Emulator for this 20-parameter model and show results for key observables. This emulator can be used to obtain Bayesian posterior estimates on the parameters, which we test by an elaborate closure test as well as a convergence study. Lastly, we employ the optimal values of the parameters found in \cite{Bayesianshort} to perform a detailed comparison to experimental data from PbPb and $p$Pb collisions. This includes both observables that have been used to obtain these values as well as wider transverse momentum ranges and new observables such as correlations of event-plane angles.
\end{abstract}
\preprint{CERN-TH-2020-175//MIT-CTP/5251}

\maketitle

{\hypersetup{hidelinks}
\tableofcontents
}

\section{Introduction}
The collisions of heavy ions at the Relativistic Heavy Ion Collider (RHIC) at Brookhaven and the Large Hadron Collider (LHC) at CERN have led to an accepted picture of a short prehydrodynamic phase followed by a relativistic fluid composed of quark-gluon plasma (QGP) with remarkably small shear viscosity~\cite{Busza:2018rrf}\@. %

The most convincing evidence for an almost perfect fluid comes from the anisotropies of the particle spectra at low momentum in combination with elaborate hydrodynamical modelling that can explain these non-trivial correlations~\cite{Heinz:2013th}\@.
This started with ideal hydrodynamics (effectively having shear viscosity $\eta=0$) \cite{Teaney:2000cw}, after which simulations including viscosity pointed to a surprisingly small value of the specific viscosity of $\eta/s \approx 0.08$, with $s$ the entropy density \cite{Romatschke:2007mq}. This value was particularly exciting, since results from string theory using a holographic dual black hole indicate that quantum matter that is infinitely strongly interacting has a universal specific viscosity equalling $\eta/s = 1/4\pi \approx 0.08$ \cite{Kovtun:2004de}.

This hydrodynamic modelling relied on two pillars. Firstly, it was assumed early on that the exploding debris left after the collision interacts so strongly that starting at a time as early as $1\,$fm$/c$ (as in \cite{Romatschke:2007mq}) a fluid would be formed that can be described by viscous hydrodynamics. This assumption is non-trivial since the full stress energy tensor has many more degrees of freedom (9 for conformal matter) as compared to hydrodynamics (the energy density plus fluid velocity gives 4 degrees of freedom). This assumption of fast `hydrodynamization' has been substantiated by studies in holography which showed that far-from-equilibrium matter can always be described by hydrodynamics within a time of the inverse temperature $1/T$ \cite{Chesler:2009cy, Heller:2011ju, Heller:2012km}. Similarly fast times have been found utilising kinetic theory at a realistic value of the coupling constant \cite{Kurkela:2015qoa, Keegan:2015avk}. 

Secondly, the hydrodynamic simulations need initial profiles for the energy density and fluid velocity. A Monte Carlo Glauber model \cite{Alver:2008aq} can provide a realistic estimate of the distribution of nucleons inside a heavy ion, but a priori it is not clear how this translates into an energy density for the plasma. Two naive pictures are to add up all left- and right-moving colliding pairs (binary collisions), as would be appropriate for a completely transparent collision where all interactions are independent (this is indeed applicable for the production of highly energetic quarks or gluons). For the low energy dynamics the collisions are not independent and it is more realistic to add up colliding nucleons, without doubly counting them if they collide several times (wounded nucleon approximation). 
In subsection \ref{sec:initialconditions} we review the more elaborate phenomenological \trento{} model \cite{Moreland:2014oya}, but here we note that none of these descriptions are based on microscopic insights.
Exceptions to this are the Color Glass Condensate (CGC) initial model \cite{Gelis:2010nm}, EKRT \cite{Niemi:2015qia} or models from holography \cite{vanderSchee:2015rta}.

Especially the uncertainty in the initial prehydrodynamic stage has led to considerable uncertainty for the hydrodynamic transport coefficients. One example is the difference in the estimate for $\eta/s \approx 0.08$ of \cite{Romatschke:2007mq} (using binary collisions) or $\eta/s \approx 0.20$ when using CGC initial conditions \cite{Gale:2012rq}. Furthermore, in these references the bulk viscosity was assumed to vanish and the comparison with experiment focused on the anisotropic flow coefficients. When also demanding a simultaneous fit of the mean transverse momentum of all identified particle species it turns out that \cite{Gale:2012rq} needed a sizeable bulk viscosity, which in turn meant that the shear viscosity was refitted to a value of $\eta/s \approx 0.095$ \cite{Ryu:2015vwa}.

All of this is to stress the significant uncertainties in the precision analysis of the QGP and furthermore the importance of having a model that describes all experimental results simultaneously, preferably among different colliding systems as well as across vastly different colliding energy scales. This can be done by a sufficiently flexible model set-up, where all free parameters are obtained from a global analysis, comparing with as much experimental data as possible. References \cite{Novak:2013bqa, Pratt:2015zsa, Sangaline:2015isa, Bernhard:2016tnd, Devetak:2019lsk} pioneered such studies using large computing resources combined with emulation and markov chain tools inspired from similar analyses in cosmology. This culminated in the most precise temperature-dependent estimate for the QGP shear viscosity to-date \cite{Bernhard:2019bmu}. 

In the current work we will build on this expertise, which was generously made available on GitHub together with an excellent documentation \cite{Bernhard:2018hnz, Moreland:2019szz}. Firstly, we enlarge the scope of the prehydrodynamic phase by allowing a free streaming parameter that can interpolate between non-interacting free streaming with the speed of light to no streaming at all. On the one hand this is motivated by possible interactions during this phase, which can slow down the free streaming. On the other hand, by enlarging the scope of the model we can gauge the sensitivity of (physical) QGP parameters such as the viscosities to the specific parametrization used for the initial condition. Secondly, we enlarge the set of parameters with three second order hydrodynamic transport coefficients. This is again motivated to study the robustness of previous results on the specifics of the hydrodynamic code, but we also hope that after the relatively successful determination of $\eta/s$, constraints can be obtained on second order transport coefficients. Thirdly, an important addition is the addition of more experimental observables, such as the transverse momentum differential observables for pions, kaons and protons, both for the yield as well as the anisotropic flow coefficients.

Of special recent interest is the debated question whether a QGP can also form in (high multiplicity) proton-ion ($p$A) collisions \cite{Nagle:2018nvi}. It is for this reason that we include $p$Pb collisions in our analysis (also done in \cite{Moreland:2018gsh}). Firstly, by obtaining results both with and without $p$Pb collisions it is possible to at least give a partial answer if hydrodynamic modelling can describe low momentum experimental data in $p$Pb. Secondly, if this is indeed possible it is the expectation that these collisions can in principle be more sensitive to several observables influenced by large gradients, which would average out for the larger and longer lived PbPb system.

In Section \ref{sec:trajectum} we will describe this enlarged model which includes 10 parameters for the initial stage, 6 first order and 3 second order transport coefficients and one final parameter at which temperature to particlize our hydrodynamic fluid into hadrons for a subsequent hadronic cascade code.
In Section \ref{sec:results} we describe its results, including a detailed study how several observables depend on any of our 20 parameters. We furthermore present several computational aspects, including elaborate closure tests.

Importantly, global analyses have never been performed with transverse momentum dependent spectra or (identified) transverse momentum dependent anisotropic flow coefficients. In our companion paper \cite{Bayesianshort} we attempt to fill this lacuna by comparing to such experimental data, both for PbPb and $p$Pb collisions. A detailed analysis of the resulting QGP properties is presented in Section \ref{sec:map}.

\begin{figure}[ht]
\includegraphics[width=\columnwidth]{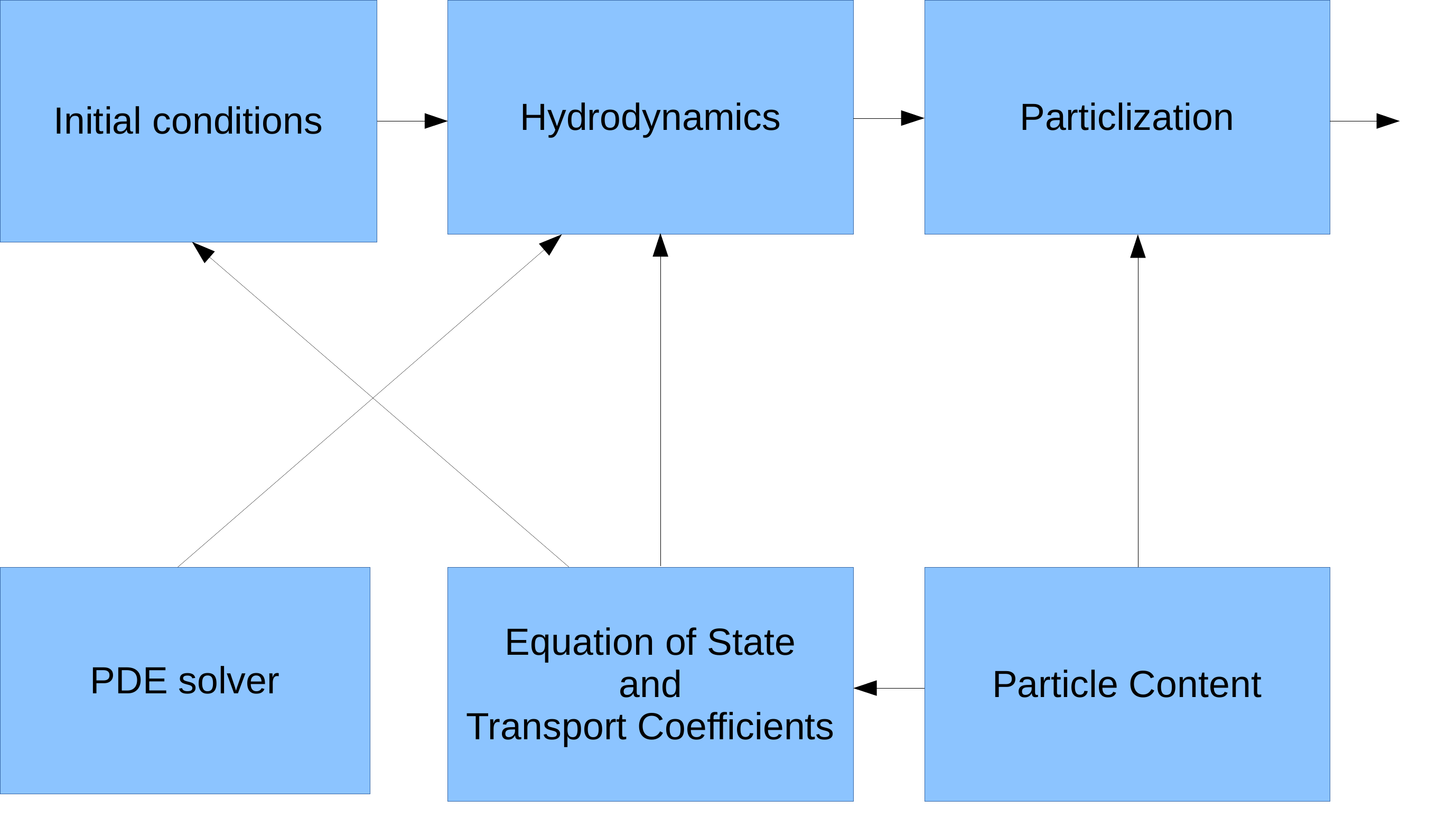}
\caption{\label{fig:Trajectumdiagram}Components implemented in \emph{Trajectum}\@. Also shown are the dependencies between the various components, where the information flow is indicated by the arrows. Note the (perhaps non-obvious) arrow between the equation of state/transport coefficients and the initial conditions. This allows the initial conditions to access information which is for example needed for Gubser flow.}
\end{figure}

\section{The \emph{Trajectum} framework}\label{sec:trajectum}
The results in this work %
have been generated using the new heavy ion code \emph{Trajectum}\@.
\emph{Trajectum} is named after the old Roman name for the city of Utrecht, where the code was developed.
The code is written in C\texttt{++} and incorporates the computation of the initial conditions, prehydrodynamic phase, hydrodynamic phase and the particlization inside a single executable.
Furthermore, for each of these components a base class specifies the interface with which the component communicates with the other components.
In this way, it becomes possible to have several versions for each component, which the user can swap out as desired.
The common interface then guarantees that whichever choice the user makes, the component will consistently interact with the others.
Additionally, \emph{Trajectum} will also query of each chosen component which parameters it requires to function properly, and will check a user-specified parameter file to read these parameters.

In \fig{fig:Trajectumdiagram} we show the various components, as well as the interactions between them.
The particle content can serve as an example of the statement that components automatically interact correctly with the others.
In the current implementation, it is possible to choose from the particle content of UrQMD \cite{Bass:1998ca,Bleicher:1999xi} or SMASH \cite{Weil:2016zrk,dmytro_oliinychenko_2020_3742965,Sjostrand:2007gs}\@.
Upon the user choosing one of these options, the choice is automatically communicated to the class handling particlization, which needs this to work out which particles to produce, but also to compute viscous corrections and amounts of particles of each type to be produced.
Upon the user choosing one of these options, the choice is automatically communicated to the class handling particlization, which uses this to compute a list of particles coming out of the hydrodynamic phase. %
In addition, the choice is automatically communicated to the class handling the equation of state, where it is needed for the Hadron Resonance Gas (HRG) part of the equation of state.

As another example, consider the choice between 2+1D and 3+1D hydrodynamics.
These models require different behaviors from the initial conditions, the PDE solvers, and the particlization algorithms.
In particular, components like the initial conditions need extra parameters to set up a 3+1D simulation \cite{Ke:2016jrd}\@.
This does not lead to incompatibility between different sets of components though, as each component is guaranteed to work well together with whichever other components the user chooses.
Instead, the components communicate, and change their behavior even according to which other components they are combined with.
As a result, in case the user specifies 3+1D hydrodynamics, the component responsible for the initial conditions will make sure that the user also specifies the extra needed parameters.

\subsection{Initial conditions}\label{sec:initialconditions}
For the initial conditions, one can choose from the following options:
\begin{itemize}
\item Monte Carlo Glauber \cite{Miller:2007ri,Barej:2017kcw,Loizides:2016djv,Bozek:2016kpf},
\item Ohio State University \cite{Welsh:2016siu,Weller:2017tsr},
\item T\raisebox{-0.5ex}{R}ENTo, both with and without nucleon substructure \cite{Moreland:2014oya,Moreland:2018gsh},
\item Gubser flow \cite{Gubser:2010ze,Gubser:2010ui,Marrochio:2013wla}.
\end{itemize}
Although conceptually separate, in \emph{Trajectum} the prehydrodynamic phase is part of the initial conditions.
Therefore the prehydrodynamic phase presented in the next subsection is only available when using T\raisebox{-0.5ex}{R}ENTo initial conditions.
The Gubser initial conditions initializes the fluid according to the Gubser solution, but the time-dependent Gubser flow will only be recovered if the ideal gas equation of state is used for the hydrodynamical evolution (see Section~\ref{sec:eosandtc}), as this is a conformal equation of state. 

In the analysis presented in this work, we use the T\raisebox{-0.5ex}{R}ENTo model with nuclear substructure \cite{Moreland:2018gsh}\@.
The T\raisebox{-0.5ex}{R}ENTo model describes nuclei using a Woods-Saxon distribution, where a minimal internucleon distance $d_\text{min}$ is imposed.
Each nucleon subsequently is made up of $n_c$ constituents, each with width given by
\[
v_\text{min} + \chi_\text{struct}\left(w - v_\text{min}\right),
\]
with $v_\text{min} = 0.2\,\text{fm}$, and $w$ and $\chi_\text{struct}$ parameters.
The constituents are sampled from a Gaussian distribution, such that the width of the matter distribution in each nucleon is $w$ on average.
After sampling the configuration of each nucleus, every pair of nucleons is labeled \emph{wounded} with a probability based on their overlap, in such a way that the nucleon-nucleon cross-section $\sigma_{NN}$ equals the proton-proton cross-section for that particular collision energy ($63\,\text{mb}$ for $2.76\,\text{TeV}$ and $70\,\text{mb}$ for $5.02\,\text{TeV}$)\@.
The constituents in each wounded nucleon each then source a thickness function with a gaussian distribution, where for each individual constituent the norm of the gaussian is given by $N\gamma/n_c$\@.
Here $\gamma$ is sampled from a gamma distrubution with width $\sigma_\text{fluct}\sqrt{n_c}$, and $N$ and $\sigma_\text{fluct}$ are parameters.
In this way, for each nucleus a thickness function is computed, $\mathcal{T}_A$ and $\mathcal{T}_B$\@.
These functions are combined together as
\begin{equation} \label{eq:trento}
\mathcal{T} = \left(\frac{\mathcal{T}_A^p + \mathcal{T}_B^p}{2}\right)^{1/p},
\end{equation}
with $p$ a parameter. This parametrizes a wide range of initial conditions, whereby in particular for $p=1$ we have $\mathcal{T} = \mathcal{T}_A + \mathcal{T}_B$ (also called wounded nucleon scaling) and for $p=0$, \eqref{eq:trento} reduces to $\mathcal{T} = \sqrt{\mathcal{T}_A \mathcal{T}_B},$ which is qualitatively similar to EKRT \cite{Niemi:2015qia,Bernhard:2016tnd} or holography \cite{vanderSchee:2015rta}\@.
This thickness function $\mathcal{T}$ is then assumed to be boost invariant and proportional to the energy density, whereby the proportionality factor is given by $1/\tau$, as is appropriated for a pressureless fluid.
\emph{Trajectum} is capable of extending this to 3+1D \cite{Ke:2016jrd}, but this was not used in this work.

\subsection{Prehydrodynamic phase}
After obtaining the initial conditions, the matter is free streamed for a time $\tau_\text{fs}$ with velocity $v_\text{fs}$\@.
This is a physically significant modification of the implementation \cite{Broniowski:2008qk,Heinz:2015arc,FreeStreamGithub} used in \cite{Bernhard:2019bmu}, where $v_\text{fs} = 1$\@.
Since the free streaming evolution starts at proper time $\tau = 0^+$, the initial velocity in the longitudinal direction is zero and we have $T^{\mu\eta}=0$, with $\eta$ here referring to spacetime rapidity.
This results in the following stress-energy tensor:
\begin{align}
T^{\mu\nu}(x,y) & = \frac{1}{2\pi\tau_\text{fs}}\int_0^{2\pi}d\phi\hat p^\mu\hat p^\nu\label{eq:freestreaming}\\
& \quad \times \mathcal{T}(x - v_\text{fs}\tau_\text{fs}\cos\phi,y - v_\text{fs}\tau_\text{fs}\sin\phi)\nonumber
\end{align}
with
\[
\hat p^\mu\hat p^\nu = \left(\begin{array}{ccc}
1 & v_\text{fs}\cos\phi & v_\text{fs}\sin\phi \\
v_\text{fs}\cos\phi & v_\text{fs}^2\cos^2\phi & v_\text{fs}^2\cos\phi\sin\phi \\
v_\text{fs}\sin\phi & v_\text{fs}^2\cos\phi\sin\phi & v_\text{fs}^2\sin^2\phi
\end{array}
\right).
\]

It is interesting to examine the small $\tau_\text{fs}$ behavior of the prehydrodynamic stress-energy tensor.
For this purpose, we expand \eqref{eq:freestreaming} for small $\tau_\text{fs}$\@.
In this case, the integral can be performed analytically, resulting in the following expression:
\[
T^{\mu\nu} = \left(\begin{array}{ccc}
\frac{\mathcal{T}}{\tau_\text{fs}} & -\frac{\mathcal{T}v_\text{fs}^2}{2}\partial_x\log\mathcal{T} & -\frac{\mathcal{T}v_\text{fs}^2}{2}\partial_y\log\mathcal{T} \\
-\frac{\mathcal{T}v_\text{fs}^2}{2}\partial_x\log\mathcal{T} & \frac{\mathcal{T}v_\text{fs}^2}{2\tau_\text{fs}} & 0 \\
-\frac{\mathcal{T}v_\text{fs}^2}{2}\partial_y\log\mathcal{T} & 0 & \frac{\mathcal{T}v_\text{fs}^2}{2\tau_\text{fs}}
\end{array}\right),
\]
where we omit the spatial arguments for brevity.
By solving the eigenvalue problem
\[
T^{\mu\nu}u_\nu = \rho u^\mu,
\]
we can then extract the fluid velocity:
\begin{equation}\label{eq:uini}
u^\mu = \left(1, -\frac{v_\text{fs}^2\tau_\text{fs}}{2 + v_\text{fs}^2}\partial_x\log\mathcal{T}, -\frac{v_\text{fs}^2\tau_\text{fs}}{2 + v_\text{fs}^2}\partial_y\log\mathcal{T}\right) + \mathcal{O}\left(\tau_\text{fs}^2\right).
\end{equation}
Using holographic simulations \cite{vanderSchee:2012qj,vanderSchee:2013pia,Romatschke:2013re} it was found in \cite{Habich:2014jna} that at strong coupling in a conformal theory an appropriate initial velocity can be initialized with $u^\perp \approx -\frac{\tau}{3.0} \nabla_\perp \log \mathcal{T}$. Quite curiously this exactly agrees with \eqref{eq:uini}
for $v_\text{fs} = 1$. This is somewhat coincidental, since the holographic evolution is quite unlike free streaming. Free streaming has zero longitudinal pressure, whereas the holographic evolution has a large negative longitudinal pressure at early times, but quickly hydrodynamizes towards a positive longitudinal pressure (see also \cite{Grumiller:2008va,Casalderrey-Solana:2013aba}). Depending on the starting time of hydrodynamics this pressure can average to the zero free streaming result.
Note also that this does not mean that the entire stress tensor agrees.
In particular, the holographic result features fast hydrodynamization, whereas we will see in subsection \ref{sec:results:prehydrodynamicphase} that in our model the prehydrodynamic phase takes the fluid away from hydrodynamics, and that a free streaming velocity of around 0.93 minimizes deviations from hydrodynamics for the bulk viscous pressure.
For the shear tensor, minimization of deviations from hydrodynamics is achieved for $v_\text{fs} \approx 0.6$\@.

\subsection{Hydrodynamic phase}
For the hydrodynamics model, one can choose between one including only first order transport coefficients, and one including also the second order transport coefficients from the 14-moment approximation \cite{Denicol:2014vaa}\@.
Both of these choices are available in both 2+1D (assuming boost invariance) and 3+1D versions.
In the analysis presented in this work, we used the second order version in 2+1D\@.
This solves the conservation equations for the stress-energy equation
\begin{equation}\label{eq:eom}
\partial_\mu T^{\mu\nu} = 0,
\end{equation}
with $\partial_\mu$ the covariant derivative.
We use the constitutive relation
\begin{equation}
T^{\mu\nu} = \rho u^\mu u^\nu - (P + \Pi)\Delta^{\mu\nu} + \pi^{\mu\nu},\label{eq:constitutiverelation}
\end{equation}
with $\Delta^{\mu\nu} = g^{\mu\nu} - u^\mu u^\nu$, $u^\mu$ the local fluid velocity and $g_{\mu\nu}$ a mostly minus metric.
The relation between the energy density $\rho$ and the pressure $P$ is given by the equation of state, which is discussed in subsection \ref{sec:eosandtc}\@.
There are also two viscous corrections, $\pi^{\mu\nu}$ and $\Pi$\@.
The shear tensor $\pi^{\mu\nu}$ is a symmetric traceless tensor which is transverse to $u^\mu$\@.
It and the bulk pressure $\Pi$ obey the second order 14-moment approximation equations of motion \cite{Denicol:2014vaa}, where we keep only the transport coefficients that have been computed explicitly in \cite{Denicol:2014vaa}:%
\begin{align}
D\Pi & = -\frac{1}{\tau_\Pi}\left[\Pi + \zeta\nabla\cdot u + \delta_{\Pi\Pi}\nabla\cdot u\Pi\right.\label{eq:intro:secondorderbulk}\\
& \qquad - \left.\lambda_{\Pi\pi}\pi^{\mu\nu}\sigma_{\mu\nu}\right],\nonumber\\
\Delta^\mu_\alpha\Delta^\nu_\beta D\pi^{\alpha\beta} & = -\frac{1}{\tau_\pi}\left[\pi^{\mu\nu} - 2\eta\sigma^{\mu\nu}\right.\label{eq:intro:secondordershear}\\
& \qquad + \delta_{\pi\pi}\pi^{\mu\nu}\nabla\cdot u - \phi_7\pi_\alpha^{\langle\mu}\pi^{\nu\rangle\alpha}\nonumber\\
& \qquad + \left.\tau_{\pi\pi}\pi_\alpha^{\langle\mu}\sigma^{\nu\rangle\alpha} - \lambda_{\pi\Pi}\Pi\sigma^{\mu\nu}\right].\nonumber
\end{align}
Here $D \equiv u^\mu\partial_\mu$, $\nabla^\mu \equiv \Delta^{\mu\nu}\partial_\nu$ and $\sigma^{\mu\nu} \equiv \frac{1}{2}(\nabla^\mu u^\nu + \nabla^\nu u^\mu) - \frac{1}{3}(\nabla \cdot u)\Delta^{\mu\nu}$\@. These constitutive equations do not include vortical transport as well as some second order transport coefficients involving the bulk viscous pressure $\Pi$. Both of these are expected to be small in boost invariant settings with approximate conformal symmetry.

\subsection{Equation of state and transport coefficients}\label{sec:eosandtc}
For the equation of state and transport coefficients, there are 2 possible choices available in \emph{Trajectum}:

{\bf Ideal gas equation of state} $P = \alpha T^4$, with $\alpha$ a constant specifiable by the user.
This equation of state comes with constant user-specifiable values for the dimensionless ratios of transport coefficients:
\[
\frac{\eta}{s}, \ \ \frac{\zeta}{s}, \ \ \frac{\tau_\pi sT}{\eta}, \ \ \frac{\tau_\Pi sT}{\zeta}, \ \ \frac{\delta_{\pi\pi}}{\tau_\pi},
\]
\[
\phi_7P, \ \ \frac{\tau_{\pi\pi}}{\tau_\pi}, \ \ \frac{\lambda_{\pi\Pi}}{\tau_\pi}, \ \ \frac{\delta_{\Pi\Pi}}{\tau_\Pi}, \ \ \frac{\lambda_{\Pi\pi}}{\tau_\Pi}.
\]
Note that the dimensionless ratios for $\tau_\Pi$ and $\lambda_{\Pi\pi}$ differ from that used by the other equation of state and transport coefficients model, by factors $(1/3 - c_s^2)^2$ and $(1/3 - c_s^2)^{-1}$, respectively.
The reason for this is that for this particular equation of state, the speed of sound $c_s$ equals $\sqrt{1/3}$, which would cause $\tau_\Pi$ to be infinite, with similar issues for $\lambda_{\Pi\pi}$\@.

{\bf Lattice QCD/HRG hybrid equation of state} with temperature dependent shear and bulk viscosities.
In this model, the equation of state is a hybrid constructed from the HotQCD lattice equation of state, together with the hadron resonance gas (HRG) constructed from the chosen particle content \cite{Huovinen:2009yb,Bazavov:2014pvz,Bernhard:2018hnz}\@.
Here the analytical fit described in \cite{Bazavov:2014pvz} to the lattice equation of state is used instead of a tabulated form, and the transition between the two parts of the equation of state lies in the region 165--200 MeV\@. A consistent switch from hydrodynamics to a HRG is therefore possible for temperatures below 165 MeV, as in this region the EoS is constructed from the HRG (see also Section~\ref{sec:particlization}.
The user can also specify mostly the same dimensionless ratios of transport coefficients as in the ideal gas model, except for the bulk relaxation time $\tau_\Pi$ and the coefficient $\lambda_{\Pi\pi}$, for which we instead specify
\[
\frac{\tau_\Pi sT(1/3 - c_s^2)^2}{\zeta}, \quad \frac{\lambda_{\Pi\pi}}{\tau_\Pi(1/3 - c_s^2)}
\]
Another difference from the previous model is that we now allow for temperature dependence in the dimensionless ratios for both the shear and bulk viscosities, which are given by the following expressions:
\begin{align*}
\frac{\eta}{s} & = \begin{cases}
a + b(T - T_c)\left(\frac{T}{T_c}\right)^{c} & T \geq T_c, \\
0.06 & T < T_c,
\end{cases}\\
\frac{\zeta}{s} & = \frac{(\zeta/s)_\text{max}}{1 + \left(\frac{T - (\zeta/s)_{T_0}}{(\zeta/s)_\text{width}}\right)^2},
\end{align*}
with $T_c = 154\,\text{MeV}$, and $a = (\eta/s)_\text{min}$, $b = (\eta/s)_\text{slope}$, $c = (\eta/s)_\text{crv}$, $(\zeta/s)_\text{max}$, $(\zeta/s)_\text{width}$ and $(\zeta/s)_{T_0}$ user-specifiable parameters.
In the analysis presented in this work, the lattice QCD/HRG hybrid equation of state with temperature dependent viscosities was used.

In the Bayesian analysis, we varied all parameters governing the shear and bulk viscosities $\eta/s$ and $\zeta/s$, as well as the following three second order transport coefficients:
\[
\frac{\tau_\Pi sT(1/3 - c_s^2)^2}{\zeta}, \ \ \frac{\tau_\pi sT}{\eta}, \ \ \frac{\tau_{\pi\pi}}{\tau_\pi},
\]
We keep the remaining coefficients
\[
\frac{\delta_{\Pi\Pi}}{\tau_\Pi} = \frac{2}{3}, \ \ \frac{\lambda_{\Pi\pi}}{\tau_\Pi(1/3 - c_s^2)} = \frac{8}{5}, \ \ \frac{\delta_{\pi\pi}}{\tau_\pi} = \frac{4}{3},
\]
\[
\phi_7P = \frac{9}{70}, \ \ \frac{\lambda_{\pi\Pi}}{\tau_\pi} = \frac{6}{5}
\]
fixed to the kinetic theory values \cite{Denicol:2014vaa} in the limit of small particle masses, which is a common choice from earlier analyses\@.
In this work we made the choice to vary the two relaxation times, which most likely have the largest influence on the hydrodynamical evolution, as well as one additional coefficient $\tau_{\pi\pi}$ that does not vanish in the conformal limit. In future work it would be important to vary also other second order coefficients, but we note that varying more parameters can come at a significant computational cost.


\subsection{PDE solvers}
There are two choices available for the PDE solvers that solve the equations of motion \eqref{eq:eom}, \eqref{eq:intro:secondorderbulk} and \eqref{eq:intro:secondordershear}:

{\bf Finite difference} \cite{Luzum:2008cw}\@.
This method discretizes spatial derivatives by simply taking a symmetric finite difference.
An advantage of this method is that it is fast since it is simple and does not need a lot of computations.
A disadvantage, though, is that this method is not always stable, and can diverge in the presence of shocks or other large spatial gradients, especially if the shear viscosity is small.
Of course, in the setting of a Bayesian analysis, this is not a good trade-off, as we need to be able to generate theoretical predictions even for very `exotic' choices of parameters.

{\bf MUSCL} \cite{Du:2019obx,Bazow:2016yra}\@.
This method separates the equations of motion into a part describing conserved quantities, plus a source term.
The conserved part is then solved by examining the fluxes between each pair of neighboring grid points, where the left- and rightmoving fluxes are combined using the Kurganov-Tadmor algorithm \cite{Du:2019obx}\@.
Crucially, the fluxes are subsequently limited using the minmod flux limiter, which guarantees the stability under all circumstances.
For small gradients the flux limiting vanishes and the result of MUSCL agrees exactly.
Stability makes this method well-suited for our purposes, as it will produce an accurate result under all circumstances.
The price to pay for stability, though, is speed, as the MUSCL algorithm is %
more complicated.
The most computationally expensive part of using MUSCL is the computation of %
the velocity and energy density from the primary variables ($T^{\tau\tau}$, $T^{\tau x}$, $T^{\tau y}$, $\pi^{\mu\nu}$ and $\Pi$)\@.
This step is required five times more often in MUSCL as compared to finite difference in order to evaluate its flux limiting feature that makes the code stable in presence of shocks.
For this reason, we wrote a `fast' version of MUSCL that applies the same flux limiting procedure directly to the velocities instead of computing the velocities from the flux limited stress tensors.
In a regime in which the velocity can be approximated as a linear function of the stress tensor, this modified flux limiting procedure is identical to the original MUSCL\@.
We have also verified that this `fast' version is just as stable as the unmodified version, and that the difference between both results is negligible for our simulations.
In this work, we used the fast version of MUSCL\@.

\subsection{Particlization and hadronic phase}\label{sec:particlization}
For the particlization, only the Cooper-Frye procedure with viscous corrections as presented in \cite{Pratt:2010jt,Bernhard:2018hnz} is available.
This model finds the isotemperature surface of the switching temperature $T_\text{switch}$, and subsequently generates particles at that surface from a thermal distribution in the local restframe with temperature $T_\text{switch}$\@.
In the Bayesian analysis we vary this particlization temperature.
For a continuous evolution the equation of state must match the one derived from the particle content used in particlization, which in our implementation is only the case for the Lattice QCD/HRG hybrid equation of state. The other equation of state is implemented mostly for testing purposes and is not used in the results presented.

For the viscous corrections at each location, the spatial momenta $p_i$ of the sampled particles are subsequently rescaled by
\[
p_i \mapsto p_i + \sum_j[(\lambda_\text{shear})_{ij} + \lambda_\text{bulk}\delta_{ij}]p_j.
\]
Here, $(\lambda_\text{shear})_{ij} \propto \pi_{ij}$, where the proportionality constant, which is species-independent, is determined by demanding that to linear order the resulting HRG has the same shear tensor as the fluid from which $\pi_{ij}$ was taken.
Likewise, $\lambda_\text{bulk}$ is computed from a similar HRG computation, however the dependence of $\lambda_\text{bulk}$ on $\Pi$ is non-linear.
Lastly, the density of sampled particles is adjusted as a function of $\Pi$ in order to match the energy density of the fluid \footnote{Without bulk viscous corrections, the energy density would be automatically satisfied by sampling from a thermal distribution, but the $\lambda_\text{bulk}$ rescaling spoils this.}\@.
In this way, the entire stress-energy tensor is by construction on average continuous across the particlization surface.
The non-linearity of the dependence of $\lambda_\text{bulk}$ on $\Pi$ guarantees that large negative values for $\Pi$ cannot flip the sign of the momenta of the sampled particles.
It is important to note that this method is only one way to generate a continuous stress-energy tensor and for future work it would be interesting to verify that different methods do not strongly affect the results, such as e.g. studied in \cite{Everett:2020yty,Everett:2021ulz}.

The sampled particles subsequently undergo a hadronic cascade.
\emph{Trajectum} does not incorporate this hadronic cascade itself, but it does offer the flexibility to work with more than one possible code for this cascade, namely UrQMD \cite{Bass:1998ca,Bleicher:1999xi} and SMASH \cite{Weil:2016zrk,dmytro_oliinychenko_2020_3742965,Sjostrand:2007gs}\@.
These two codes require their inputs to be in slightly different formats, but importantly for the whole simulation, their resonance %
content is different.
This means that in order for \emph{Trajectum} to consistently interact with the hadronic cascade, it must use the exact same resonance content, to indeed ensure continuity across the particlization surface.
In this work, we have solely used SMASH, as that contains a few more resonances as compared to UrQMD\@.

\subsection{Validation of \emph{Trajectum}}
Because \emph{Trajectum} is a new code, it is important to validate its results.
We performed several checks to verify the correct functioning of the code:

\begin{itemize}
\item To test the equations of motion, we compared the numerical computation of Gubser flow to the analytically known result \cite{Gubser:2010ze,Gubser:2010ui,Marrochio:2013wla}\@.
This leads to good agreement, but in Gubser flow there is no bulk viscosity, and there is also little dependence on second order coefficients.
For this reason we performed an additional check, by comparing the implementation of the equations of motion for generic events to an independently performed computation in Mathematica.
The two computations agree to machine precision.
\begin{figure}[ht]
\includegraphics[width=0.8\columnwidth]{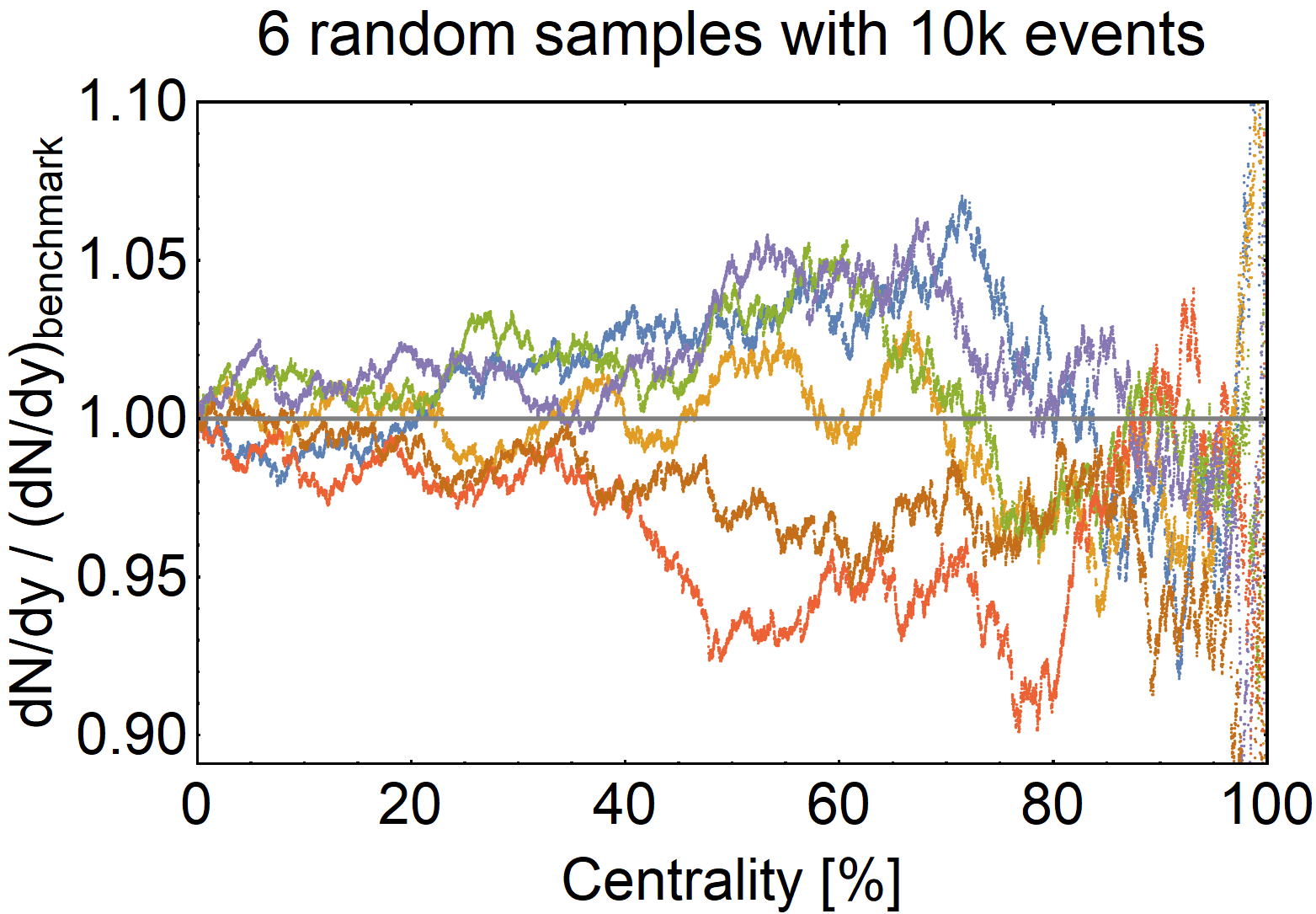}
\caption{\label{fig:centralitymult}Particle multiplicity where centrality has been determined from 10k events, relative to the true distribution from which the samples were obtained.}
\end{figure}
\item The particlization procedure we use is designed so that on average the stress-energy tensor is continuous across the particlization surface.
We checked that this is indeed true to good precision, by oversampling the number of produced particles, producing a statistically significant sample.
\item We compared various quantities such as the initial state eccentricities $\epsilon_n$ between our implementation of T\raisebox{-0.5ex}{R}ENTo and the original version \cite{TrentoGithub}, finding good agreement.
\item We computed the observables from \cite{Bernhard:2019bmu} using the Maximum a Posteriori (MAP) values for the parameters specified there.
The results from \emph{Trajectum} agree well with theirs.
\end{itemize}

\subsection{Centrality classes}\label{sec:centralities}

\begin{figure*}[ht]
\centering
\includegraphics[width=\textwidth]{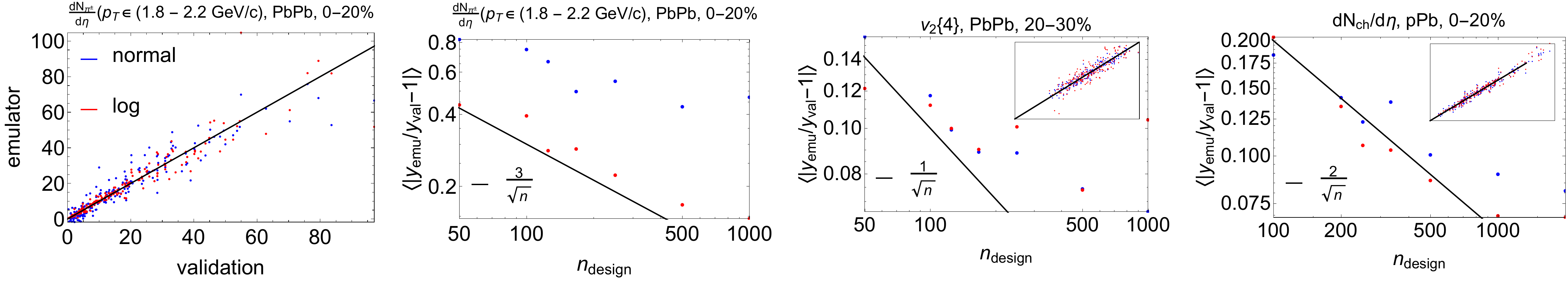}
\caption{\label{fig:emulator}
(left) Emulator prediction shown against model output (validation) for charged pion multiplicity in the $1.8 - 2.2\,$GeV $p_T$-bin for the 0 - 20\% centrality class using 1000 design points. When emulating the logarithm of the yield (labelled log, red) the accuracy of the emulator is much improved. The right three plots compare the emulator error, defined as $\langle |y_{\rm emu}/y_{\rm val} -1|\rangle$ versus the number of design points for the left plot, $v_2\{4\}$ at $20-30\%$ centrality and $dN_{\rm ch}/d\eta$ in $p$Pb at $0-20\%$ centrality respectively (validation versus emulator results are shown as insets). For $v_2\{4\}$ the statistical error becomes dominant at about 250 design points and the log scaling does not improve. For $p$Pb the emulator error is relatively large, which is the reason why this work uses 2000 design points for $p$Pb.}
\end{figure*}

A common but relatively large problem is the determination of centrality classes. When using a finite number of minimum bias events ($\mathcal{O}(10^4)$) the boundary between centrality classes can easily be the dominant error, especially for observables that depend strongly on centrality. This is illustrated in \fig{fig:centralitymult}, where we show the ratio of the multiplicity of six random samples to the `true' distribution (taken from Section~\ref{sec:map}, we also used this distribution to draw the random samples). For a realistic simulation using 10k minimum bias events this results in errors up to 10\%, which is especially unfortunate as this particular observable has low experimental uncertainties. 

For the theoretical centrality determination it is however possible to use a trick to resolve this statistical problem without running extra hydrodynamic simulations.
Firstly, 
it is known that for PbPb the initial entropy correlates very well with final state multiplicity.
In turn, the initial entropy correlates very well with the spatial integral of the initial thickness function in \trento{}, for which one does not need to perform the computationally expensive free streaming stage.
The way in which one can use these facts is to generate a large number $\mathcal{O}(10^5)$ of \trento{} simulations, and in this way generate an accurate distribution of initial thickness functions.
We then sort the simulations into bins, and we save the upper and lower initial thickness functions for each bin.
Subsequently, for each bin we run the entire hydrodynamical evolution for some fixed number of events with initial thickness function within the bin boundaries.
In this way, we guarantee that for these events the initial thickness functions are distributed the same way as for the large set described above.
Since the initial thickness function correlates well with the final multiplicity, we generate final multiplicities with the correct distribution as well, and hence get a determination of centrality as if we had used $\mathcal{O}(10^5)$ events.
In the limit of a large number of events, the distribution still converges to the true distribution as long as the number of initial condition simulations is larger than the number of full hydrodynamics computations.
The only difference is that it converges to the true distribution faster than without this optimization, so the resulting events are still minimal bias.
In particular, since our centrality determination is based on final state multiplicity this means that even if the final state multiplicity does not correlate well with the initial thickness function the final distribution will still converge to the true distribution, albeit at a slower rate than with a strong correlation and our improved algorithm.

One can also use this method to generate weighted samples, by simply generating more events for the desired bins.
For our $p$Pb simulations, we bias the selection towards higher multiplicities, and subsequently weigh events in the analysis accordingly.
In this way we obtain better statistics for (relatively rare) higher multiplicity events.

A related, but separate issue is the experimental centrality selection. In our boost invariant simulations our only choice is to bin events according to their multiplicity at mid-rapidity. Experimentally, however, centrality classes are often selected by binning at relatively high rapidity (ALICE often uses the V0 detectors, located at pseudorapidities $-3.7$ up to $-1.7$ and at $2.8$ up to $5.1$). For PbPb collisions this bias is not so severe, since the V0 amplitude correlates strongly with multiplicity at mid-rapidity (see Fig.~15 in \cite{Abelev:2013qoq}), but for $p$Pb this can introduce a serious bias \cite{Adam:2014qja}. Ideally we would compare to experimental data that also determines centrality at mid-rapidity, but unfortunately this is rarely available. This is part of the reason why we chose to independently fit the normalizations in PbPb and $p$Pb collisions.

\section{Results}\label{sec:results}

\begin{figure}[h]
\includegraphics[width=0.49\textwidth]{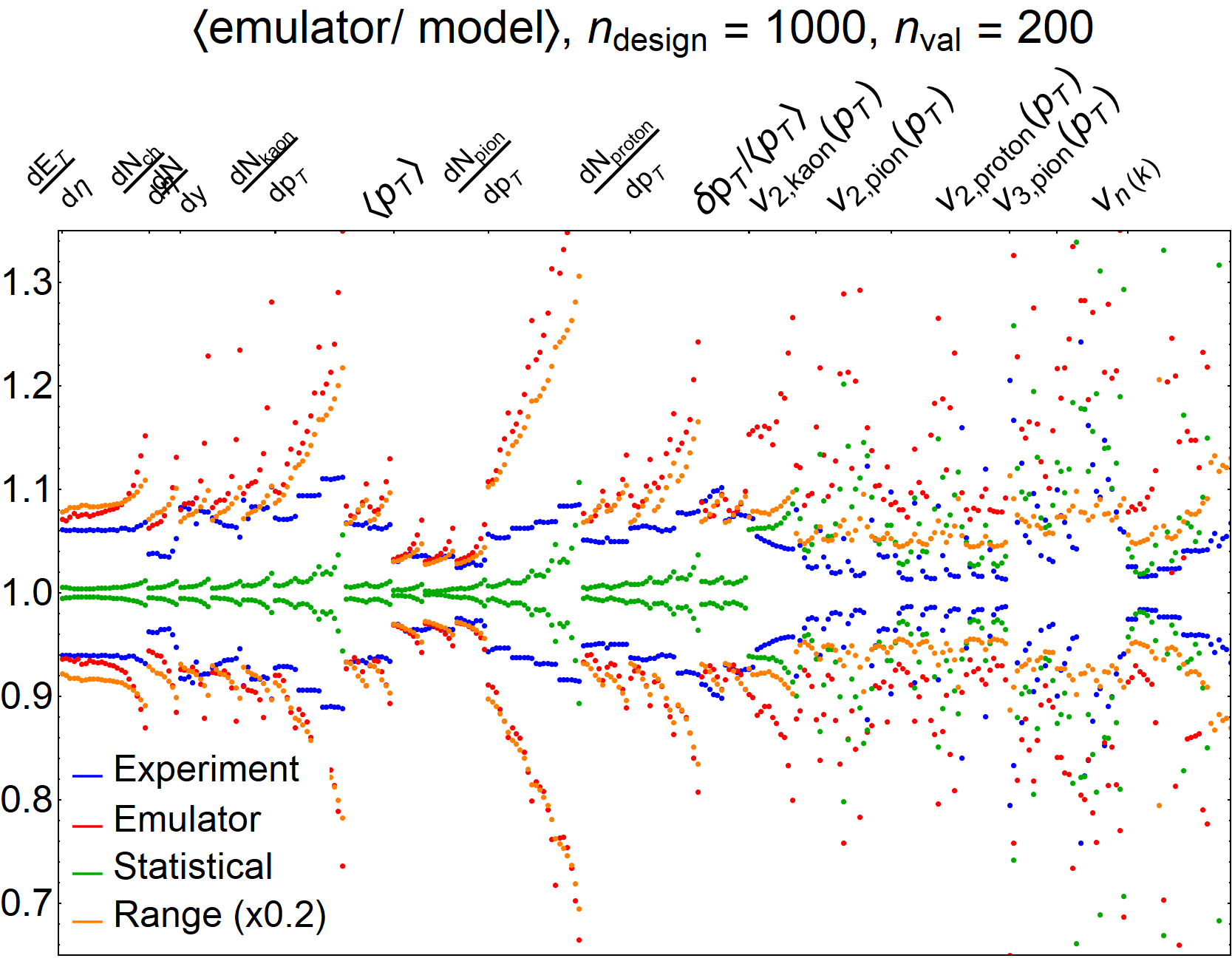}
\includegraphics[width=0.49\textwidth]{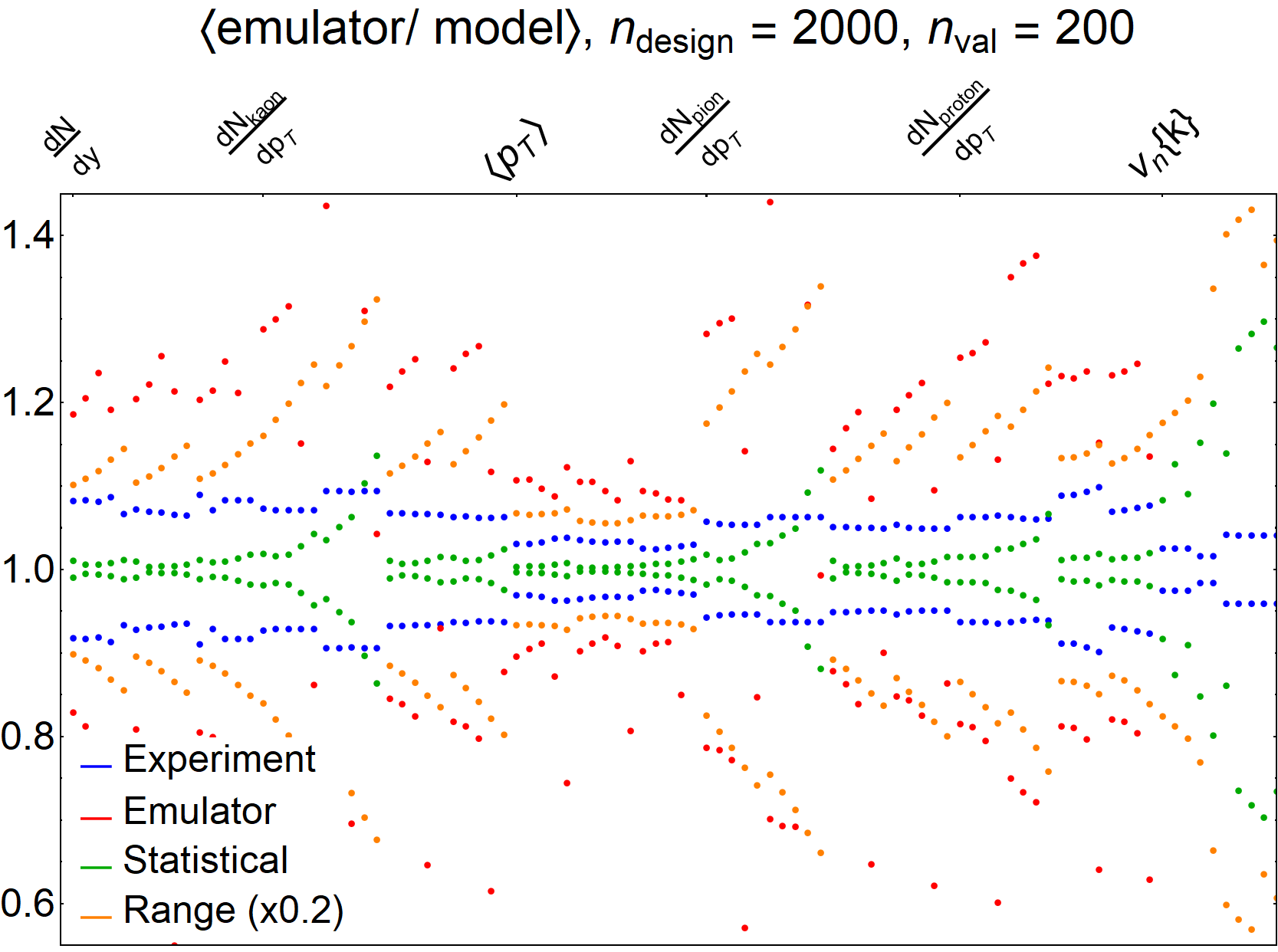}
\caption{\label{fig:validation} We show the validation of our emulator for PbPb (top) and $p$Pb (bottom) systems in red, whereby we show the average of the emulator discrepancy with the true model result for 200 validation points (red). Experimental uncertainties are shown and often of the same order as the emulator uncertainty, which in turn is much smaller than the parameter range probed (orange). Statistical model uncertainty is dominant for a few statistically difficult observables (green). The $p$Pb system is much harder to emulate, as also found in \cite{Moreland:2019szz}\@.}
\end{figure}

\subsection{Emulator}\label{sec:emulator}

As our aim is to perform a global analysis of heavy ion collisions depending on all parameters specified in Section \ref{sec:trajectum} it is necessary to obtain a large sample of events for many points in this parameter space. Even for a single parameter point it is however computationally expensive to obtain sufficient statistics for interesting observables, and it is therefore standard practice to evaluate the model at a number $n_{\rm design}$ of `design' points in the parameter space and employ an emulator to interpolate those results to any point in parameter space. These design points are typically distributed on a latin hypercube with reasonable (physical) limits on the parameter range \cite{stein1987large, Bernhard:2018hnz}, that in our case will equate with the range of prior probabilities.

To further reduce computational time we decided to leave the nucleon-nucleon cross section $\sigma_{\rm NN}$ as a parameter. The major advantage of doing so is that a single scan over the design points can be used to compare to experimental data from several colliding energies. The addition of one extra parameter is much more efficient than performing two separate scans. Each energy will have its own normalization and cross section, whereby we can choose to either fix the cross section to its independently measured proton-proton value, or we can leave it up to the Bayesian analysis to estimate the cross section. %
In the current work we fix $\sigma_{\rm NN} = 63\, (70)\,$mb for collisions at $\sqrt{s_{\rm NN}} = 2.76\, (5.02)\,$TeV.

For the emulator we train a Gaussian Process Emulator with a squared exponential (a Gaussian) plus a noise kernel \cite{williams2006gaussian,Bernhard:2018hnz,Moreland:2019szz}, as implemented in the scikit-learn library for python.
Even though the emulator itself estimates its own uncertainty it is important to validate its output, which is shown in \fig{fig:emulator} for three representative observables. The uncertainty of the emulator is not shown for clarity, but corresponds well with the actual validation error (equal to the deviation from the black diagonal line). For some observables such as transverse momentum spectra the spread of points is such that for points with small values a small emulator error can make a large relative change. In these situations it is often better to train the emulator on the logarithm of the datapoints, and afterwards exponentiate the emulator results. This is shown as the `normal' versus `log' result in \fig{fig:emulator}, and gains can be as large as a factor three reduction in emulator error. In this work we decided to designate $dN/d\eta$, $dE_T/d\eta$ and the transverse momentum spectra of pions, kaons and protons as observables on which to perform the log transformation.

For an accurate emulation with limited computing resources it is important to balance the number of design points versus the number of events per design point. For some observables, such as the $v_2\{4\}$ shown in \fig{fig:emulator} (The observable $v_n\{k\}$ is defined in Section \ref{sec:mapanisotropicflow}.), statistics is often the limiting factor and increasing the number of design points does not improve the emulator uncertainty. For statistically easier observables, such as the multiplicity, the emulator error decreases as approximately $1/\sqrt{n_{\rm design}}$ for  longer. Secondly, we noticed that $p$Pb observables are more difficult to emulate. For these reasons we chose to model PbPb and $p$Pb with 1000 and 2000 design points, using 6k and 40k events per design point respectively.

To emulate the full dataset it is important to perform a linear transformation on the datapoints, as these are often highly correlated. This is done using a Principal Component Analysis (PCA), which extracts the few Principal Components (PCs) that contain independent information. In our dataset we found that the first 25 PCs per colliding system suffice, capturing over 97\% of the variation in the data. An important criterion on the number of PCAs to use is the estimated (statistical) noise found by the emulator training. For our case only from the 9th PC onward the noise levels can surpass 10\%, and only at the 22nd and 24th the noise was dominant (over 40\%), which led us to conclude that 25 PCs is sufficient to capture all non-trivial information. Performing the PCA also automatically gives a good estimate of theoretical correlations among the data points, which will be important later (the log transformation above also needs to be applied to the correlation matrix).

After performing the PCA it is again necessary to validate the emulator, which is shown in \fig{fig:validation} for PbPb (top) and $p$Pb (bottom) in red. This is estimated by taking the model values $m$ and emulator predictions $e$ for the same 200 validation points for all our observables, where the red points then correspond to $\langle e/m \rangle \pm s$, with $s$ the standard deviation of $e/m$\@. For comparison we also show the experimental error of the data point (blue), the average statistical error of our model (green) and the range of model predictions given our prior range, defined as the standard deviation over the mean (orange). Ideally the emulator uncertainty should be smaller than the experimental uncertainty, but this is only the case for a  few observables and adding further design points is hence likely to improve our results. The statistical uncertainty is only dominant for statistically difficult observables, such as $p_T$-differential $v_2$, and can be improved by running design points with more events. Lastly, we find that the range of model output is often roughly five times the emulator uncertainty, which gives a good discriminating power. As is clear from the plots the emulator performs worse for $p$Pb collisions than for PbPb collisions (consistent with \cite{Moreland:2019szz}), which is the reason why we ran $p$Pb with 2000 design points and used only 1000 design points for PbPb.

\subsection{Parameter dependence of observables}

\begin{figure*}[ht]
\includegraphics[width=0.99\textwidth]{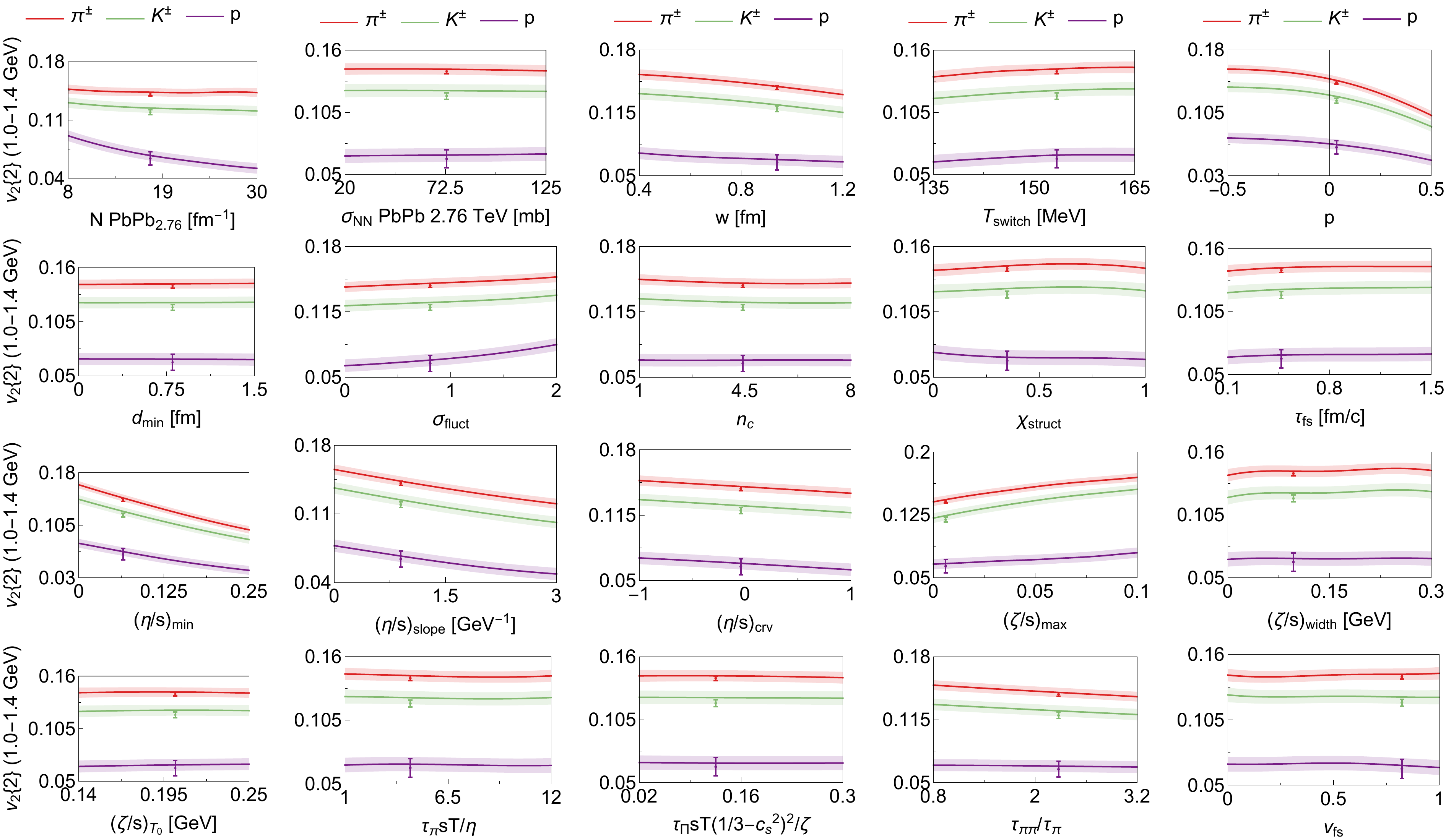}
\caption{\label{fig:emuv2}We show $\vtwo{}$ for the $1.0 - 1.4\,$GeV $p_T$-bin for pions, kaons and protons in the $20 - 30\%$ centrality class as a function of our model parameters. All parameters are fixed to the optimal point found in \cite{Bayesianshort}, with the exception of the varying parameter displayed. The relevant datapoints \cite{Adam:2016nfo} are displayed at its optimal point. As expected the elliptic flow decreases as the shear viscosity is increased. Similarly strong dependencies are seen for the \trento{} parameter $p$ (affecting the initial geometric anisotropy, see \eqref{eq:trento}) and for protons the norm and fluctuations $\sigma_{\rm fluct}$ are important.}
\end{figure*}

\begin{figure*}[ht]
\includegraphics[width=0.99\textwidth]{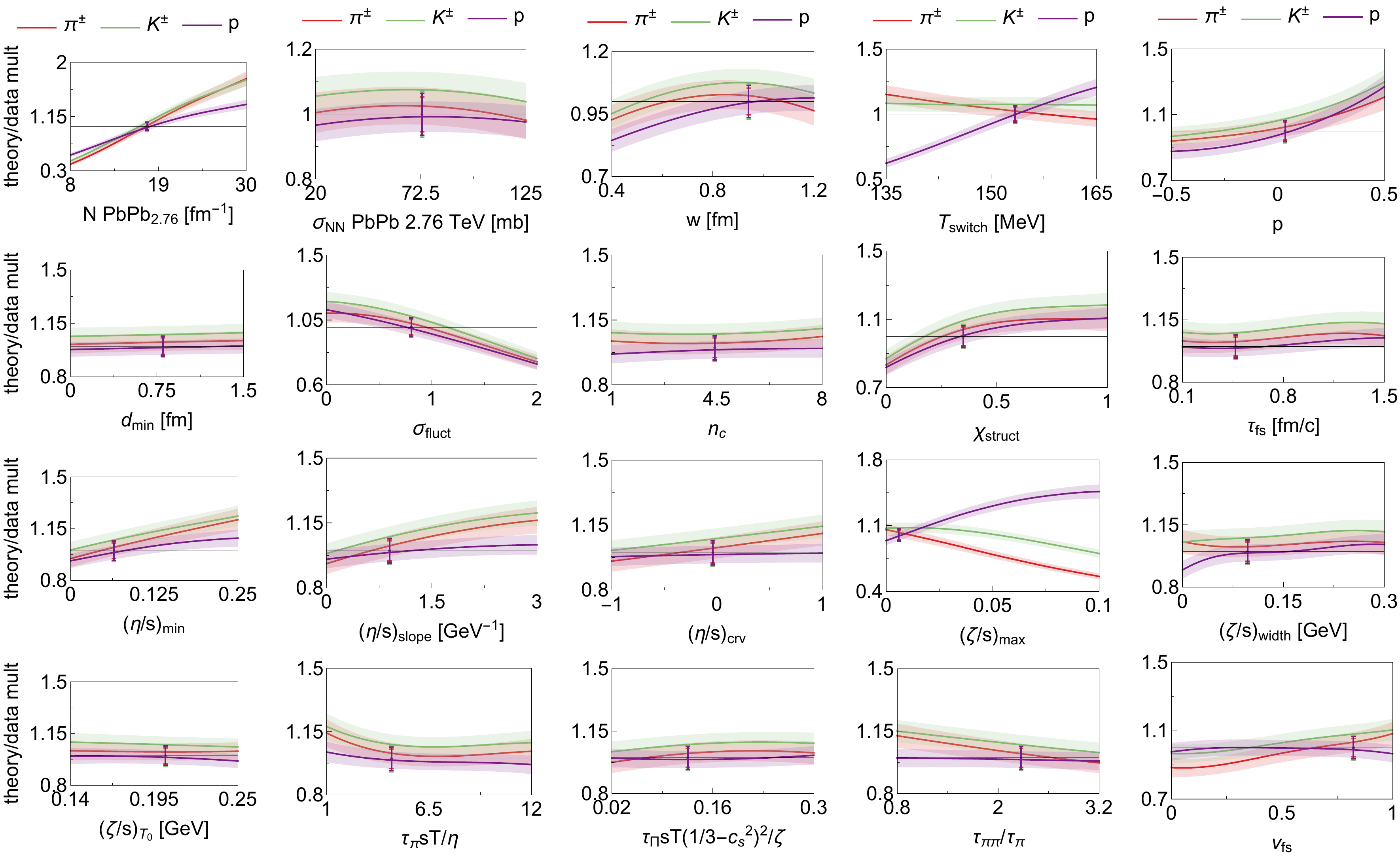}
\caption{\label{fig:emuspec} Similar to \fig{fig:emuv2} we show the multiplicity for the $1.0 - 1.4\,$GeV $p_T$-bin for pions, kaons and protons in the $20 - 30\%$ centrality class as a function of our model parameters  normalized to the relevant datapoints \cite{Abelev:2013vea}. As expected the multiplicity depends strongly on the norm. Strong dependencies for the proton to pion ratio are seen for the switching temperature (as expected) as well as the bulk viscosity.}
\end{figure*}

\begin{figure}[ht]
\includegraphics[width=0.49\textwidth]{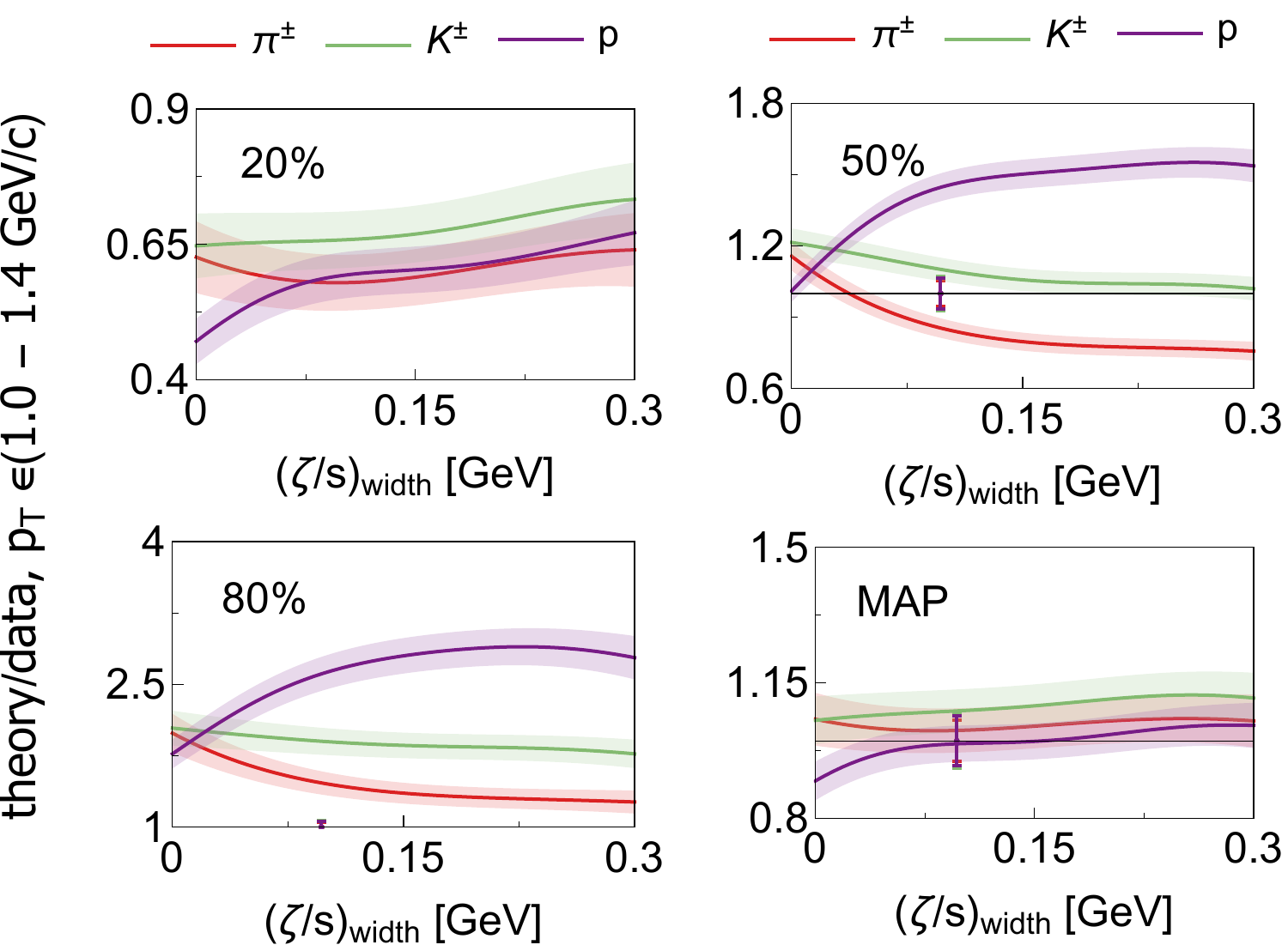}
\caption{\label{fig:bulkwidthexample}To illustrate the interaction of the norm and width of the bulk viscosity we fix all parameters at respectively 20\%, 50\%, 80\% of their range and at the MAP value. Clearly, when the norm is at 80\% there is a much stronger dependence on the width. Only if either the norm or the width are close to zero it is possible to simultaneously fit the pion, kaon and proton multiplicities in this intermediate $p_T$-bin.}
\end{figure}

\begin{figure}[ht]
\includegraphics[width=0.49\textwidth]{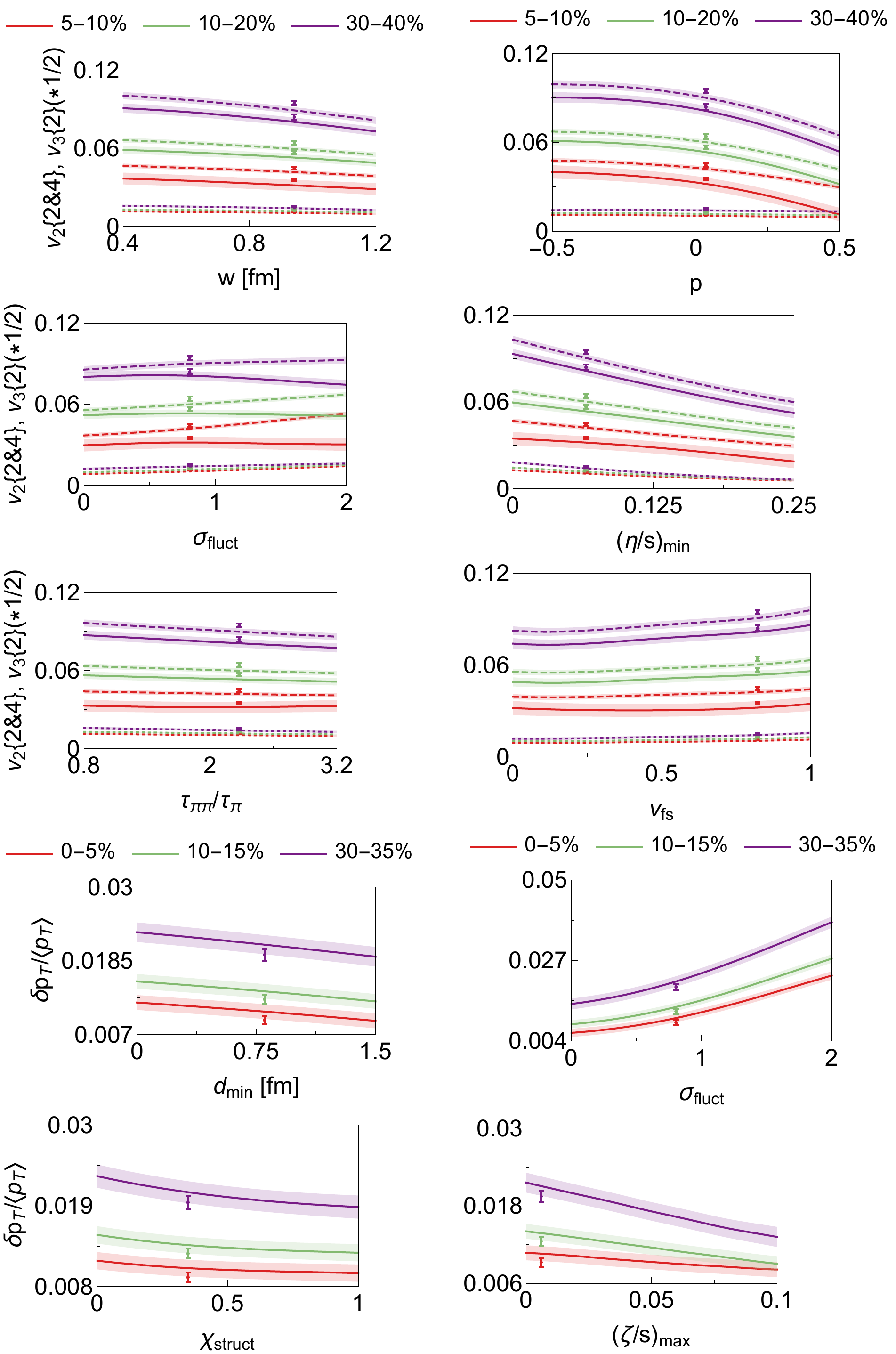}
\caption{\label{fig:vnkemulatorcentrality}To illustrate the centrality dependence more clearly we show $\vtwofour{}$ (solid), $\vtwo{}$ (dashed) and $v_3\{2\}$ (dotted) coefficients (top) and $\delta p_T$ fluctuations (bottom) for several centrality classes as a function of representative parameters for 2.76 TeV PbPb collisions. There is a clear increase in both $\delta p_T$ as well as the difference between $\vtwofour{}$ and $\vtwo{}$ when increasing fluctuations through $\sigma_{\rm fluct}$.}
\end{figure}

\begin{figure}[ht]
\includegraphics[width=0.49\textwidth]{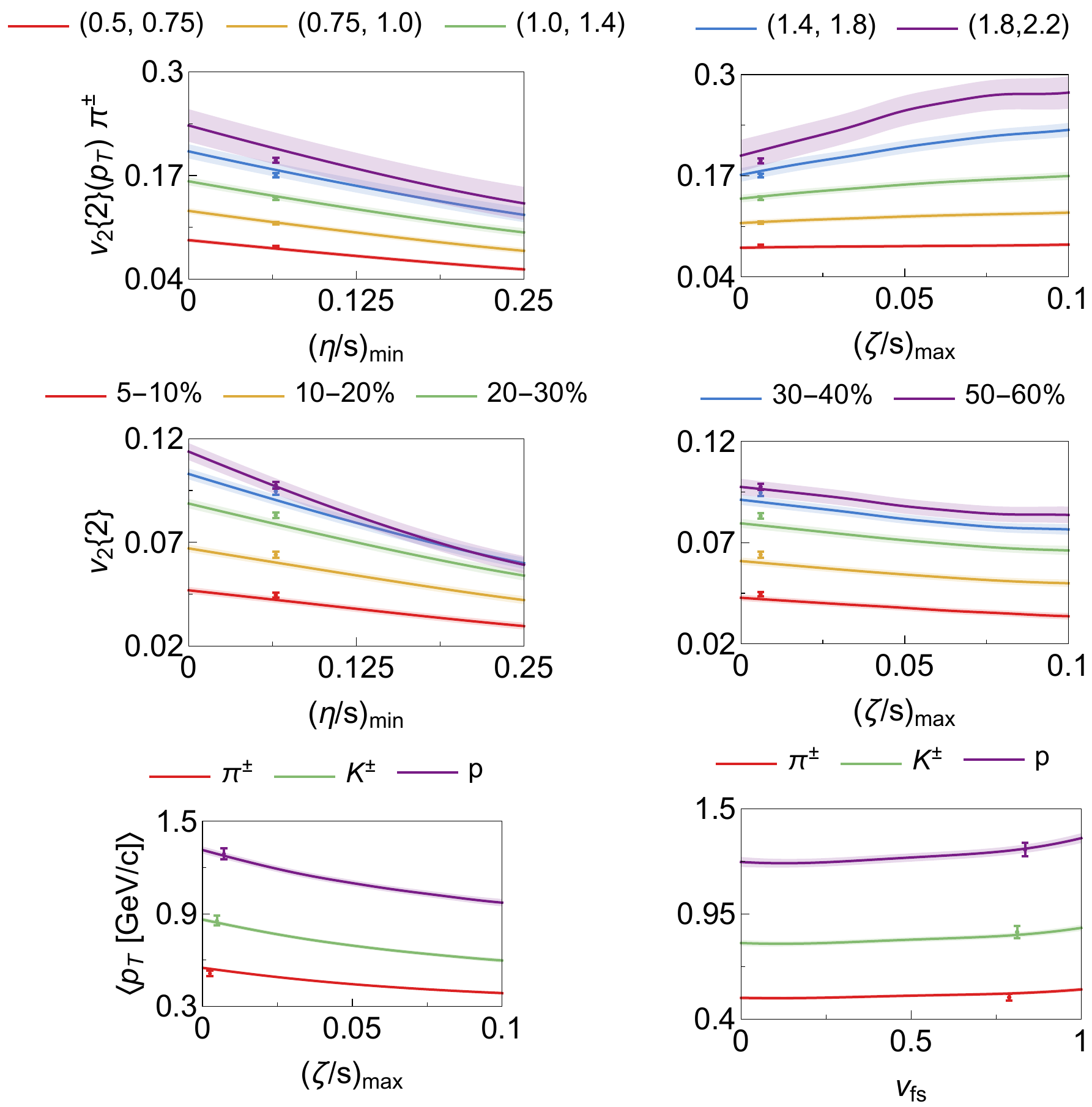}
\caption{\label{fig:emuvnvisc}We show the $p_T$-differential $v_2\{2\}$ for five $p_T$ bins as a function of the shear and bulk viscosity for the 20-30\% centrality class (top), as well as the integrated $v_2\{2\}$ for several centrality classes (middle). Quite interestingly the bulk viscosity causes an increase of $v_2\{2\}$ in all $p_T$ bins, even though the integrated $v_2\{2\}$ decreases. This can be understood by the fact that the mean $p_T$ decreases as a function of the bulk viscosity (bottom left). We also note an increase in mean $p_T$ as a function of $v_{\rm fs}$ (bottom right).}
\end{figure}

Once we have a fully trained emulator we can evaluate the emulator on a set of parameters and immediately obtain all observables, reducing the computational time from days to fractions of a second. As we have 514 observables as a function of a 20-dimensional parameter space it is impossible to display all this information here, but in \fig{fig:emuv2} and \fig{fig:emuspec} we respectively show the representative $v_2\{2\}$ and multiplicity in the intermediate $p_T$-bin of $1.0 - 1.4\,$GeV for pions, kaons and protons in the $20 - 30\%$ centrality class. All panels use the optimal parameters found in \cite{Bayesianshort}, with the exception of the parameter being displayed. The experimental data point is shown at the optimal value for the parameter itself, such that at this point all curves agree exactly with each other and are indeed also close to the actual datapoint.

A few things are as qualitatively expected: $\vtwo{}$ decreases as we increase either $\etamin{}$ or $\etaslope{}$. $\vtwo{}$ depends sensitively on the \trento{} parameter $p$. The multiplicity depends strongly on the norm (but perhaps surprisingly less so for protons) and the proton to pion ratio increases for higher $T_{\rm switch}$. As also expected the dependence on second order transport coefficients is much less pronounced, and it is not possible to extract a preferred value by just looking at these plots; we will come back to a global analysis shortly. We also see a strong dependence on the bulk viscosity through $\zetamax$ on both the particle ratios as well as $\vtwo{}$. This is partly due to the influence of the bulk viscosity on hydrodynamics, and partly due to its influence on the stress tensor at the moment of particlization (see \cite{Everett:2020yty} for a more detailed study on the influence of particlization; we use what they call the PTB prescription). In the Appendix we show similar results for $p$Pb collisions, this time showing the elliptic flow coefficient and the mean kaon transverse momentum for several centrality classes. As already mentioned emulator uncertainty is larger for this smaller system, but it is still clear that dependence on our parameters is increasingly non-trivial as we go to these smaller systems.

Of course the sensitivity of observables on a parameter may (strongly) depend on the location in the parameter space. From Fig.~\ref{fig:emuv2} and \ref{fig:emuspec} it may for instance seem that observables are rather insensitive to the width of the bulk viscosity. This, however, is due to the low value of the norm for the bulk viscosity for the optimal parameters, which in turn (physically) means there is little to no effect of $\zetawidth$. Note though that this (physical) correlation is not explicitly put in for the emulator, and even for $\zetamax = 0$ some dependence on the width is still expected, at least within the error band of the emulator. All this can be clearly seen in \fig{fig:bulkwidthexample}, which shows the sensitivity of the number of pions, kaons and protons with $p_T \in [1.0 - 1.4]\,\text{GeV}$ to $\zetawidth$ where all parameters are chosen to lie at 20\%, 50\%, 80\% of the prior range or at the MAP point respectively. In this case we hence have that $\zetamax = (0.06, 0.15, 0.24, 0.0060)$. Clearly the largest variation is present for $\zetamax = 0.24$, which incidentally also has the highest overall yield since the norm $N$ is then also fixed at 80\%. The figure also shows that unless either $\zetamax$ or $\zetawidth$ are close to zero it is hard to obtain simultaneous agreement with the experimental data points for all identified particles.

Some observables have a strong centrality dependence. For this reason we show in \fig{fig:vnkemulatorcentrality} the anisotropic flow coefficients $v_2\{2\}$, $v_2\{4\}$ and $v_3\{2\}$ as well as the transverse momentum fluctuations for three different centrality classes as a function of a few representative parameters.
As can be expected the transverse momentum fluctuations depend sensitively to the fluctuating \trento{} parameter $\sigma_{\rm fluct}$ (see \ref{sec:initialconditions})\@. A large bulk viscosity on the other hand reduces the $p_T$ fluctuations. The difference between $\vtwofour{}$ and $\vtwo{}$ increases strongly as $\sigma_{\rm fluct}$ increases, especially for central collisions. This is in agreement with the statistical interpretation for the difference between these two estimates of elliptic flow~\cite{Miller:2003kd}.

Lastly, it can be surprising that the $v_2\{2\}$ in the $(1.0 - 1.4)\,$GeV bin in \fig{fig:emuv2} increases as a function of the bulk viscosity, whereas from the dissipative character of transport coefficients the viscosity is expected to isotropize the plasma. In \fig{fig:emuvnvisc} we show this in more detail, both for the shear and bulk viscosity, and indeed for all $p_T$ bins the $v_2\{2\}$ increases, but nevertheless the integrated $v_2\{2\}$ decreases. This can be understood by the fact that the $v_2\{2\}$ is larger for larger tranverse momentum and the fact that the bulk viscosity reduces the mean transverse momentum (\fig{fig:emuvnvisc} bottom).

\subsection{Bayesian posterior estimate}\label{sec:mcmc}

After having obtained an emulator and hence model predictions as a function of our input parameters the next step is to estimate the probability distributions for our parameter space $\boldsymbol{x}$. Using Bayes' theorem this is done by firstly assuming a prior probability distribution, which is typically flat within a given (physically motivated) range and zero outside this range, and by secondly reweighing this distribution taking into account new evidence from comparing the model output with some set of experimental data. Mathematically, this amounts to a posterior distribution given by
\begin{equation}
    \mathcal{P}(\boldsymbol{x}|\mathbf{y}_{\exp})
    = \frac{e^{-\Delta^2/2}}{\sqrt{(2\pi)^{n} \det\left(\Sigma(\boldsymbol{x})\right)}} \mathcal{P}(\boldsymbol{x}) 
    \label{eq:bayes}
\end{equation}
with $\mathcal{P}(\boldsymbol{x})$ the (flat) prior probability density and where
\begin{equation}
    \Delta^2
    = \left(\mathbf{y}(\boldsymbol{x})-\mathbf{y}_{\rm exp}\right)\cdot \Sigma(\boldsymbol{x})^{-1} \cdot \left(\mathbf{y}(\boldsymbol{x})-\mathbf{y}_{\rm exp}\right),
    \label{eq:delta}
\end{equation}
with $\mathbf{y}(\boldsymbol{x})$ the predicted data for parameters $\boldsymbol{x}$, $\mathbf{y}_{\rm exp}$ the $n$ experimental data points and $\Sigma(\mathbf{x}) = \Sigma_{\rm emu}(\mathbf{x}) + \Sigma_{\rm exp}$ is the sum of the emulator and experimental covariance uncertainty matrices. For the emulator this follows from the correlations in the theoretical model, as determined by a Principal Component Analysis (PCA)\@. In principle the experimental matrix should be provided by the relevant experiment, but as this is rarely available we follow the simple prescription from \cite{Bernhard:2018hnz}. Numerically it is often more convenient to work with the logarithm of \eqref{eq:bayes}, which we will refer to as the log-likelihood (LL) and we note that \eqref{eq:bayes} should still be normalized.

\begin{figure*}[ht]
\includegraphics[width=\textwidth]{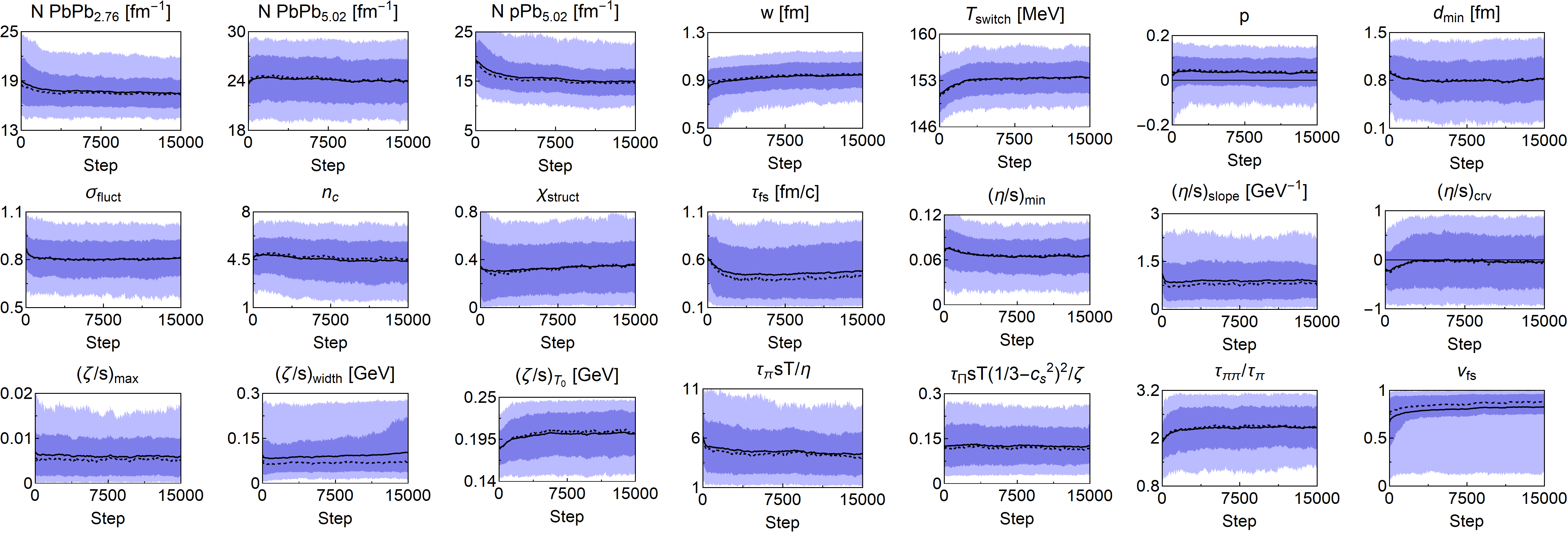}
\caption{\label{fig:mcmcmean}Mean (solid) and median (dashed) combined with $68\%$ and $95\%$ confidence intervals (shaded dark and light respectively), computed over all 600 walkers as a function of the step number. After about 9k steps the distributions are converged (apart from some statistical fluctuations) and the remainder of the chain points can be used as an approximation for the true posterior distribution.}
\end{figure*}

For the experimental data points we use the same set of 514 observables as in \cite{Bayesianshort}, which built upon \cite{Bernhard:2018hnz} and \cite{Moreland:2019szz}. These include 
PbPb charged particle multiplicity $dN_{ch}/d\eta$ at 2.76 \cite{Aamodt:2010cz} and 5.02 TeV \cite{Adam:2015ptt}, transverse energy $dE_T/d\eta$ at 2.76 TeV \cite{Adam:2016thv}, identified yields $dN/dy$ and mean $p_T$  for pions, kaons and protons at 2.76 TeV \cite{Abelev:2013vea}, integrated anisotropic flow $v_n\{k\}$ for both 2.76 and 5.02 TeV \cite{Adam:2016izf} and $p_T$ fluctuations \cite{Abelev:2014ckr} at 2.76 TeV\@. All of these use centrality classes of width 2.5\% (for central transverse energy) up to 20\% (for peripheral $v_n\{k\}$)\@.
On top of this we added identified transverse momentum spectra using six coarse grained $p_T$-bins separated at $(0.5, 0.75, 1.0, 1.4, 1.8, 2.2, 3.0)\,$GeV both for  PbPb at 2.76 \cite{Abelev:2013vea} and $p$Pb at 5.02 TeV \cite{Adam:2016dau}, anisotropic identified flow coefficients using the same $p_T$ bins (statistics allowing) at 2.76 \cite{Adam:2016nfo} and 5.02 TeV \cite{Acharya:2018zuq}. As in \cite{Moreland:2019szz} we use $\tilde v_n\{k\}$ anisotropic flow coefficients for $p$Pb at 5.02 TeV \cite{Aaboud:2017acw}. 
Since in $p$Pb $v_n\{k\}$ can become imaginary we here use $\tilde v_n\{k\} \equiv {\rm sgn}(v_n\{k\}^k) |v_n\{k\}|$, which equals $v_n\{k\}$ when it is real. 
For $p$Pb we furthermore compare to the mean $p_T$ for pions, kaons and protons at 5.02 TeV \cite{Abelev:2013haa}. Here we typically use the $0-5\%$, $5-10\%$, $10-20\%$ up to $50-60\%$ centrality classes, except for the $p$Pb flow coefficients, where high multiplicity events are specifically interesting. In that case we follow \cite{Moreland:2019szz} and use five multiplicity classes, separated according to charged particle multiplicity over the mean particle multiplicity. With the bias introduced in subsection \ref{sec:centralities} this was possible with reasonable statistics up to the bin with multiplicity $4$ to $5$ times the average particle multiplicity. In total this leads to 418 and 96 datapoints for PbPb and $p$Pb collisions respectively.

To compute the actual posterior distributions we use a Markov Chain Monte Carlo, which produces a chain of parameters with a distribution that converges to the true distribution. For this we use the EMCEE code \cite{ForemanMackey:2012ig}, with
a chain of 600 walkers initialized using a uniform distribution and subsequently evolved for 15k steps.
It is important to run the computation for a sufficient number of steps to allow the distribution to converge to the true distribution.
\fig{fig:mcmcmean} shows the mean and median over all walkers for all parameters as a function of step number.
The $68\%$ and $95\%$ confidence intervals are shaded, whereby we note that we take these intervals symmetrically with for the second region 2.5\% probability left on either side of the distribution (see \cite{Bernhard:2018hnz} for an alternative definition that minimizes the width of the distribution).
For the distribution to have converged it is necessary that not only the mean but also the entire distribution has converged, apart from statistical fluctuations.
As can be seen in \fig{fig:mcmcmean}, some parameters satisfy this condition faster than others.
For example, $\sigma_\text{fluct}$ converges quite fast, while $w$ only ends its downward trend after around 9k steps, so that for all analyses only the part of the chain after this point should be used.

\begin{figure*}[ht]
\includegraphics[width=\textwidth]{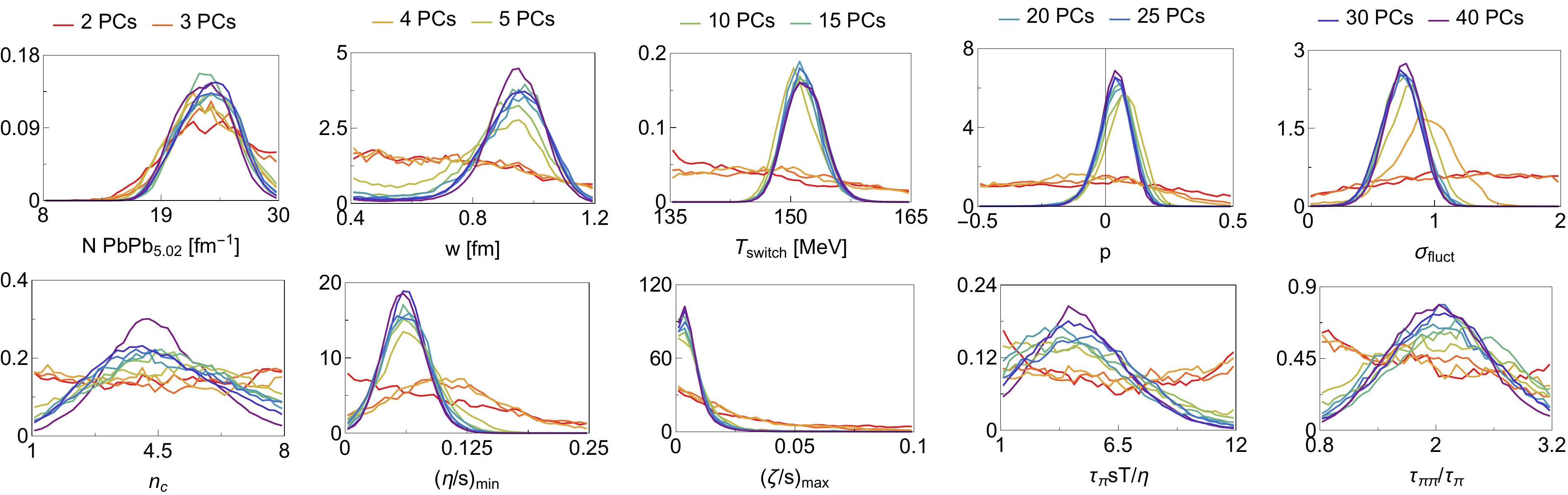}
\caption{\label{fig:changepcas}For 10 representative posteriors we show the influence of including a varying number of Principal Components (PCs) for the posteriors of PbPb at 2.76 and 5.02 TeV. Some parameters converge more smoothly than others, but from 10 PCs onward the precision gains are modest.}
\end{figure*}

As another test on the convergence of the chains we varied the number of PCs in Fig.~\ref{fig:changepcas}, shown for PbPb at 2.76 and 5.02 TeV only. For some parameters the posterior changes quite suddently, for instance the switching temperature is clearly most constrained by the 5th PC, after which not much precision is gained anymore. Throughout this work we chose to show results using 25 PCs.

\subsection{Closure tests}

\begin{figure*}[ht]
\includegraphics[width=\textwidth]{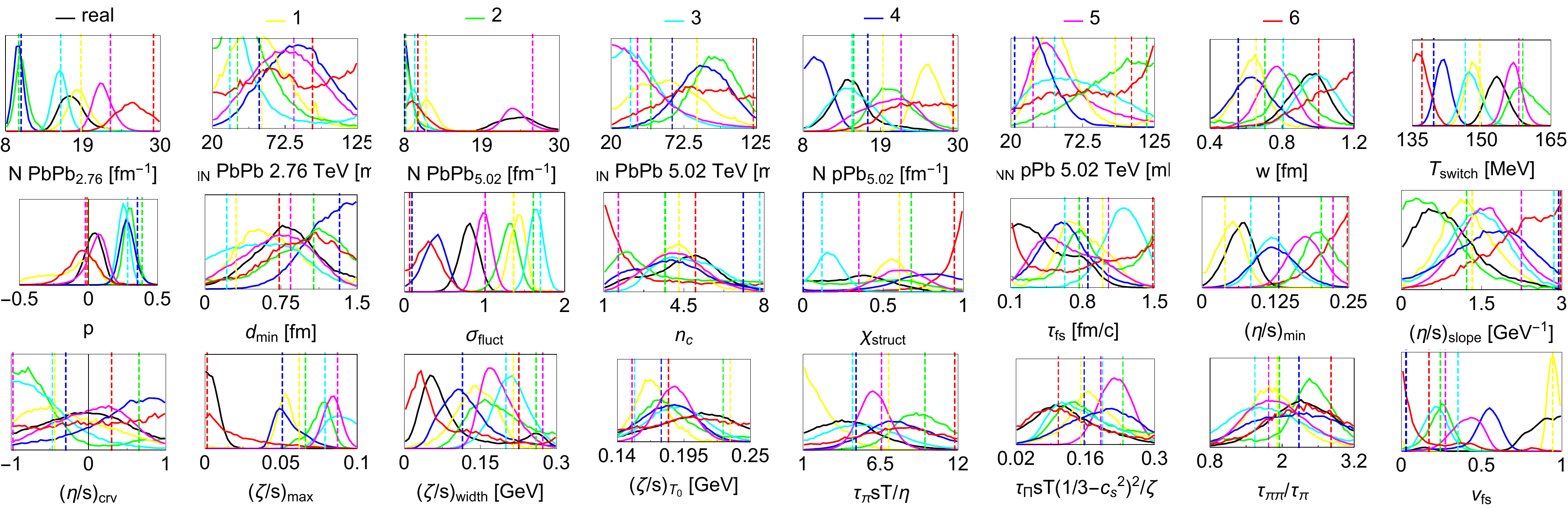}
\caption{\label{fig:closures} For all parameters that we vary we show the posterior distributions obtained from model output at six random locations in parameter space (shown as dashed lines). For comparison we also show the posterior comparing to real data from \cite{Bayesianshort} (black). The posterior distributions match well with the input parameter, and for many parameters strong constraints are possible using the experimental data as presented in Section \ref{sec:mcmc}.}
\end{figure*}

\begin{figure}[ht]
\includegraphics[width=0.7\columnwidth]{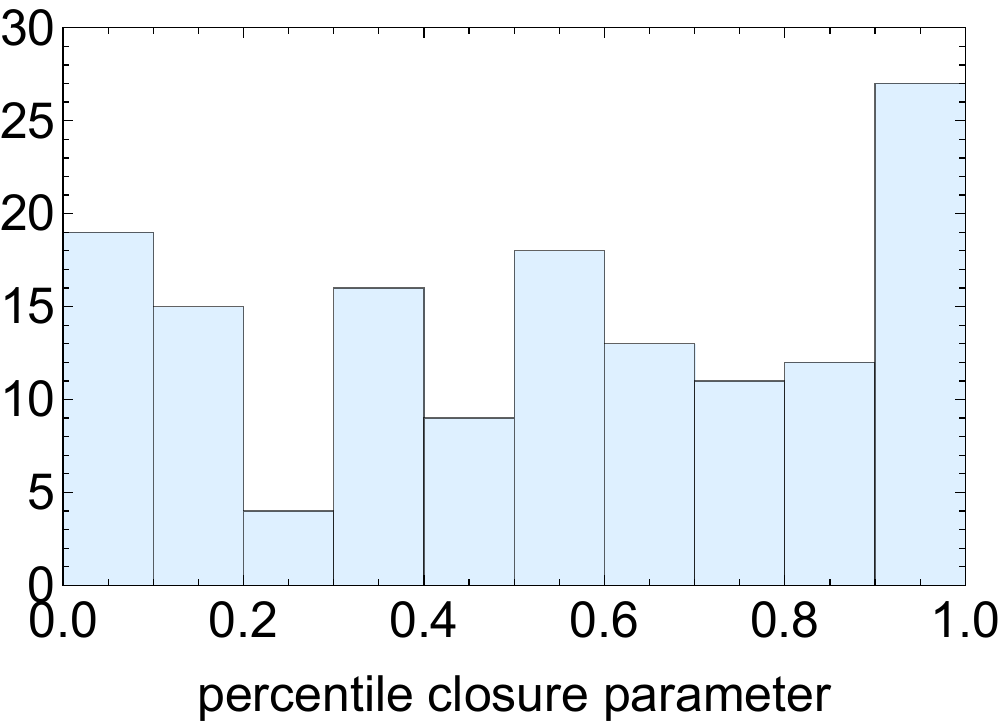}
\caption{\label{fig:closurespercentiles} We show the distribution of percentiles of the input parameter in the posterior probability distribution for all $6\times24$ parameters displayed in \fig{fig:closures}. Ideally this should be a flat distribution, with each bin containing $14.4 \pm 3.8$ points. The highest bin is overrepresented, which is partly due to our choice of random input parameters that can lie at the edge of the prior probability distribution.}
\end{figure}

\begin{figure*}[ht]
\includegraphics[width=\textwidth]{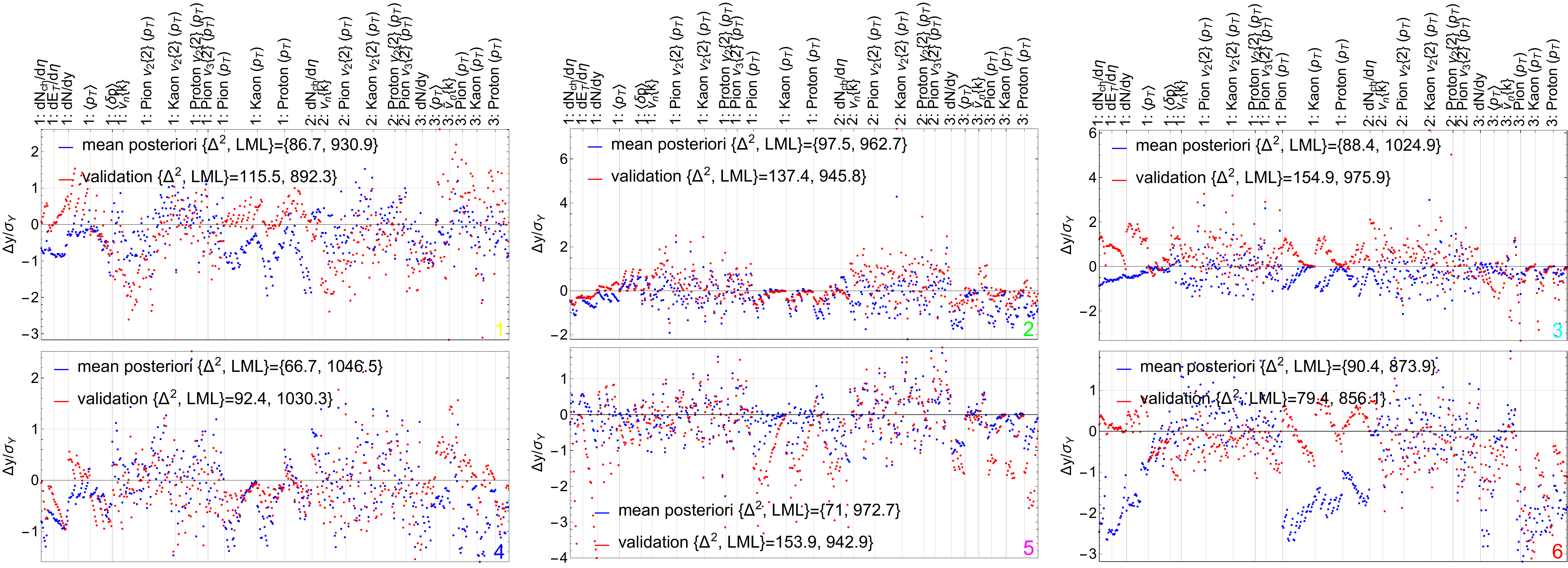}
\caption{\label{fig:closuresdeltay}For the closure points presented in \fig{fig:closures} we present the average of 100 posterior datasets, as compared to the `true' data point with the estimated uncertainty $\sigma_y$ (including both emulator uncertainty and statistical uncertainty from obtaining the datapoint in the model). The validation dataset is obtained from the emulator prediction at the input parameters. Data are labelled by class, where the number refers to the system (PbPb at 2.76 and 5.02 TeV and $p$Pb at 5.02 TeV respectively) and subclasses or centralities are not displayed. In total we included 514 datapoints.}
\end{figure*}

\begin{figure}[ht]
\includegraphics[width=0.9\columnwidth]{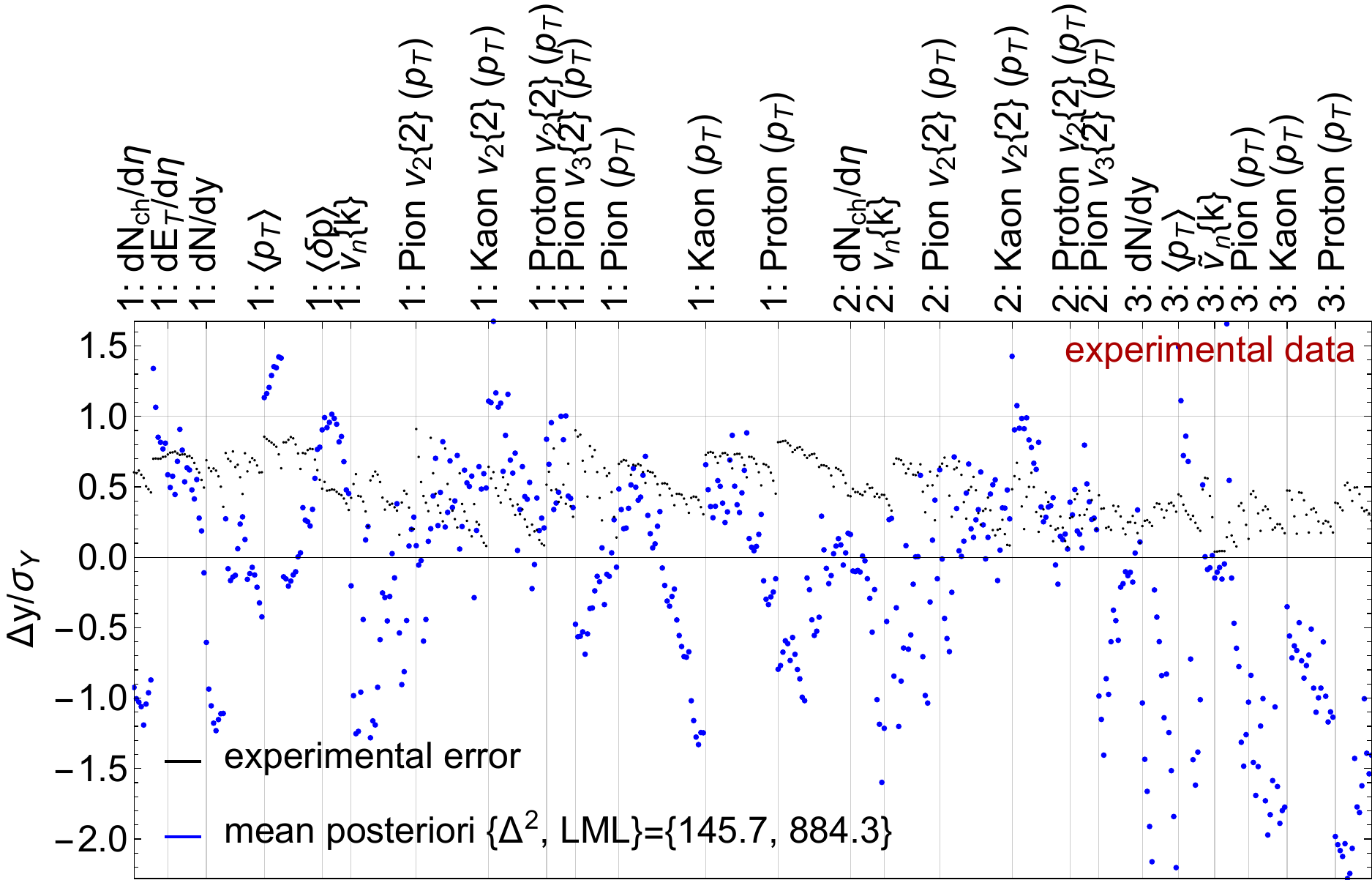}
\caption{\label{fig:realdeltay}We show the equivalent plot of \fig{fig:closuresdeltay} for the real experimental data. Black points represent the part of $\sigma_y$ due to the experimental error, where for $p$Pb collisions (labelled 3) the emulator error is clearly dominant. All posterior averaged fit our 514 experimental data points nicely according to a normal distribution.}
\end{figure}

For any model including emulators and Bayesian estimates it is crucial to implement a closure test: given a set of parameters one determines the model output (`experimental data') and from this output one obtains the posterior estimates on the parameters, which naturally should be consistent with the input parameters (see also \cite{Paquet:2020rxl}). This test is useful for two reasons: firstly it gives a strong test that all steps from emulation to the MCMC are working correctly, and importantly that the uncertainties obtained are realistic. Secondly, since the closure test uses the same dataset as the `real' model it shows how constraining this dataset is on each of the varying parameters. It could for instance be that the model is not sensitive to one of the input parameters, and the closure test should then result in a flat distribution regardless of the input parameter. This latter effect can also depend on where we are in the parameter space; on physical grounds it is for instance clear that the width of the bulk viscosity is unconstrained if the norm of the bulk viscosity vanishes. It is therefore important to perform the closure test at several points in the parameter space.

Fig.~\ref{fig:closures} shows the posterior distribution for six such points; the dashed lines indicate the (random) input parameters for each closure test using different colors. Note that both the norm as well as the nucleon-nucleon cross section are chosen randomly for each system separately, whereby we did not constrain the norm to be larger for 5.02 TeV as compared to 2.76 TeV collisions. For reference we include the `real' posterior distributions as obtained in \cite{Bayesianshort} (black). A few parameters are clearly well constrained, including the norms, the width, the switching temperature, the \trento{} parameter $p$, the amount of fluctuations, the free streaming velocity as well as the minimum of $\eta/s$ and the norm of $\zeta/s$. Other parameter distributions are more similar to the priors (being flat over the range), or depend more sensitively on the point in parameter space. $\tau_\pi s T/\eta$ is for instance quite well constrained for points $1$, $2$ and $5$, but less so for the others.

Most posterior distributions align well with the input parameter. This can be quantified by looking at the percentile in the posterior distribution of the input parameter, of which a histogram is shown for all $6\times24$ parameters in Fig~\ref{fig:closurespercentiles}. Clearly the distribution of the percentiles is approximately flat, as expected. There is a small excess around 100\%, which can indicate that we should not have put (random) input parameters close to the edge of the parameter range. In that case the prior probabilities, which are zero outside the parameter range, guarantee a percentile that is close to zero or a hundred percent. For the real dataset this issue is not present, as all prior probabilities have been chosen to include a parameter range that should reasonably include the physical point. Lastly, in rare cases the posterior is clearly off, such as for instance for $v_{\rm fs}$ for point 4. In this case there is a strong correlation with $\chi_{\rm struct}$ (see also \cite{Bayesianshort}), which is also off, thereby at least providing part of the explanation.

A more detailed and refined look can be obtained by showing the average posterior deviations of the actual 514 datapoints for each closure point, as shown in blue in Fig.~\ref{fig:closuresdeltay} as a ratio with respect to the uncertainty that includes both emulator uncertainty as well as statistical uncertainty in computing the `data' point. Systems are numbered one, two and three for PbPb at 2.76 and 5.02 and $p$Pb at 5.02 TeV respectively. The red points show the deviation when using the emulator on the input parameters directly, without using the posterior distribution. Ideally all these points would agree within their normal errors, whereby the red points test solely the emulator and the blue points also test the posterior distributions. In the text we also display the (average) values of the deviation computed by \eqref{eq:delta} as well as the LL of \eqref{eq:bayes}. The red validation is supposed to give the best estimates of the data points and naively should have the highest LL. Nevertheless, the posterior distributions are optimized to maximize this likelihood, so it is no surprise that these often give a higher LL. Nevertheless, their relative closeness gives further credibility that all works well. Lastly, it is worthwhile to study a few outliers more specifically. Point 2 has a 4 (blue) or 6 (red) standard deviation difference for a pion $v_2$ in system 2. Looking more carefully, this is a high $p_T$ off-central datapoint whereby the norm for this closure point is furthermore small (8.21 in this case). This discrepancy is hence a result of a lack of statistics for this particular somewhat unlucky combination. Point 5 has a rather strong deviation in red for the centrality dependence in system one already at the emulator level (as seen from red points becoming consistently worse for $dN_{\rm ch}/d\eta$ and $dE_{\rm T}/d\eta$), which would be interesting to investigate further.

Finally, we can compare to the equivalent result using the real experimental data, as shown in Fig.~\ref{fig:realdeltay}. Here we do not have knowledge of the true physical parameters, so there are no validation points. Instead we do show the part of $\sigma_y$ due to experimental error (in black). We see that the mean of all posterior distributions nicely follows a normal distribution with no clear outliers. The $p$Pb results have larger deviations, especially when compared to the experimental error. This is because it is much harder to simulate this smaller system, as already noticed in Fig.~\ref{fig:validation}.

\subsection{Discretization error}\label{sec:discretizationerror}
\begin{figure*}[ht]
\includegraphics[width=\textwidth]{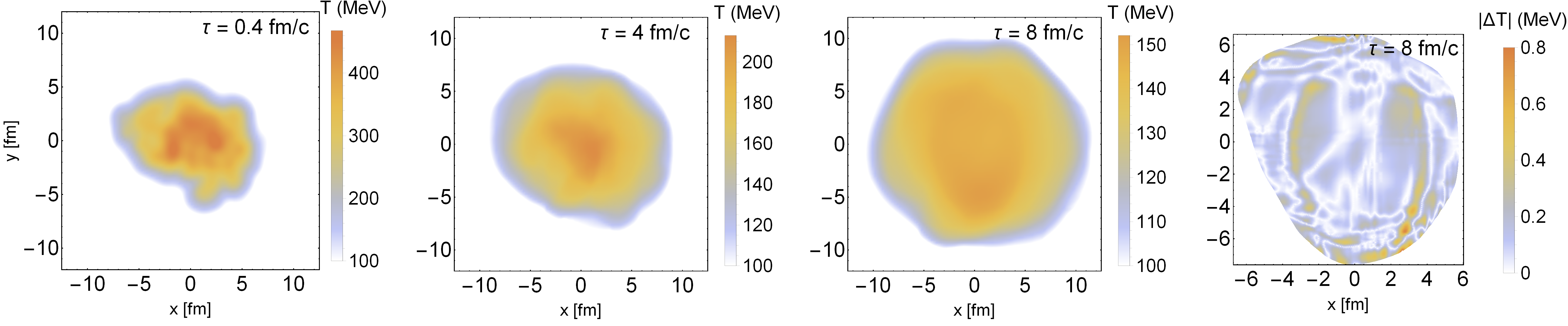}
\caption{\label{fig:convergence} We show the temperature of a typical event at the initial time $\tau = 0.4\,$fm$/c$ and at $4$ and $8\,$fm$/c$, together with the difference in temperature at the end of the evolution comparing resolutions of $N_\text{sites}=200$ or $300$. The difference is shown only where there is plasma (defined by $T > 140\,$MeV) and shows that the discretization error is small.}
\end{figure*}

\begin{figure}[ht]
\includegraphics[width=\columnwidth]{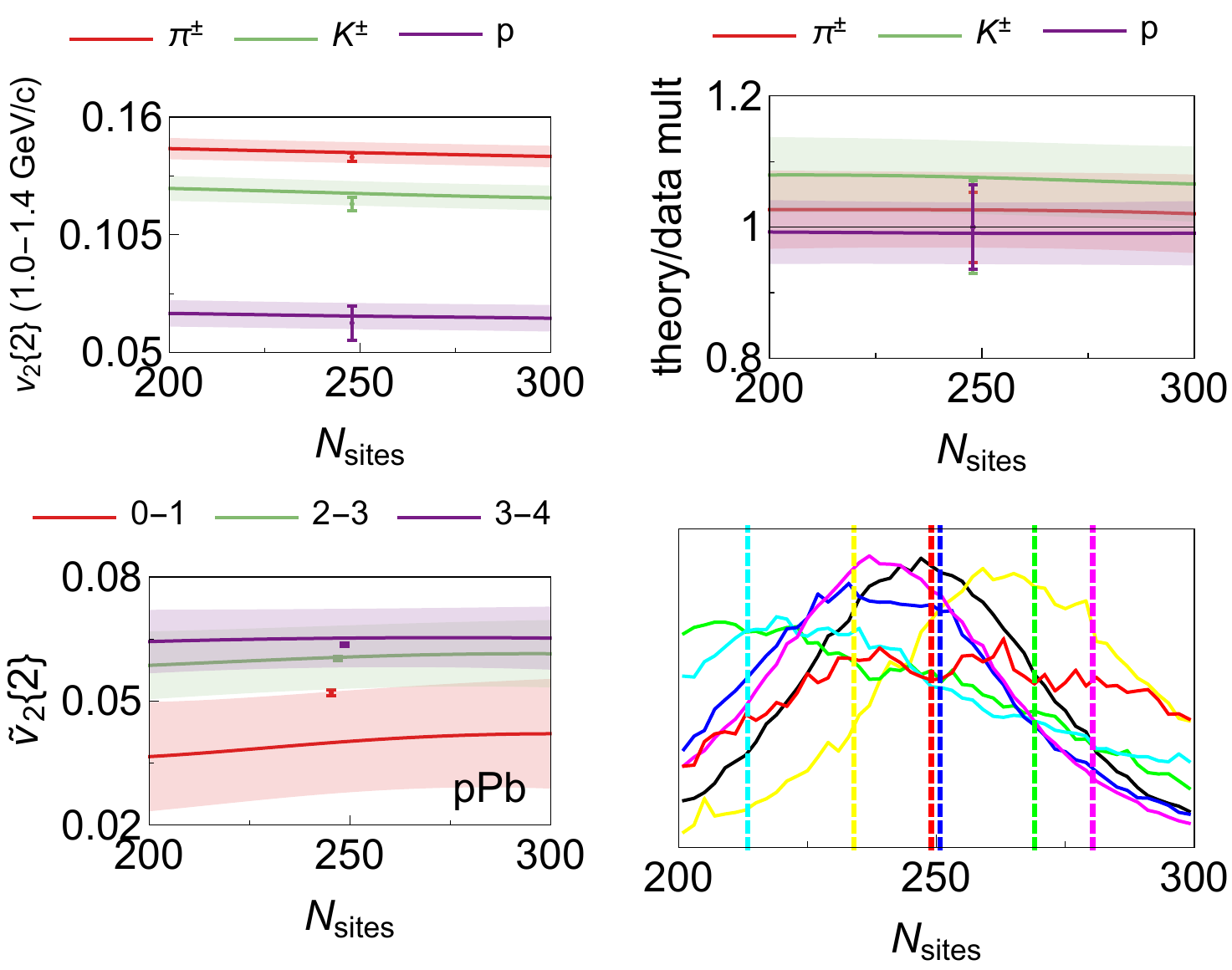}
\caption{\label{fig:nsitescombo}Top left: $v_2\{2\}$ (1.0--1.4 GeV) for PbPb at $2.76\,$TeV, shown for pions, kaons and protons in the $20-30\%$ centrality class. Top right: charged particle multiplicity, normalized by the experimental data, for PbPb at 2.76 TeV, shown for pions, kaons and protons in the same centrality class. Bottom left: $\tilde v_2\{2\}$ for $p$Pb at 5.02 TeV, in the multiplicity classes $N_{\rm ch}/\langle N_{\rm ch}\rangle \in [0,1]$, $N_{\rm ch}/\langle N_{\rm ch}\rangle \in [2,3]$ and $N_{\rm ch}/\langle N_{\rm ch}\rangle \in [4,5]$\@. %
Bottom right: posterior distribution for $N_\text{sites}$, shown together with 6 posterior distributions for the closure chains. Colors are as in \fig{fig:closures}\@.%
}
\end{figure}

An important source of potential error is the discretization error in the simulation.
Our simulations are performed on a square lattice with lengths of $50\,\text{fm}$, containing $N_\text{sites} \times N_\text{sites}$ lattice sites.
A straightforward way to examine the effect of discretization error is to perform the hydrodynamic simulation on the same event with multiple values for $N_\text{sites}$ and comparing the results.
We performed this check for $N_\text{sites} = 200, 300, 400$, and as shown in \fig{fig:convergence} we find no significant differences between the results.

In addition to this, we also employed a novel technique, by which we vary $N_\text{sites}$ in the Bayesian analysis as if it were a physical parameter.
In particular, one looks for two important indications that the error is indeed under control, i.e.~that the smallest value for $N_\text{sites}$ in the chosen range is large enough.
Firstly, it is possible to utilize the emulator to verify that our observables do not depend on $N_\text{sites}$. 
This is shown for a few representative observables in \fig{fig:nsitescombo} (top left, top right, bottom left).
All observables shown are indeed consistent with being independent of $N_\text{sites}$ when taking into account the uncertainty associated with the emulator (shown as a band).
$\tilde v_2\{2\}$ for $p$Pb shows the strongest dependence, whereby the relatively wide uncertainty band is both due to the statistical difficulty obtaining accurate results for (low multiplicity) $p$Pb flow coefficients, as well as the more difficult emulation (see also section \ref{sec:emulator}).
Secondly, the posterior distribution treating $N_\text{sites}$ as a physical parameter should not show any preference or correlation, which is implied if indeed observables do not depend on $N_\text{sites}$.
\fig{fig:nsitescombo} (bottom right) shows the result of the six closure tests that we performed for $N_\text{sites}$, together with the posterior from the comparison with real data.
One can see that while none of the distributions are completely flat, the locations of the maxima do not correlate with the known input values used in the closure tests.
This means that the non-flat distributions are unlikely to reflect any true correlations found through the posterior estimates.

We also checked the correlations of all 24 parameters with $N_\text{sites}$ for the six closure and the real posterior distributions. For all parameters the average correlation never exceeded 10\%, from which we conclude that also the correlations with $N_\text{sites}$ are insignificant.
Lastly, we note that this technique could also be used to estimate a minimum number of required grid points, or point to observables and parameter settings that need particularly accurate simulations.

Another source of discretization error can come from the fact that $\tau_\pi$ or $\tau_\Pi$ can become of the same order as the time step $\Delta t$.
For the simulations used in this work and in \cite{Bayesianshort}, $\Delta t = 0.08 \Delta x$, where $\Delta x$ varies between 0.17 and $0.25\,$fm (for $p$Pb more refined simulations are necessary, whereby we used a lattice of $12\,$fm in size with at least 70 lattice sites such that $0.11 <\Delta x < 0.17\,$fm)\@.
This leads to $\Delta t \leq 0.02\,$fm$/c$ for our simulations.
For this reason we require that $\tau_\pi > 2\Delta t$ and $\tau_\Pi > 0.1\,\text{fm}/c$ throughout the simulation.
Due to the temperature dependence in both $\eta/s$ and $\zeta/s$, whether this rule is invoked often depends in a relatively complicated way on the parameters.

We verified that in less than 0.2\% of our posterior distribution we needed to invoke this requirement for $\tau_\pi$ at any temperature below $400\,$MeV and hence we can trust the resulting distribution not to be affected by the minimum time step.
For $\tau_\Pi$ this is however different, as our analysis keeps the ratio $\tau_\Pi sT(1/3 - c_s^2)^2/\zeta$ fixed. As our posterior distribution obtains a small specific bulk viscosity $\zeta/s$ this implies that often $\tau_\Pi$ is small as well. We hence verified that in most of our simulations the constraint $\tau_\Pi > 0.1\,$fm$/c$ is dominant. This means we should not put much trust in our posterior distributions for $\tau_\Pi$, which in our case did not give strong constraints anyway.

\subsection{Prehydrodynamic phase}\label{sec:results:prehydrodynamicphase}
    
\begin{figure*}[ht]
\includegraphics[width=0.49\textwidth]{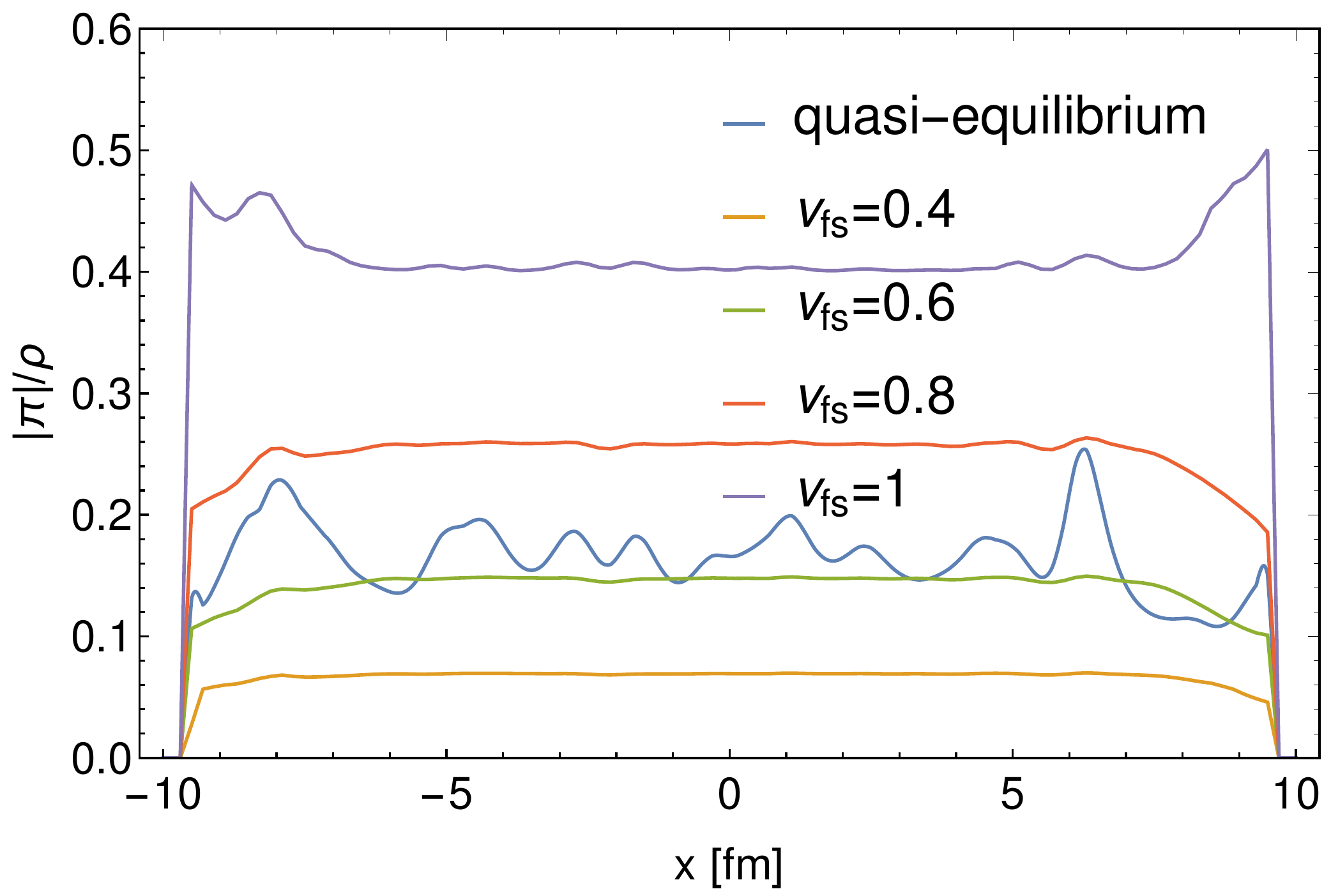}
\includegraphics[width=0.49\textwidth]{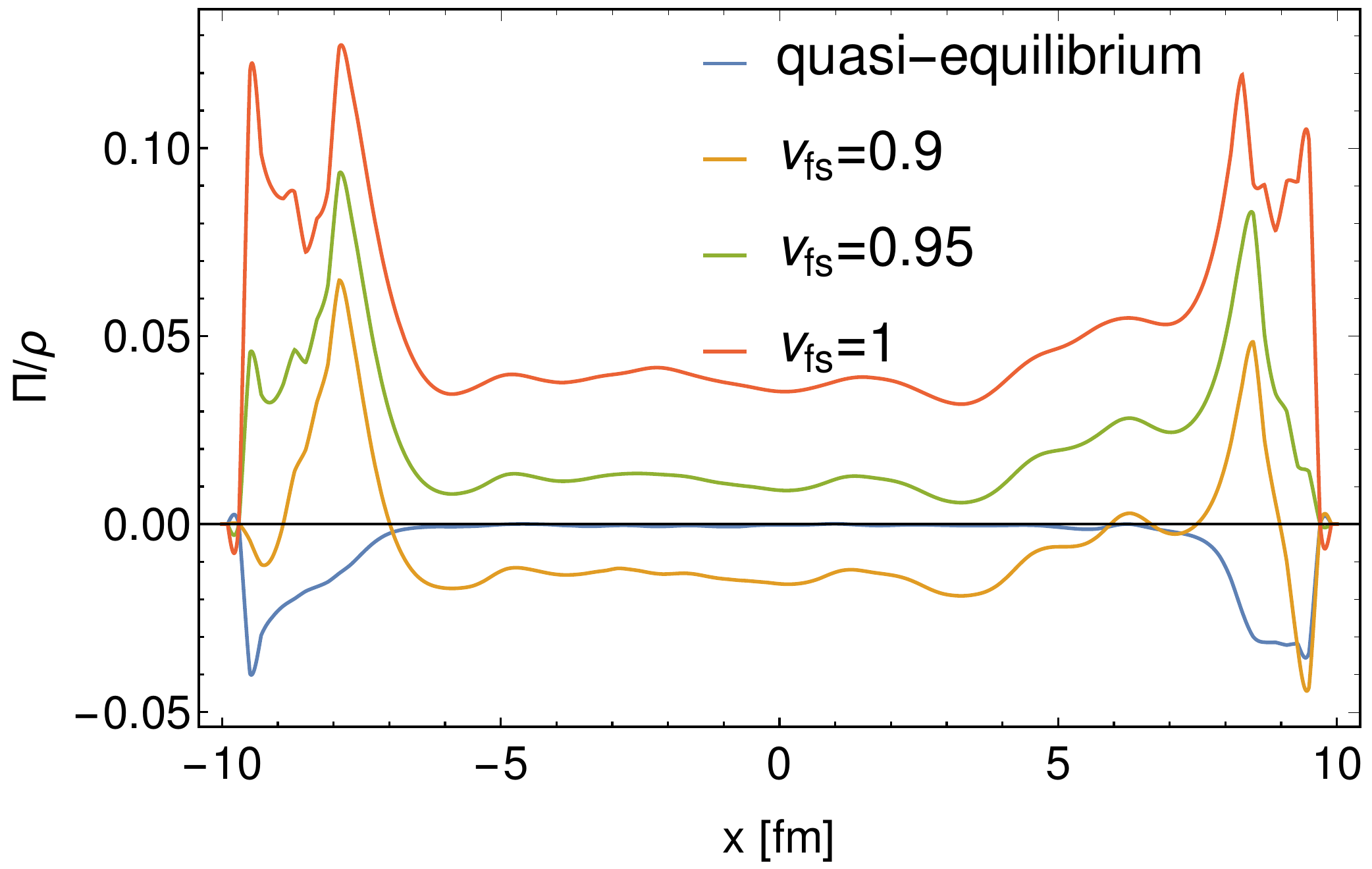}
\caption{\label{fig:vfscheck}Left: Ratio of the absolute shear viscous pressure $|\pi|$ and the energy density $\rho$, where we define $|\pi| = \sqrt{g_{\mu\nu}g_{\rho\sigma}\pi^{\mu\rho}\pi^{\nu\sigma}}$\@. Right: Ratio of the bulk viscous pressure $\Pi$ and the energy density $\rho$, computed at the time of initialization in a cross-section through the plasma. 
The free streaming has been performed on the same event with three different free streaming velocities (different colours). We also show the `quasi-equilibrium' value, defined as the zeros of the right hand sides of \eqref{eq:intro:secondordershear} and \eqref{eq:intro:secondorderbulk}.
}
\end{figure*}

This section attempts a more refined understanding of the prehydrodynamic phase and more in particular the influence of the free streaming velocity $\vfs{}$. First of all we note that the free streaming velocity directly affects the pressure of the free streaming stress tensor \eqref{eq:freestreaming}, whereby $\vfs{}=0$ leads to a pressureless fluid and $\vfs{}=1$ corresponds to a conformal stress tensor for which the equation of state (EOS) equals $P=\rho/3$. This pressure is in turn matched to the viscous shear tensor $\pi^{\mu\nu}$ and the bulk viscous pressure $\Pi$ in order to guarantee a continuous stress tensor. Secondly, it is important to realise that the EOS at the relevant temperatures is not yet fully conformal, and in fact at $T = 0.4\,$GeV we have $P \approx 0.85 \rho/3$ \cite{Huovinen:2009yb}. From the equation of state one may therefore suspect that physically a free streaming velocity around $0.85$ may be the most realistic.%

To quantify this effect we are interested in the effects of changing the free streaming velocity $v_\text{fs}$ on the initialization of the shear stress $\pi^{\mu\nu}$ and the bulk viscous pressure $\Pi$\@.
These are shown in respectively the left and right panel of \fig{fig:vfscheck}, in a cross-section through the plasma, relative to the energy density $\rho$.
Since the shear tensor is traceless, we define the absolute value of the shear tensor as the simplest nonvanishing scalar
\begin{equation}
|\pi| \equiv \sqrt{g_{\mu\nu}g_{\rho\sigma}\pi^{\mu\rho}\pi^{\nu\sigma}}.\label{eq:abspidef}
\end{equation}
The event shown is generated using the maximum a posteriori (MAP) values for the parameters presented in \cite{Bayesianshort}, which in particular means that these figures are shown at $\tau_{\rm fs} = 0.47\,\text{fm}/c$\@.
Furthermore, this one event is computed multiple times, each time using a different value for $v_\text{fs}$\@.
In addition, the `quasi-equilibrium' values of $|\pi|/\rho$ and $\Pi/\rho$ are also shown.
These are obtained by respectively setting the left hand sides of \eqref{eq:intro:secondordershear} and \eqref{eq:intro:secondorderbulk} to zero.
In other words, these are the values that the shear stress and bulk viscous pressure would take if the fluid was initialized close to the value that \eqref{eq:intro:secondordershear} and \eqref{eq:intro:secondorderbulk} would relax to given enough time.

In the left panel of \fig{fig:vfscheck}, one can see that the value closest to the `quasi-equilibrium' value for the shear stress is $v_\text{fs} \approx 0.6$\@.
However, because $\pi^{\mu\nu}$ is a tensor it is not clear that these configurations resemble each other.
To investigate this further we furthermore computed the absolute value of the difference, defined analogously as $|\Delta\pi| \equiv \sqrt{g_{\mu\nu}g_{\rho\sigma}\Delta\pi^{\mu\rho}\Delta\pi^{\nu\sigma}}$, with $\Delta\pi \equiv \pi_\text{fs}^{\mu\nu} - \pi_\text{quasi-eq}^{\mu\nu}$\@.
This shows that  $v_\text{fs} \approx 0.6$ indeed minimizes the deviation from the quasi-equilibrium value, whereby $|\Delta\pi|/\rho \approx 0.05$.

In the right panel of \fig{fig:vfscheck}, one can see that the `quasi-equilibrium' value for the bulk viscous pressure is slightly negative, in agreement with the small value found for the bulk viscosity $\zeta$.
However, when initializing the plasma with $v_\text{fs} = 1$, the bulk viscous pressure is much larger and positive, as is indeed expected for an EOS that has a pressure below the conformal value.
As one lowers $v_\text{fs}$, the bulk viscous pressure decreases, and eventually changes sign around $v_\text{fs} \approx 0.93$\@.
The posterior distribution for $v_\text{fs}$ is peaked around $v_\text{fs} \approx 0.9$, suggesting that data perhaps shows a preference for initialization near quasi-equilibrium.
Such an initialization is also what is suggested by holography, where the system quickly \emph{hydrodynamizes} after the initial stage \cite{vanderSchee:2013pia}\@.

The fact that the $v_\text{fs}$ for a quasi-equilibrium initialization of $\pi^{\mu\nu}$ and $\Pi$ are different means that within the free streaming model it is impossible to initialize both the shear stress and bulk pressure near quasi-equilibrium. This implies that while the stress-energy tensor itself is continuous at the switch to hydrodynamics, its derivative will not be continuous. This is in marked contrast with e.g. holographic models \cite{vanderSchee:2013pia}, which would hence be interesting to study further in future work.


The behavior of $\Pi$ as a function of $v_\text{fs}$ can be understood as follows:
For $v_\text{fs} = 0$, the stress tensor has the form $T^{\mu\nu} = \rho\delta_0^\mu\delta_0^\nu$\@.
Using the constitutive relation \eqref{eq:constitutiverelation}, one can decompose this stress tensor, yielding $\pi^{\mu\nu} = 0$ and $\Pi = -P$\@.
This therefore leads to $\Pi$ being negative.
At early times we can use \eqref{eq:freestreaming}, which for $v_\text{fs} = 1$ yields a traceless stress tensor, i.e.~$\rho - 3(P + \Pi) = 0$\@.
For our equation of state the speed of sound is below the conformal bound, which implies $3P < \rho$ and hence that $\Pi$ is positive.
For intermediate values of $v_\text{fs}$, the behavior of $\Pi$ interpolates between these two cases.

\begin{figure*}[ht]
\includegraphics[width=0.49\textwidth]{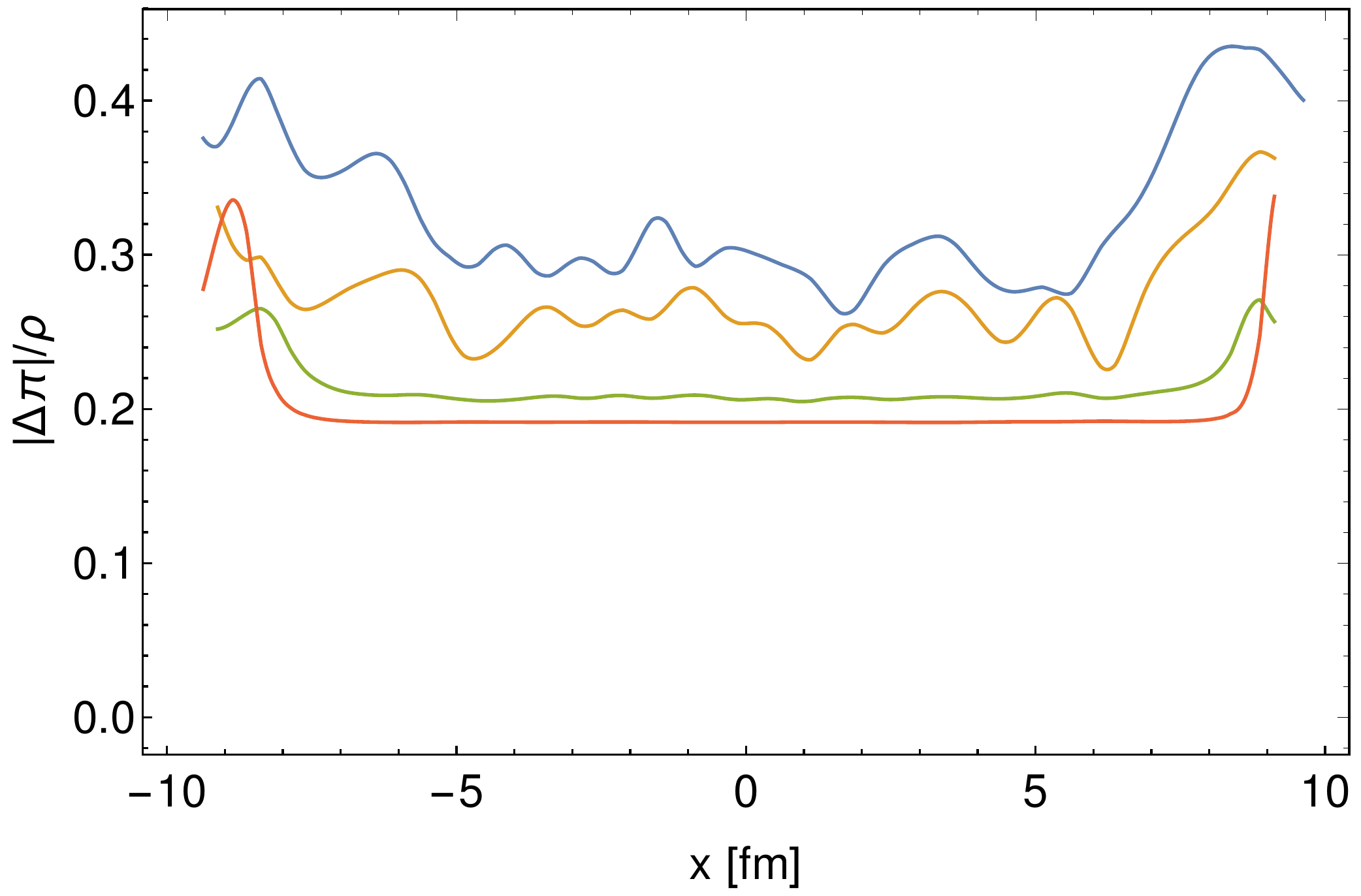}
\includegraphics[width=0.49\textwidth]{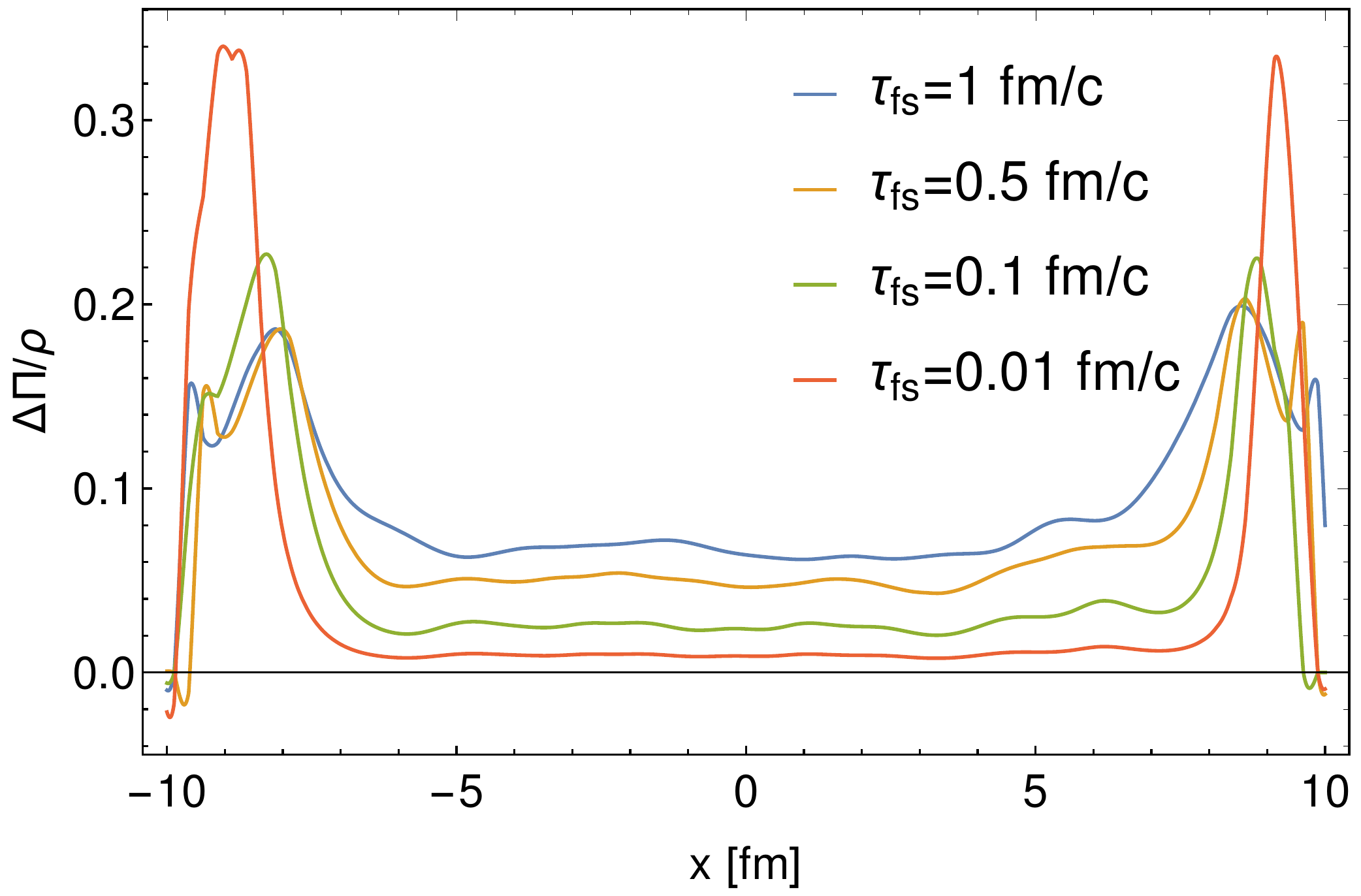}
\caption{\label{fig:taufscheck}Difference from quasi-equilibrium relative to the energy density $\rho$ for different values of the free streaming time $\tau_\text{fs}$, shown for the shear stress (left) and the bulk viscous pressure (right)\@. Here we define $|\Delta\pi| \equiv \sqrt{g_{\mu\nu}g_{\rho\sigma}\Delta\pi^{\mu\rho}\Delta\pi^{\nu\sigma}}$, with $\Delta\pi^{\mu\nu} \equiv \pi_\text{fs}^{\mu\nu} - \pi_\text{quasi-eq}^{\mu\nu}$ and $\Delta\Pi = \Pi_\text{fs} - \Pi_\text{quasi-eq}$\@.}
\end{figure*}

\begin{figure}[ht]
\includegraphics[width=\columnwidth]{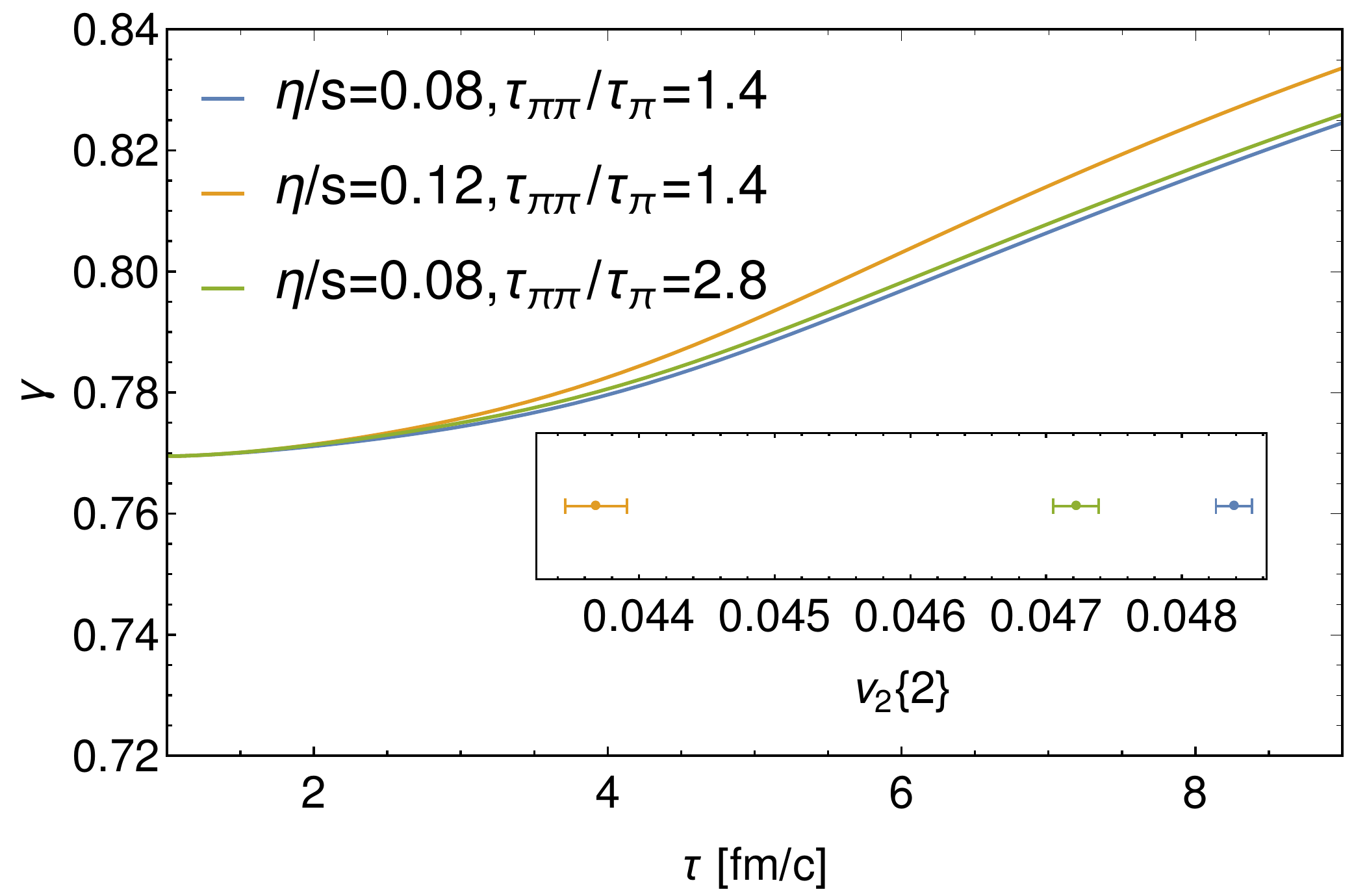}
\caption{\label{fig:v2proxyplot}$\gamma$, defined by \eqref{eq:gammadefinition}, as a function of proper time $\tau$ for the idealized plasma initialized according to \eqref{eq:idealizedplasma}, for different values of $\eta/s$ and $\tau_{\pi\pi}/\tau_\pi$\@. Inset: the resulting values for the elliptic flow $v_2\{2\}$\@.}
\end{figure}

Another interesting dependence of the prehydrodynamic phase is that on the free streaming time $\tau_\text{fs}$\@.
In \fig{fig:taufscheck}, the deviation from quasi-equilibrium is shown for the shear stress (left) and the bulk viscous pressure (right)\@.
As in \fig{fig:vfscheck}, the same event is computed, but here we used $v_\text{fs} = 1$ together with various values of $\tau_\text{fs}$\@.
It can clearly be seen that the shear tensor moves away from quasi-equilibrium during the prehydrodynamic phase, even though the $\tau_\text{fs}$-dependence is mild.
For the bulk viscous pressure, the interior of the plasma moves away from quasi-equilibrium during the prehydrodynamic phase, whereas the edges move towards quasi-equilibrium.
These edges contain only a small part of the total energy though, so we can conclude that
the prehydrodynamic stage moves the fluid away from quasi-equilibrium.%

\subsection{Correlation between $\tau_{\pi\pi}/\tau_\pi$ and $(\eta/s)_\text{min}$}
As shown in \cite{Bayesianshort}, $\tau_{\pi\pi}/\tau_\pi$ and $(\eta/s)_\text{min}$ are negatively correlated.
The reason for this is that $\eta/s$ is mostly determined by the measurement of $v_2\{2\}$, and that both $\eta/s$ and $\tau_{\pi\pi}$ tend to lower this observable, making the effects of these two transport coefficients more or less interchangeable.
The mechanism ultimately causing this is that both of these transport coefficients are dissipative, and hence increase the entropy by making the fluid less anisotropic.
For this reason, we expect this correlation to be generically present, but in the following we will look at a specific configuration, to examine whether the mechanism by which the anisotropy decreases is similar in both cases.

In the case of the shear viscosity, the decrease in anisotropy is well understood to work through the mechanism that in the `short' direction, pressure gradients are larger, driving the build-up of momentum in that direction, while the shear viscosity counters this momentum build-up \cite{Teaney:2003kp}. In a similar spirit, we look at an idealized plasma, where we initialize $\pi^{\mu\nu}$ and $\Pi$ to zero, set $u^\mu = (1,0,0,0)$, and take the energy density to be
\begin{equation}
\rho(x,y) = \frac{\alpha}{1 + \exp\left(\frac{\sqrt{x^2 + (1.3y)^2} - R}{\theta}\right)},\label{eq:idealizedplasma}
\end{equation}
with $R = 5\,\text{fm}$, $\theta = 1\,\text{fm}$ and $\alpha = 50\,\text{fm}^{-4}$\@.
Initializing the plasma as such at $\tau = 0.98\,\text{fm}/c$, we let it evolve with $\eta/s$ independent of temperature, and the bulk viscosity equal to that found in \cite{Bernhard:2018hnz}\@.
We then define the quantity $\gamma$ as follows:
\begin{equation}
\gamma(\tau) = \frac{\int_0^\infty\mathrm{d}x\int_0^\infty\mathrm{d}y\,T^{01}(x,y,\tau)}{\int_0^\infty\mathrm{d}x\int_0^\infty\mathrm{d}y\,T^{02}(x,y,\tau)},\label{eq:gammadefinition}
\end{equation}
which measures the ratio of the total $x$-momentum and $y$-momentum present in one quadrant.
In an isotropic setting $\gamma$ equals unity, and by using the departure from unity we can analyze how the dynamics of the fluid convert the spatial anisotropy into an anisotropic momentum distribution.
In \fig{fig:v2proxyplot}, the result of this analysis is shown for three different choices of $\eta/s$ and $\tau_{\pi\pi}/\tau_\pi$ (note that increasing $\eta/s$ by 50\% also increases $\tau_{\pi\pi}$ by 50\% since we fix our ratios such that $\tau_{\pi} \propto \eta/s$)\@.
One can clearly see that increasing $\eta/s$ decreases the amount of anisotropy, thereby decreasing $v_2\{2\}$\@.
One can also see that indeed $\tau_{\pi\pi}/\tau_\pi$ has a similar, albeit smaller, effect.
Hence we can indeed conclude that both of these dissipative corrections tend to lower the amount of momentum anisotropy.

In the inset of \fig{fig:v2proxyplot}, we see the associated $v_2\{2\}$ of these three simulations.
It can clearly be seen that the observed momentum anisotropy %
translates into $v_2\{2\}$, explaining the observed correlation.%

\section{Maximum a posteriori}\label{sec:map}

\begin{figure}[h!]
\includegraphics[width=0.9\columnwidth]{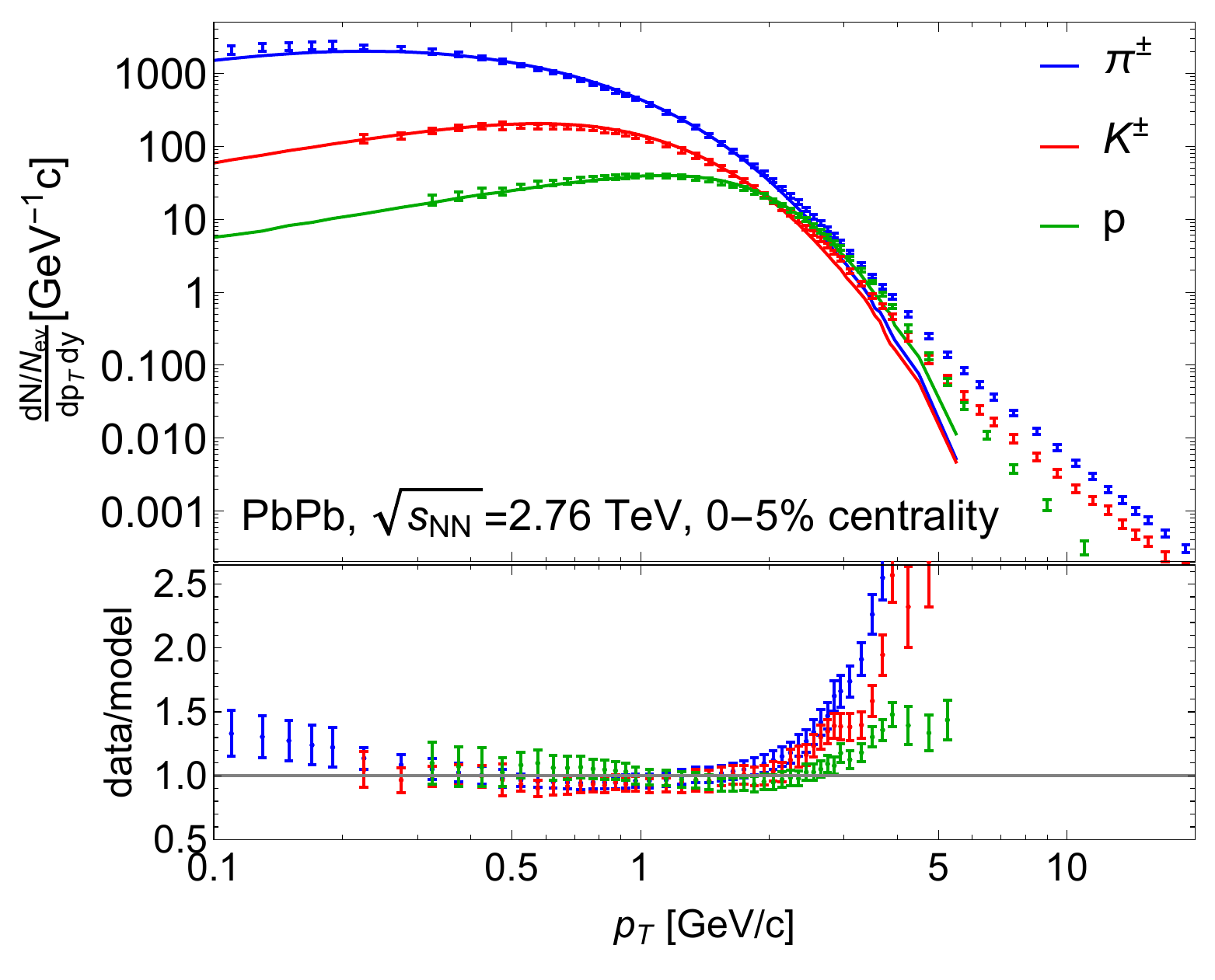}
\includegraphics[width=0.9\columnwidth]{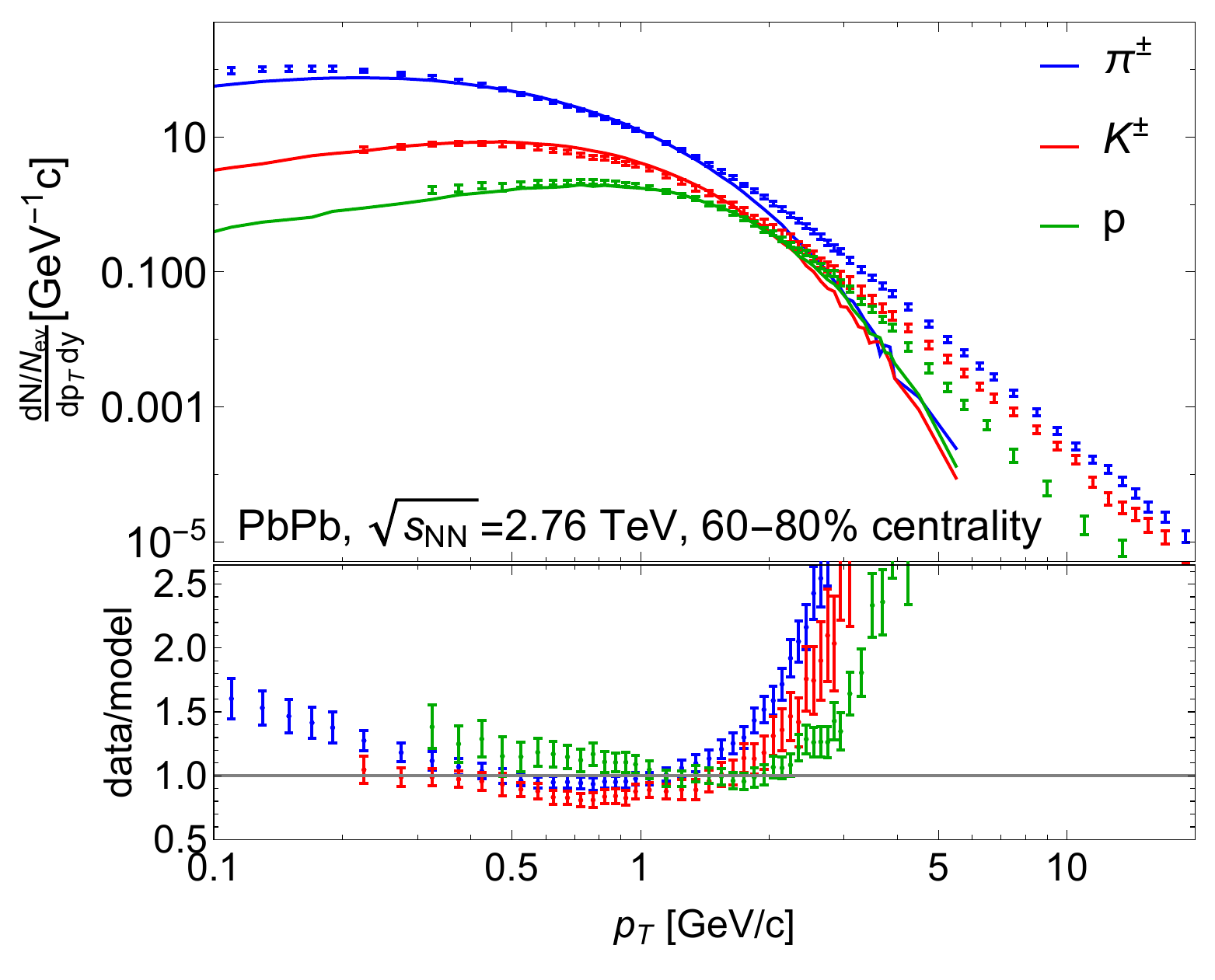}
\includegraphics[width=0.9\columnwidth]{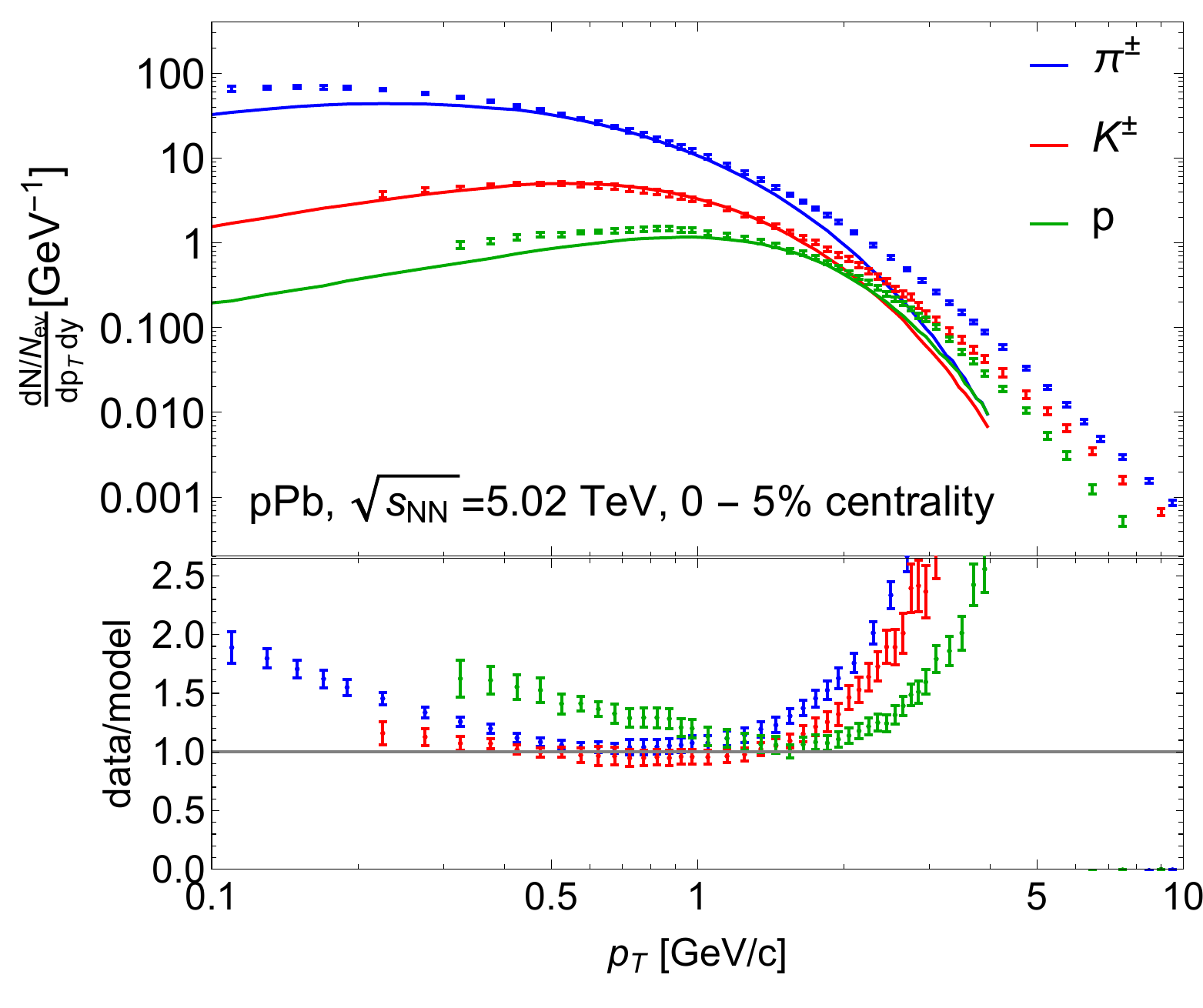}
\caption{\label{fig:mapspectra}We compare the MAP results for the transverse momentum spectra for charged pions, kaons and protons for central collisions (top), peripheral (middle, data from ALICE \cite{Abelev:2014laa}) and central $p$Pb (bottom, data in \cite{Adam:2016dau}) collisions. The model somewhat underpredicts the low $p_T$ pions (also seen in \cite{Devetak:2019lsk}), as well as the spectra at high $p_T$. The latter is more pronounced in peripheral and $p$Pb collisions and can possibly be explained by polynomial processes in pQCD.}
\end{figure}

\begin{figure}[ht]
\includegraphics[width=\columnwidth]{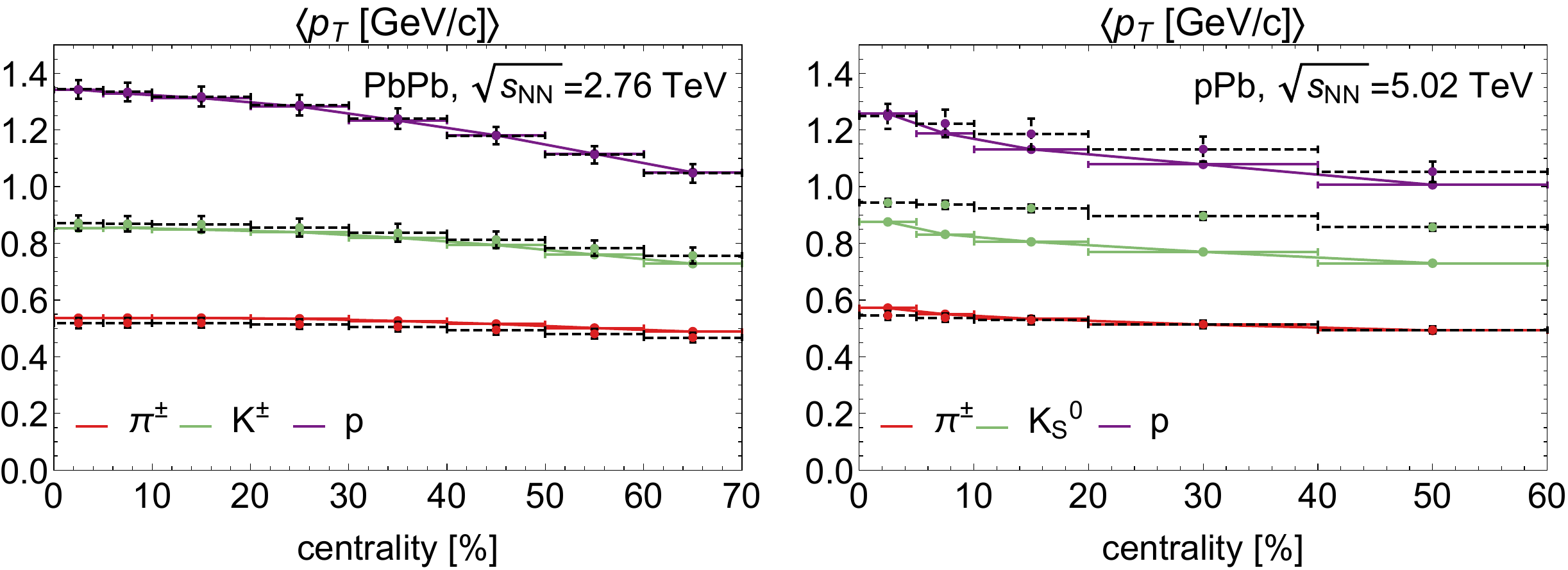}
\caption{\label{fig:mapmeanpt}We compare the mean of the transverse momentum for pions, kaons and protons as a function of centrality (colored) to experimental data from ALICE (black, \cite{Adam:2016dau}). The hydrodynamic results match perfectly, with the possible interesting exception of the mean kaon transverse momentum in peripheral $p$Pb collisions that possibly requires a non-hydrodynamic explanation.}
\end{figure}

With the emulator and MCMC together with 514 data\-points we obtained optimal (Maximum a posteriori, MAP) values for our varying parameters \cite{Bayesianshort}\@. Strictly speaking these values are just the mean expectation value for each parameter and non-trivial correlations could mean that such a combination is not maximizing the LL of \eqref{eq:bayes}\@.
We verified, however, that this method led to a LL that is comparable with the maximum LL obtained otherwise.
In this section we analyze a high statistics run (400k and 4M events for PbPb and $p$Pb respectively) using these parameters, which allows us both to verify the emulator
at this point specifically and importantly compare to experimental data that was statistically not feasible to include in the Bayesian optimization.

\subsection{Transverse momentum spectra}\label{sec:mapspectra}

In \fig{fig:mapspectra} we show the transverse momentum spectra for central and very peripheral PbPb collisions as well as central $p$Pb collisions including ALICE data \cite{Abelev:2014laa,Adam:2016dau} for the entire range where particle identification is possible. These spectra have been included in the posterior estimate using $p_T$ bins separated at $(0.5, 0.75, 1.0, 1.4, 1.8, 2.2, 3.0)\,$GeV for centralities up to the $40 - 50\%$ class. At high $p_T$ the spectra are dominated by hard processes described by perturbative QCD, which have a characteristic polynomial fall-off as a function of $p_T$. Our hydrodynamic thermal model falls off exponentially due to the Boltzmann factor with the switching temperature, and hence it is expected that around $p_T \approx 3\,$GeV the hydrodynamic prediction starts to deviate from the full experimental result. This effect is more pronounced for very peripheral collisions as well as $p$Pb collisions. At low $p_T$ below 500 MeV (not included in the posterior) there is a small surplus of pions, especially for very peripheral collisions where this deviation exceeds two standard deviations (see also \cite{Devetak:2019lsk} for a similar result).

From the spectra it is possible to obtain the average transverse momentum, as shown in \fig{fig:mapmeanpt}\@. Overall this provides an excellent fit, but the experimental mean transverse momentum for kaons in peripheral $p$Pb collisions is surprisingly high (we show the neutral kaons here, as their experimental uncertainty is much smaller than that of the charged kaons~\cite{Adam:2016dau}). The hydrodynamic results seem consistent with \cite{Moreland:2019szz} and given the significantly lower mean kaon momentum in PbPb collisions it is possible that this requires a non-hydrodynamic explanation, which is also apparent from the deviation of the kaon spectrum at intermediate $p_T$ in \fig{fig:mapspectra} (bottom)\@. Also the spectra of pions and protons deviate from the experimental data at intermediate $p_T$, but this is compensated at lower $p_T$, which then leads to the correct mean transverse momentum. This stresses once more why it is important to include the full identified transverse spectra. Since the average kaon transverse momentum has perhaps the most apparent deviation we chose this observable when showing its dependence on our parameters in \fig{fig:emusmeanptppb} in the Appendix.

\begin{figure}[ht]
\includegraphics[width=0.95\columnwidth]{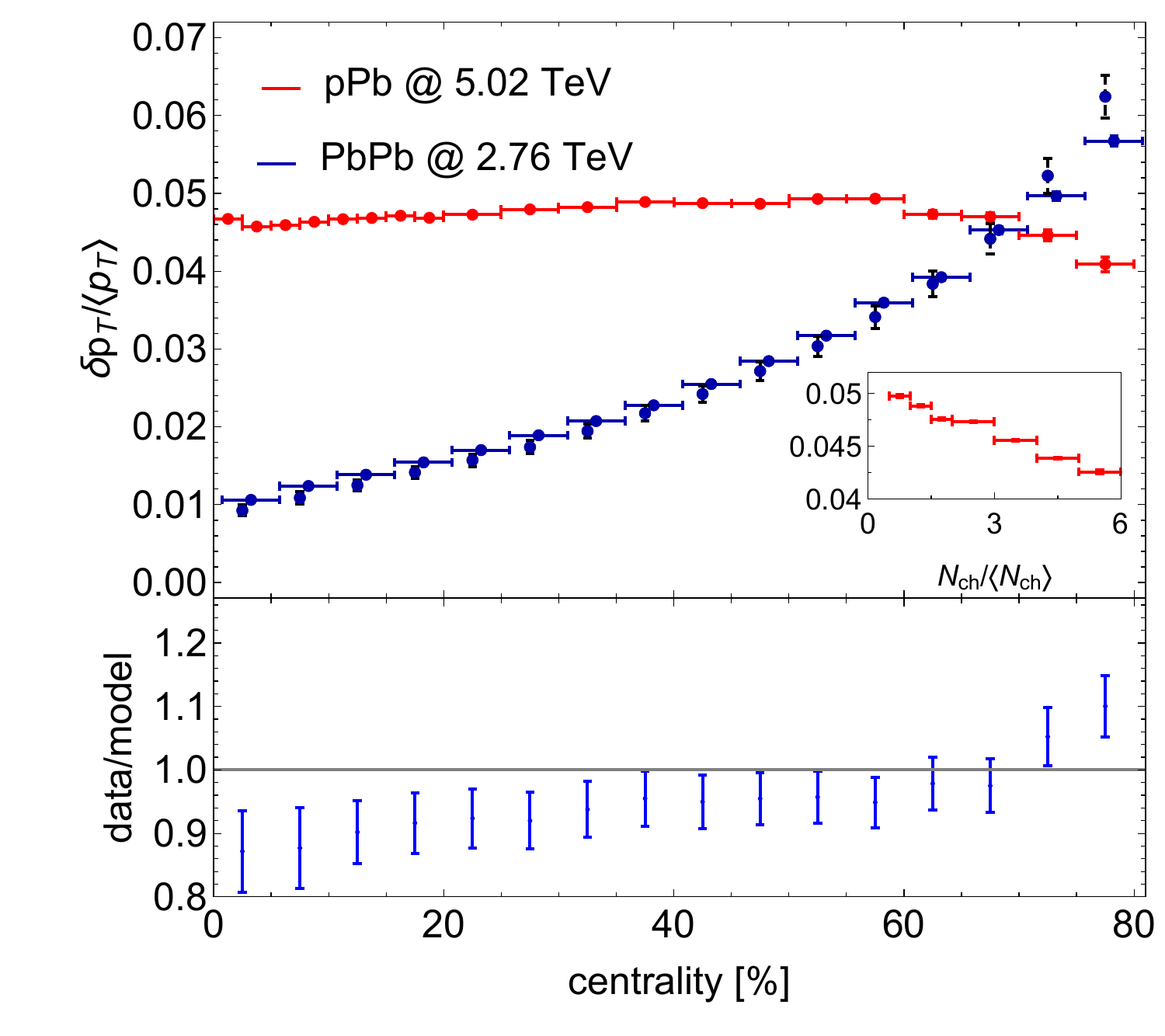}
\caption{\label{fig:mapptfluct}We compare the transverse momentum fluctuations of charged particles as a function of centrality (blue) to experimental data from ALICE (black, \cite{Abelev:2014ckr}). Especially for peripheral collisions this is a difficult observable \cite{Bernhard:2018hnz} and even though we did not use the centrality bins beyond 60\% to obtain the MAP values there is still a satisfactory match with the model. The red points are predictions for $p$Pb collisions at $\sqrt{s_{\rm NN}}=5.02\,$TeV for the same $0.15 < p_T [{\rm GeV}/c] < 2.0$ range. The inset shows the $p$Pb fluctuations for ultra-central events with multiplicities up to 6 times the average multiplicity.}
\end{figure}

We also stress that the MAP results are obtained using the expectation values found in \cite{Bayesianshort}, but importantly they do not include the posterior uncertainties. Especially for $p$Pb this is relevant, as the emulator has a significant uncertainty, and indeed the posterior mean kaon transverse momentum has a wide band that may just about include the experimental result (see Fig.~1 in \cite{Bayesianshort})\@. It could hence be that certain parameters especially relevant for $p$Pb can still be found that provide a better fit to the kaon mean $p_T$\@. One likely option here would be to have a larger free streaming velocity $v_{\rm fs}$ (see \fig{fig:emusmeanptppb}).

Lastly, we show the transverse momentum fluctuations in \fig{fig:mapptfluct}, defined as $\delta p_T^2 = \Bigl\langle\!\Bigl\langle(p_{T, i} - \langle p_T\rangle) (p_{T, j} - \langle p_T\rangle)\Bigr\rangle\!\Bigr\rangle$, where the double brackets denote an average over all pairs in the same event as well as averaging over the centrality class. Up to 60\% in centrality this observable is also used to obtain the posteriors (as also done in \cite{Bernhard:2018hnz}), but even then it turns out to be difficult to capture the peripheral bins (see \cite{Bernhard:2018hnz}). It is therefore satisfactory that we manage to obtain a good agreement, with perhaps the exceptions of the most central and most peripheral bins.

Transverse momentum fluctuations have not been measured for $p$Pb collisions and it is therefore interesting to make a prediction, which is shown as red points in \fig{fig:mapptfluct} (see also \cite{Moreland:2018gsh} for a similar prediction). The fluctuations do not depend strongly on centrality, but a relatively strong decrease is seen for ultra-peripheral collisions (see inset). Remarkably we see a similar decrease for very peripheral collisions.

\begin{figure}[ht]
\includegraphics[width=\columnwidth]{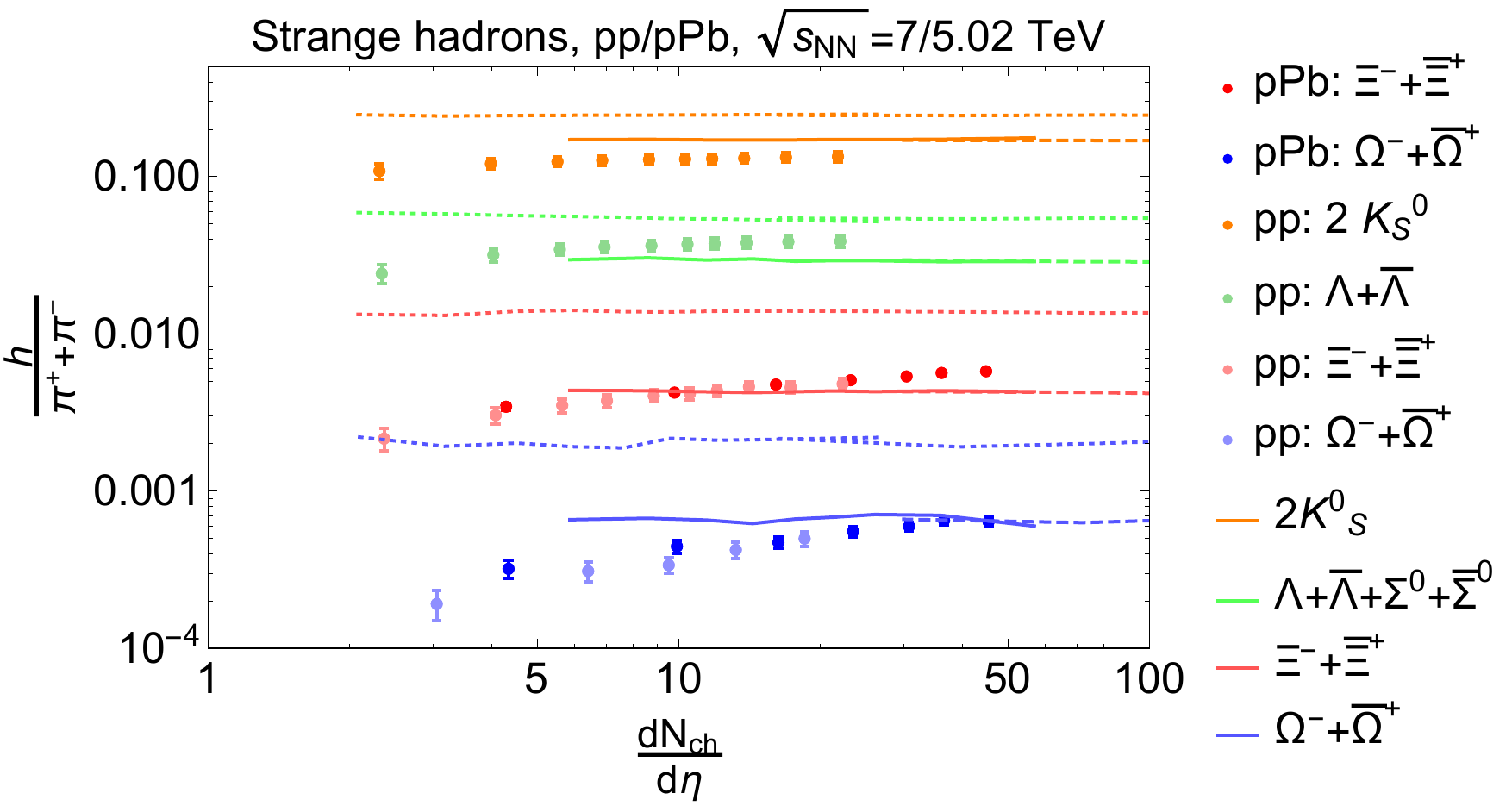}
\caption{\label{fig:strangeness}We show the ratio of hadrons containing strange quarks with respect to the pion yield as a function of multiplicity for $pp$ (experiment only \cite{ALICE:2017jyt}), $p$Pb (experiment \cite{Adam:2015vsf} and model (solid)) and PbPb collisions (model only, dashed). To show the importance of the resonances and the afterburner we include results without applying the hadronic afterburner (dotted). Experimental strangeness increases with multiplicity, while the theoretical curves do not depend on multiplicity. Small but significant deviations exist even at for high multiplicity events for both kaons and lambda hadrons.}
\end{figure}

\subsection{Strangeness enhancement}

The fraction of strange quarks in the final particle spectra is enhanced as one goes to higher and higher multiplicities~\cite{ALICE:2017jyt}, which was originally proposed as a signature of QGP formation.
The naive explanation for this is that in nucleon-nucleon collisions the fraction of strange quarks is canonically suppressed as compared to the fraction that would exist in thermal equilibrium.
In \fig{fig:strangeness} we show the ratio of multiplicities of strange hadrons with respect to pions versus event activity including experimental data for both $pp$ and $p$Pb collisions \cite{Adam:2015vsf,ALICE:2017jyt}. The theoretical curves contain results from our $p$Pb  (solid) and PbPb (dashed) collisions. To stress the importance of resonance decays as well as hadronic interactions we also include the corresponding results without using the SMASH afterburner as dotted lines.

In general the hydrodynamic results align well with the high-multiplicity $pp$ and $p$Pb results, indeed confirming that a thermal model like ours can explain the saturation of the strangeness fraction. There is no significant multiplicity dependence in our hydrodynamic model. There is also a small but significant discrepancy even at high multiplicities: the kaons are overpredicted and the lambdas are underpredicted. In \cite{Vovchenko:2019kes,Cleymans:2020fsc} fits have been performed in the canonical ensemble, which can become important as the system size becomes smaller; see also \cite{Bellini:2018khg} for thermal fits performed by the ALICE collaboration.

\subsection{Anisotropic flow}\label{sec:mapanisotropicflow}

\begin{figure*}[ht]
\includegraphics[width=\textwidth]{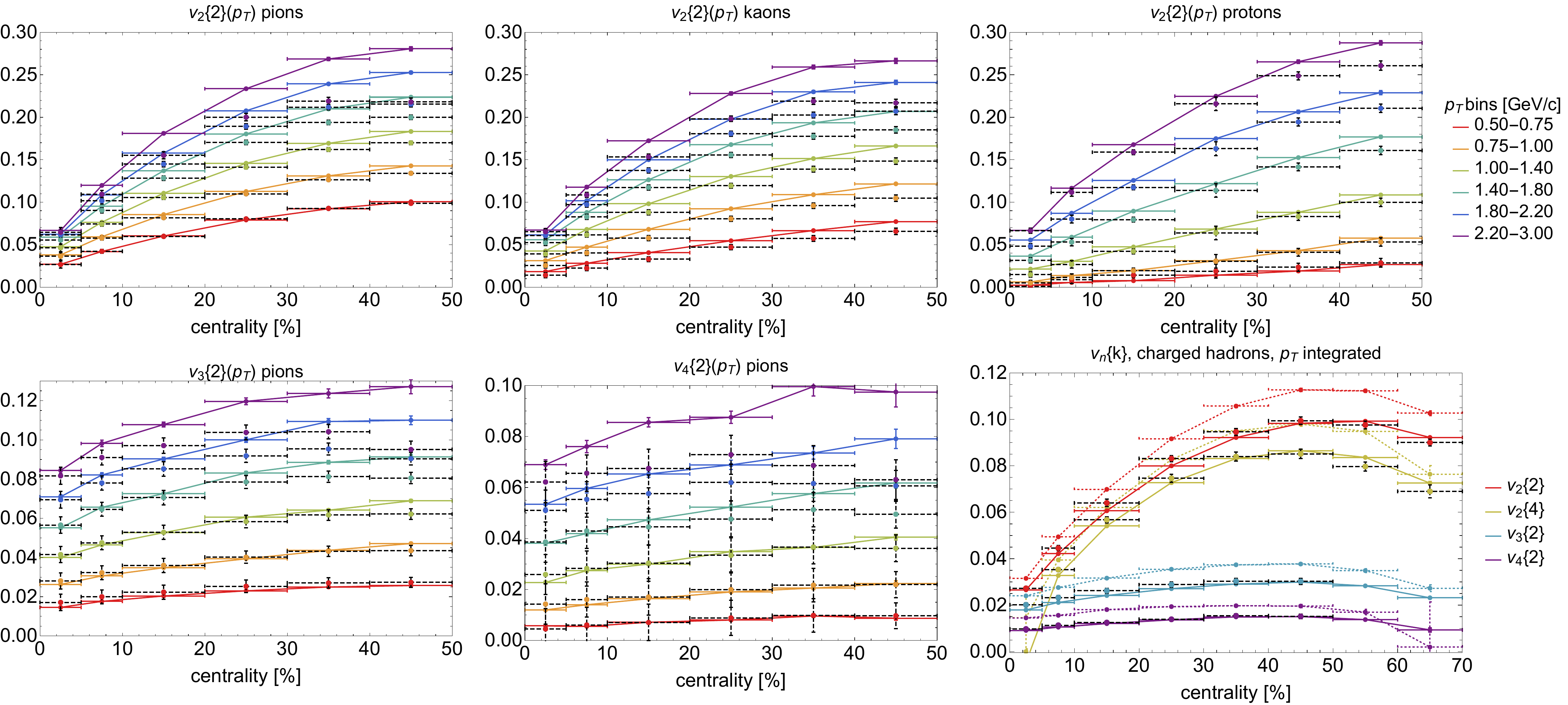}
\caption{\label{fig:mapvnk}We show the anisotropic flow coefficients for PbPb collisions at 2.76 TeV using our MAP values. The top row shows for six $p_T$ bins $v_2\{2\}$ for pions (left), kaons (middle) and protons (right) as a function of centrality. On the bottom row, the left (middle) panel shows $v_3\{2\}$ ($v_4\{2\}$) for pions, and the rightmost panel shows integrated flow for $v_2\{2\}$, $v_2\{4\}$, $v_3\{2\}$ and $v_4\{2\}$\@. The solid colored lines indicate the full model result, the dotted lines are the result without using the afterburner. The black data points are from \cite{Adam:2016nfo,Adam:2016izf}\@.}
\end{figure*}

\begin{figure}[ht]
\includegraphics[width=0.9\columnwidth]{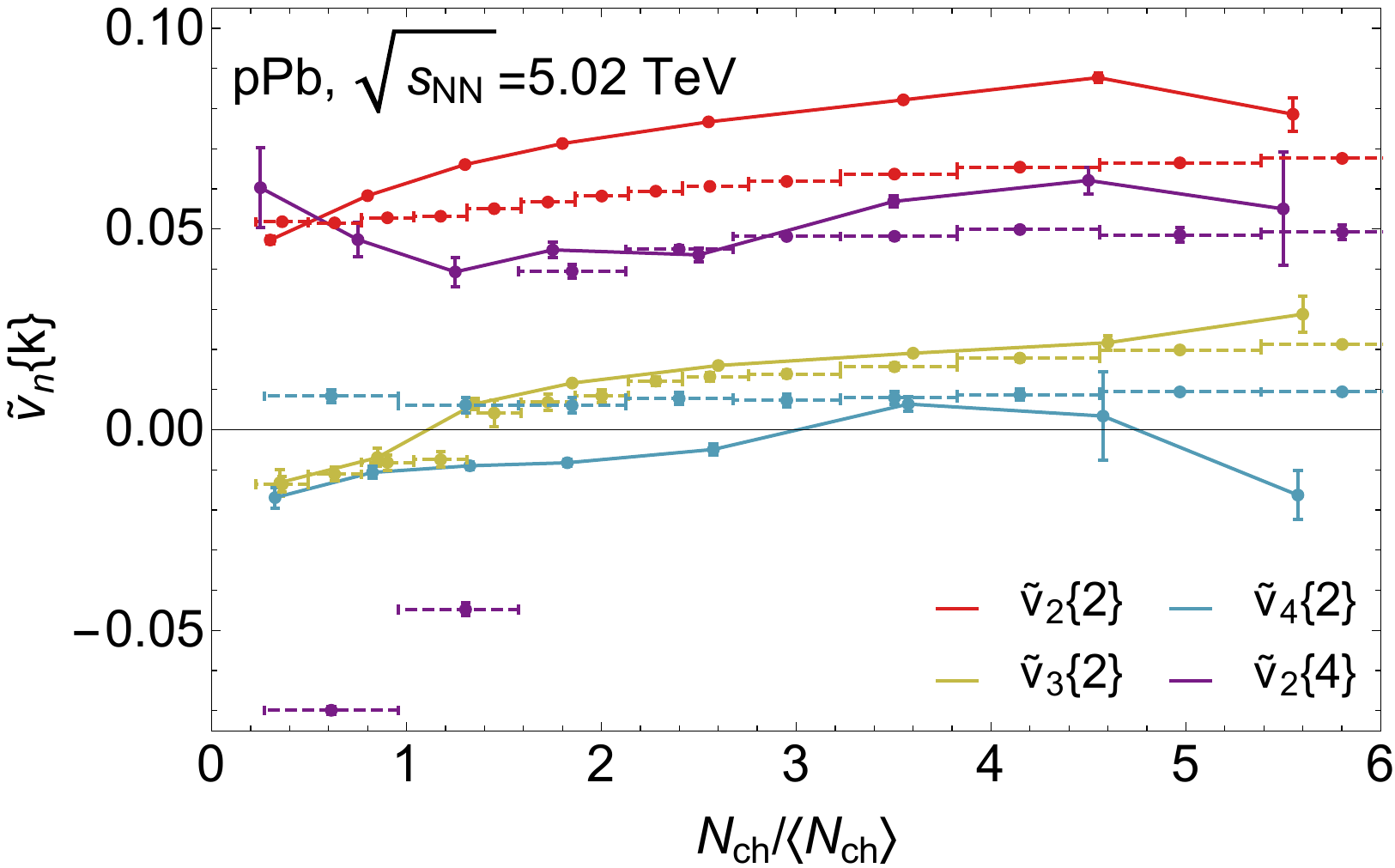}
\caption{\label{fig:vnppb}We show the anisotropic flow coefficients for $p$Pb collisions at 5.02 TeV obtained for our MAP values. Given the difficulty obtaining accurate $p$Pb results and hence posterior distributions the model compares well with the data (dashed, \cite{Aaboud:2017acw})\@. Interestingly our method also reproduces the flow coefficients for negative $\tilde v_n\{k\}$, which corresponds to imaginary values for the standard $v_n\{k\}$\@.}
\end{figure}

Anisotropic flow coefficients are an interesting family of observables that encode the anisotropy of the particle distributions in the final state.
Examples are integrated ($v_n\{k\}$) over a certain $p_T$- and $\eta$-range, and $p_T$-differential ($v_n\{k\}(p_T)$)\@.
The integrated flow coefficients are computed by first computing the $Q_n$-vector for each event \cite{Bilandzic:2010jr,Bilandzic:2012wva}:
\begin{equation}\label{eq:Qn}
Q_n = \sum_{i=1}^Me^{in\phi_i},
\end{equation}
where $\phi_i$ is the azimuthal angle of particle $i$ and the sum runs over a certain set of $M$ particles.
Each different set of chosen particles defines a different observable.
For example, one can choose to use only charged particles, or to also include neutral particles.
Also, one can choose to incorporate only particles of a certain species, leading to identified flow, or to include all species, leading to unidentified flow.
In this work, we chose to compute unidentified charged particle flow coefficients.
For this, we used all charged particles with $0.2\,\text{GeV} \leq p_T \leq 5\,\text{GeV}$ and $|\eta| \leq 0.8$ in accordance with \cite{Adam:2016izf}\@.
After having computed the $Q_n$-vector, one can compute the two- and four-particle correlations for that particular event as
\[
\langle2\rangle_n = \frac{|Q_n|^2 - M}{M(M - 1)}
\]
\begin{align*}
\langle4\rangle_n & = \frac{|Q_n|^4 + |Q_{2n}|^2 - 2\re\left[Q_{2n}Q_n^*Q_n^*\right]}{M(M - 1)(M - 2)(M - 3)},\\
& \quad - 2\frac{2(M - 2)|Q_n|^2 - M(M - 3)}{M(M - 1)(M - 2)(M - 3)}.
\end{align*}
These two- and four-particle correlations are then averaged over all events in a given centrality class, with per-event weights given by $M(M - 1)$ for the 2-particle correlations, and $M(M - 1)(M - 2)(M - 3)$ for the 4-particle correlations.
This yields the averaged 2- and 4-particle correlations $\langle\langle2\rangle\rangle_n$ and $\langle\langle4\rangle\rangle_n$\@.
The integrated flow coefficients are then given by
\[
v_n\{2\} = \sqrt{\langle\langle2\rangle\rangle_n},
\]
\[
v_n\{4\} = \sqrt[4]{2\langle\langle2\rangle\rangle_n^2 - \langle\langle4\rangle\rangle_n}.
\]

The differential flow is computed in a similar way, but now we define two groups of particles for each observable: the reference flow particles (RFP) and the particles of interest (POI)\@.
In this way, one can reconstruct the $p_T$-dependence of the flow.
First, one defines the $Q_n$-vector in the same way as for the integrated flow, where the sum now runs over all RFP\@.
In addition, one defines
\[
q_n = \sum_{i=1}^{m_q}e^{in\phi_i},
\]
where the sum runs over only the $m_q$ POI\@.
One then defines the two- and four-particle correlations:
\[
\langle2'\rangle_n = \frac{q_nQ_n^* - m_q}{m_qM - m_q},
\]
\begin{align*}
\langle4'\rangle_n & = \left[q_nQ_nQ_n^*Q_n^* - q_{2n}Q_n^*Q_n^* - q_nQ_nQ_{2n}^* - 2Mq_nQ_n^*\right.\\
& \quad - 2m_q|Q_n|^2 + 7q_nQ_n^* - Q_nq_n^* + q_{2n}Q_{2n}^*\\
& \quad + 2q_nQ_n^* + \left.2m_qM - 6m_q\right]\\
& \quad / \left[(m_qM - 3m_q)(M - 1)(M - 2)\right].
\end{align*}
In similar fashion to the integrated flow coefficients, these are averaged over all events, with weights $m_qM - m_q$ and $(m_qM - 3m_q)(M - 1)(M - 2)$ for the 2- and 4-particle correlations, respectively.
The resulting averages over all events, $\langle\langle2'\rangle\rangle_n$ and $\langle\langle4'\rangle\rangle_n$, can then be used to obtain the differential flow coefficients for the $p_T$-bin the POI were taken from:
\[
v_n\{2\}(p_T) = \frac{\langle\langle2'\rangle\rangle_n}{\sqrt{\langle\langle2\rangle\rangle_n}},
\]
\[
v_n\{4\}(p_T) = \frac{2\langle\langle2'\rangle\rangle_n\langle\langle2\rangle\rangle_n - \langle\langle4'\rangle\rangle_n}{\left(2\langle\langle2\rangle\rangle_n^2 - \langle\langle4\rangle\rangle_n\right)^{3/4}}.
\]

In \fig{fig:mapvnk} we show the anisotropic flow coefficients for pions, kaons and protons in different $p_T$ bins as a function of centrality, as well as the integrated $v_n\{k\}$ for both a simulation including (solid) and without (dotted) the hadronic afterburner SMASH. By showing the flow coefficients before applying the afterburner we clearly show how the afterburner can in this case significantly decrease anisotropic flow.
The figure shows an impressive agreement with the data, especially considering how both the $p_T$ bin as well as the centrality affects these different hadrons differently. The only significant deviations are found in the three highest $p_T$ bins of pions and the highest $p_T$ bin of kaons. It is interesting that for protons all curves agree, even though in the Bayesian analysis only the $(1.0, 1.4)$ and $(1.4, 1.8)$ bins were used due to the more limited statistics for the protons.

\fig{fig:vnppb} shows the equivalent figure for the flow coefficients in $p$Pb collisions. Recall that we defined $\tilde v_n\{k\} \equiv {\rm sgn}(v_n\{k\}^k) |v_n\{k\}|$, for which $\tilde v_n\{k\}$ is negative when $v_n\{k\}$ is imaginary. Only the second and third harmonic are used for the posterior estimates. 
As also mentioned in Section \ref{sec:mapspectra} the emulator error for $p$Pb is large for this observable, which is not included in the theoretical errors shown.
It is however comforting to see a reasonable description of the data, including statistically significantly negative values for $\tilde v_3\{2\}$ and $\tilde v_4\{2\}$ (for $\tilde v_4\{2\}$ this is opposite from both ATLAS \cite{Aaboud:2017acw} and ALICE \cite{Acharya:2019vdf} experimental data), which in particular shows that even within a purely hydrodynamic model with hadronic cascade $v_n\{2\}$ is not necessarily real.

\subsection{Event-plane correlations}
Event-plane angle correlations are another interesting family of observables that we did not use in the Bayesian analysis.
It is hence interesting to compare the results of \emph{Trajectum} to experimental data.
We used the definition from the ATLAS collaboration \cite{Aad:2014fla}\@.
For two-plane correlations, this starts by splitting each event into two subevents.
Subevent N contains all charged particles with pseudorapidity $-2.5 \leq \eta \leq -0.5$, while subevent P contains all charged particles with pseudorapidity $0.5 \leq \eta \leq 2.5$\@.
In addition to the pseudorapidity requirement, both subevents require the transverse momentum to satisfy $0.5\,\text{GeV} \leq p_T \leq 5\,\text{GeV}$\@.
For each subevent, we then compute the $Q$-vectors
\[
Q_n = \sum_{j=1}^Me^{in\phi_j},
\]
where the sum runs over all $M$ particles in the subevent, $n$ is an integer, and $\phi_i$ the azimuthal angle of particle $i$\@.
We then decompose each $Q$-vector into a magnitude $v_n$ and event plane angle $\Psi_n$:
\begin{equation}
Q_n = Mv_ne^{in\Psi_n}\label{eq:vnpsidef}
\end{equation}
The event-plane correlations are then defined as correlations between the $\Psi_n$ as follows:
\begin{align}
& \langle\cos(k(\Phi_n - \Phi_m))\rangle_\text{SP}\label{eq:eventplanecorrelations}\\
& \quad = \frac{\left\langle\left(v_n^\text{P}\right)^{k/n}\left(v_m^\text{N}\right)^{k/m}\cos(k(\Psi_n^\text{P} - \Psi_m^\text{N}))\right\rangle + \text{P} \leftrightarrow \text{N}}{2\res\{k\Psi_n\}\res\{k\Psi_m\}},\nonumber
\end{align}
where the average is over all events in the centrality class.
For the observable to be properly defined  $k$ must be a multiple of both $n$ and $m$\@.
This correlation aims to measure the correlation between the true event plane angles $\Phi_n$, as defined from a (hypothetical) smooth distribution emitting a large number of particles.
However, due to finite statistics the correlation between the measured event plane angles $\Psi_n$ is different.
To compensate for this difference, the definition includes the resolution factors
\begin{align*}
\res\{k\Psi_n\} = \sqrt{\left\langle\left(v_n^\text{N}v_n^\text{P}\right)^{k/n}\cos(k(\Psi_n^\text{P} - \Psi_n^\text{N}))\right\rangle}.
\end{align*}

The three-plane correlations are defined in a similar fashion.
In this case, each event is subdivided into three subevents.
Subevent A contains all charged particles with $-2.5 \leq \eta \leq -1.5$, subevent B contains charged particles with $-1 \leq \eta \leq 1$, and subevent C contains the charged particles with $1.5 \leq \eta \leq 2.5$\@.
Transverse momenta satisfy $0.5\,\text{GeV} \leq p_T \leq 5\,\text{GeV}$ for each subevent.
The three-plane correlations are then defined by
\begin{align}
& \langle\cos(i\Phi_l + j\Phi_m + k\Phi_n)\rangle_\text{SP}\label{eq:threeplanecorrelations}\\
& \quad = \frac{\left(c_{lmn}^{ijk}\right)^\text{BAC} + \left(c_{lmn}^{ijk}\right)^\text{BCA}}{\res\{i\Phi_l^\text{B}\}\res\{j\Phi_m^\text{A}\}\res\{k\Phi_n^\text{C}\} + \text{A} \leftrightarrow \text{C}},\nonumber
\end{align}
where
\begin{align*}
& \left(c_{lmn}^{ijk}\right)^\text{ABC} =\\
& \quad \left\langle\left(v_l^\text{A}\right)^{l/i}\left(v_m^\text{B}\right)^{m/j}\left(v_n^\text{C}\right)^{n/k}\cos\left(i\Psi_l^\text{A} + j\Psi_m^\text{B} + k\Psi_n^\text{C}\right)\right\rangle
\end{align*}
and
\begin{align*}
\left(\res\{i\Phi_l^\text{A}\}\right)^2 & = \left\langle\left(v_l^\text{A}v_l^\text{B}\right)^{i/l}\cos\left(i\left(\Psi_l^\text{A} - \Psi_l^\text{B}\right)\right)\right\rangle\\
& \quad \times \left\langle\left(v_l^\text{A}v_l^\text{C}\right)^{i/l}\cos\left(i\left(\Psi_l^\text{A} - \Psi_l^\text{C}\right)\right)\right\rangle\\
& \quad / \left\langle\left(v_l^\text{B}v_l^\text{C}\right)^{i/l}\cos\left(i\left(\Psi_l^\text{B} - \Psi_l^\text{C}\right)\right)\right\rangle.
\end{align*}
For the observable to be properly defined, $i/l$, $j/m$ and $k/n$ must be integer, and $i$, $j$ and $k$ must add up to zero.
Again, resolution factors are employed to compensate for the difference between the idealized true event plane angles $\Phi_n$ and the measured event plane angles $\Psi_n$\@.

\begin{figure}[t]
\includegraphics[width=\columnwidth]{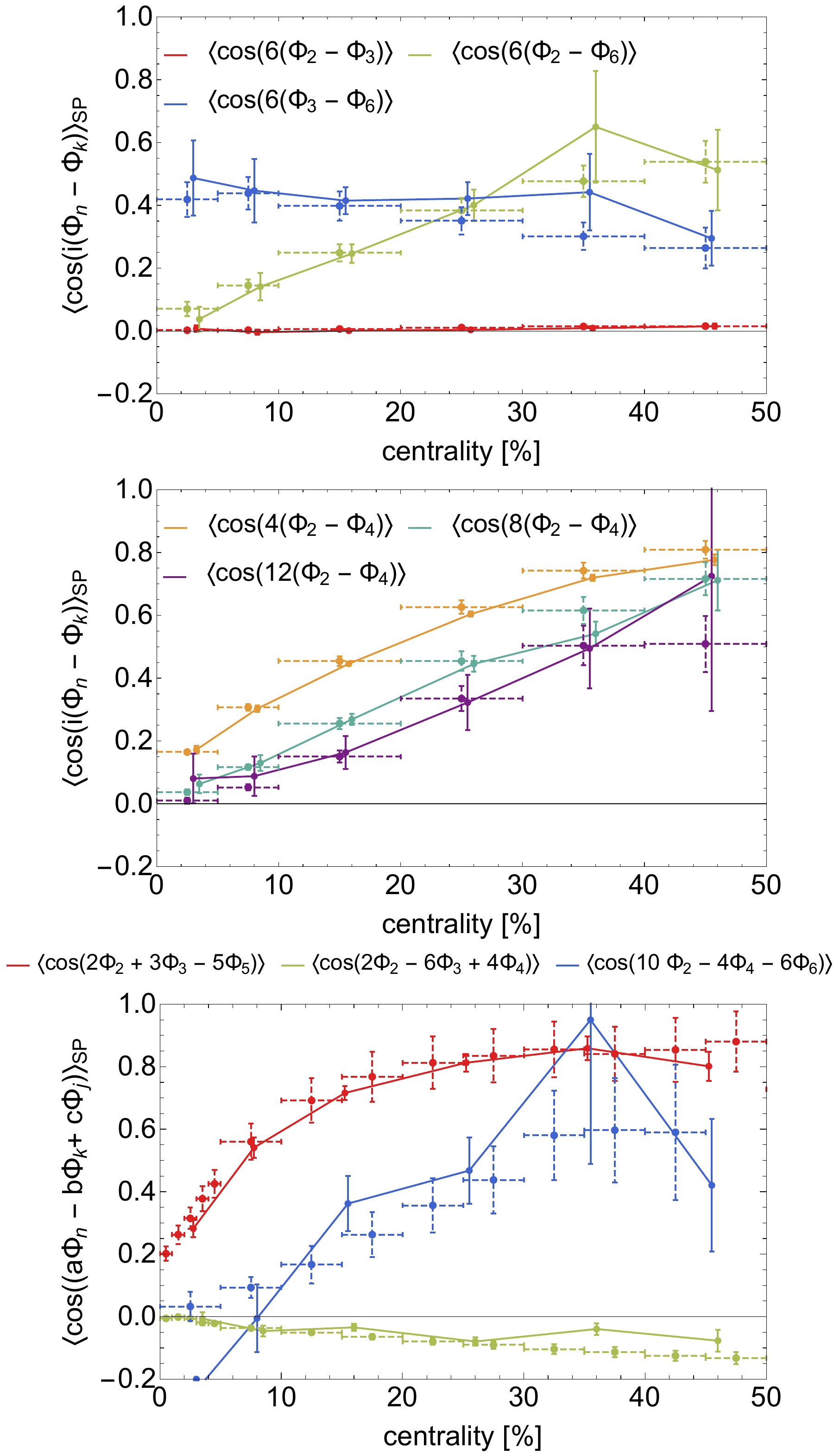}
\caption{\label{fig:psicor}Event-plane correlations using the MAP parameters (solid), for PbPb at 2.76 TeV\@. ATLAS data (dashed) is shown for comparison \cite{Aad:2014fla}\@.}
\end{figure}

Because the cosines always evaluate between $-1$ and $1$, both the numerator and the denominator of \eqref{eq:eventplanecorrelations} and \eqref{eq:threeplanecorrelations} become quite accurate rather fast.
However, for many observables both the numerator and the denominator are also close to zero, leading to large uncertainties in the final result.
\fig{fig:psicor} shows several event-plane correlations that have reasonably small final uncertainties for 200k events generated with \emph{Trajectum} using the MAP values. 

The experimental agreement with ATLAS is excellent for the two-plane correlations, which is non-trivial since this observable is in a different class than the ones used to obtain the MAP values. We note that \cite{Niemi:2015qia} obtained a similarly good agreement, but their agreement was only possible when specifically choosing $\eta/s = 0.20$\@. Reference \cite{Niemi:2015qia} also obtained the event plane angles directly from the hydrodynamic profile, whereas it is clear from \eqref{eq:eventplanecorrelations} that significant modifications can come from both the finite statistic sample as well as the hadronic afterburner used in the current work.
For the three-plane correlations the statistics is more limited, but for those observables that are feasible to compute we also obtain good agreement.

\begin{figure*}[ht]
\includegraphics[width=0.98\textwidth]{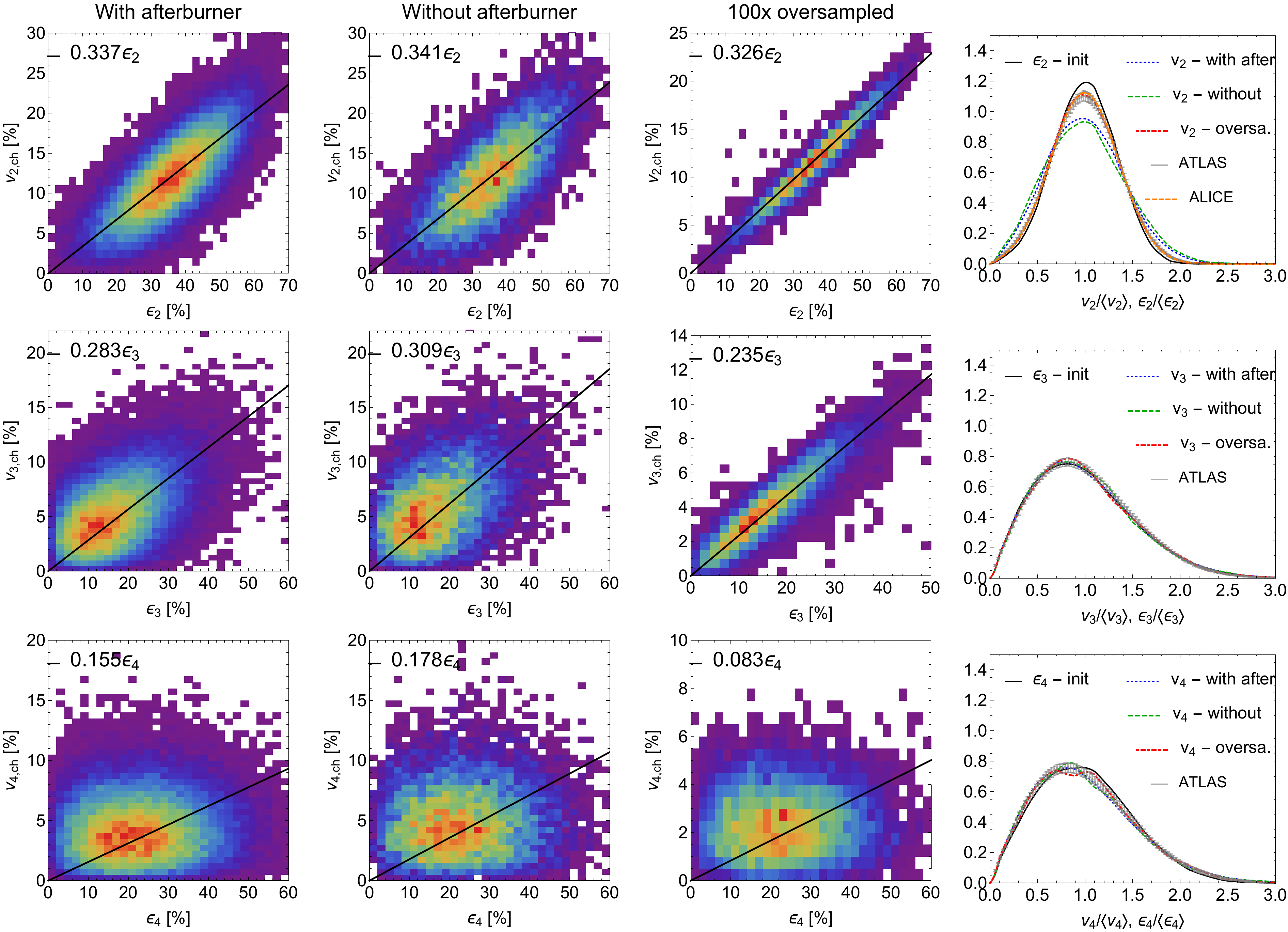}
\caption{\label{fig:vnepsn}Event-by-event anisotropic flow $v_n$ (defined by \eqref{eq:vnpsidef}) as a function of initial state eccentricities $\epsilon_n$, for $n = 2$ (top row), $n = 3$ (middle row) and $n = 4$ (bottom row), at 20--40\% centrality for PbPb collisions at 2.76 TeV\@. The left column shows the result with the hadronic afterburner, the center left column shows the result where the particles produced at particlization do not have any further interactions. The center right column is similar to the middle column, except that the number of produced particles at particlization has been artificially multiplied by a factor 100\@. A linear fit through the origin is also shown for each of these three panels. The rightmost panel shows projections on the $\epsilon_n$ and $v_n$ axes of the three other columns together with experimental data from ATLAS (gray, \cite{Aad:2013xma}). The data is unfolded to match the oversampled distribution, which indeed compares well with these MAP results. ALICE has a similar measurement (orange, dashed, \cite{Acharya:2018lmh}), which uses a different method and agrees with \cite{Aad:2013xma}\@.}
\end{figure*}

\subsection{Eccentricities and anisotropic flow}\label{sec:eccentricities}
It is well known that hydrodynamics translates initial state spatial anisotropies into final state momentum anisotropies \cite{Ollitrault:1992bk,Niemi:2012aj,Giacalone:2017uqx,Bhalerao:2019uzw}\@.
In particular, there is an approximately linear relation between initial state ellipticity $\epsilon_2$ and the final state elliptic flow $v_2$ and similarly for the triangular flow $v_3$. The proportionality constant depends on quantities such as the shear viscosity.
Here, $v_n$ are defined on an event-by-event basis through \eqref{eq:vnpsidef}, and $\epsilon_n$ can be defined from the initial state through \cite{Niemi:2015qia}
\[
\epsilon_n = \left|\frac{\int dx\,dy\,r^ne^{in\phi}\rho(x,y)}{\int dx\,dy\,r^n\rho(x,y)}\right|,
\]
where $r^2 = x^2 + y^2$, $\phi = \arctan(y/x)$ and $\rho(x,y)$ the energy density in the transverse plane.
From a computational point of view, this is interesting, because it opens up the possibility of obtaining $v_2$ %
by running a small number of full hydrodynamic simulations to determine the proportionality constant, 
after which it sufficient to obtain the initial state $\epsilon_2$ to determine $v_2$.

In principle there is then the initial anisotropy $\epsilon_2$, the anisotropy of the freeze-out surface which leads to anisotropic flow and finally the anisotropy of the final particle spectra $v_n$. A priori it is not clear how to relate the final particle spectra to the hydrodynamic freeze-out surface, first of all because of the influence of the afterburner and secondly since the finite multiplicity will lead to statistical fluctuations. In \cite{Aad:2013xma} these effects were unfolded using a Bayesian method, leading to a $v_n$ distribution that is constructed to be similar to the freeze-out surface. In our code we have direct access to the initial eccentricity and the final particle spectra, and with a trick by generating 100x more particles (oversampling) we can obtain an accurate estimate of the freeze-out surface anisotropy.

These results are illustrated in \fig{fig:vnepsn}, where we show $v_n$ as a function of $\epsilon_n$ for $n = 2, 3$ and $4$ for three different cases.
In the left panels, we perform a full hydrodynamics simulation with afterburner, in the center left, we perform the same calculation without the afterburner, and in the center right panels we perform a computation without afterburner, but increasing the number of sampled particles in the output by a factor 100\@.
Different from the rest of this section, the results for `without afterburner' and `100x oversampled' were performed using 50k events.
All panels also show linear fits through the distribution.
One can clearly see that the correlation for $n = 2$ and $3$ is much better when using oversampling.
For $n = 4$, the correlation is not more pronounced with oversampling, indicating that the absence of the correlation is not due to statistics.
The reason for this is actually physical, as $v_4$ is determined by a non-linear combination of $\epsilon_2$ and $\epsilon_4$ \cite{Yan:2015jma}\@.

A more detailed view can be obtained from the $v_n$ and $\epsilon_n$ distributions, normalized to their mean values, as shown in the right column of \fig{fig:vnepsn}. For $\epsilon_2$ and $v_2$ the oversampled distribution differs significantly from the final $v_n$ distributions, both with and without the hadronic afterburner, and is also different from the $\epsilon_2$ distribution. For the third and fourth harmonics all distributions are similar, which reflects the fact that these harmonics originate from fluctuations, whereas $v_2$ has a large mean contribution from the ellipticity from the geometry. It is hence no surprise that the statistical fluctuations from the finite number of particles in the afterburner widen the $v_2$ distribution. Experimentally it is possible to unfold this statistical effect, and after this unfolding the data indeed agrees with the oversampled result \cite{Aad:2013xma}. From a theoretical point of view this has the disadvantage that a direct comparison with the output of the hydrodynamic code with afterburner is not possible, but on the other hand it has the advantage that without using an afterburner the hydrodynamic profiles themselves can be directly compared to the experimental distributions, such as for instance done in \cite{Niemi:2015qia}.

\section{Discussion}

We performed a detailed closure test, including carefully comparing the posterior distributions with the `true' parameters, which gives convincing evidence that the emulation and Monte Carlo work well. Nevertheless, we have to stress that a closure test does not tell if our model is reasonable as a physical description of heavy ion collisions. After all, the closure test only compares to output from the model itself and hence has to work by construction. It is hence always important to study weaknesses in the physical model in higher detail, especially where part of the physics may be missing. Clear examples can be intermediate $p_T$ particles, which are not necessarily described by a hydrodynamic or thermal model, especially in $p$Pb systems. Nevertheless, as shown in Fig.~\ref{fig:realdeltay} and verified in Section~\ref{sec:map}, our model comfortably fits almost all data, thereby giving confidence that our (phenomenological) model scope is wide enough to capture most of the physics presented.

As far as we are aware we are the first to obtain imaginary values for $v_3\{2\}$ and $v_4\{2\}$ using a purely hydrodynamic model (see \fig{fig:vnppb})\@. This is interesting, since without statistical fluctuations both $v_n\{4\}$ and especially $v_n\{2\}$ are expected to be real in hydrodynamics. Even though our posterior probabilities for $p$Pb collisions are relatively uncertain, our MAP parameters give a significantly imaginary value for $v_3\{2\}$, which is in agreement with ATLAS data \cite{Aaboud:2017acw}\@.

There are several avenues for improvement. First of all this can include adding more experimental data. Especially important to further constrain the temperature dependence of the transport coefficients will be the addition of data at lower energies from RHIC (also done in \cite{Auvinen:2017fjw,Everett:2020yty})\@. One could also add additional system sizes, such as XeXe at the LHC or ${}^3$HeAu at RHIC, or one could even study what would be gained by performing OO or ArAr collisions. It is computationally relatively difficult to add more PbPb and $p$Pb observables at LHC energies, as many of these would require a much larger number of events per design point. One way forward in this scheme is to use simpler observables, such as initial state eccentricities, that correlate well with (more difficult) final state flow correlations. Initial investigations into such methods were presented in Section \ref{sec:eccentricities} (see also \cite{Niemi:2015qia, Aad:2013xma})\@. Another opportunity for such observables could be to perform a smaller run with better statistics.

\begin{figure*}[p]
\includegraphics[width=0.99\textwidth]{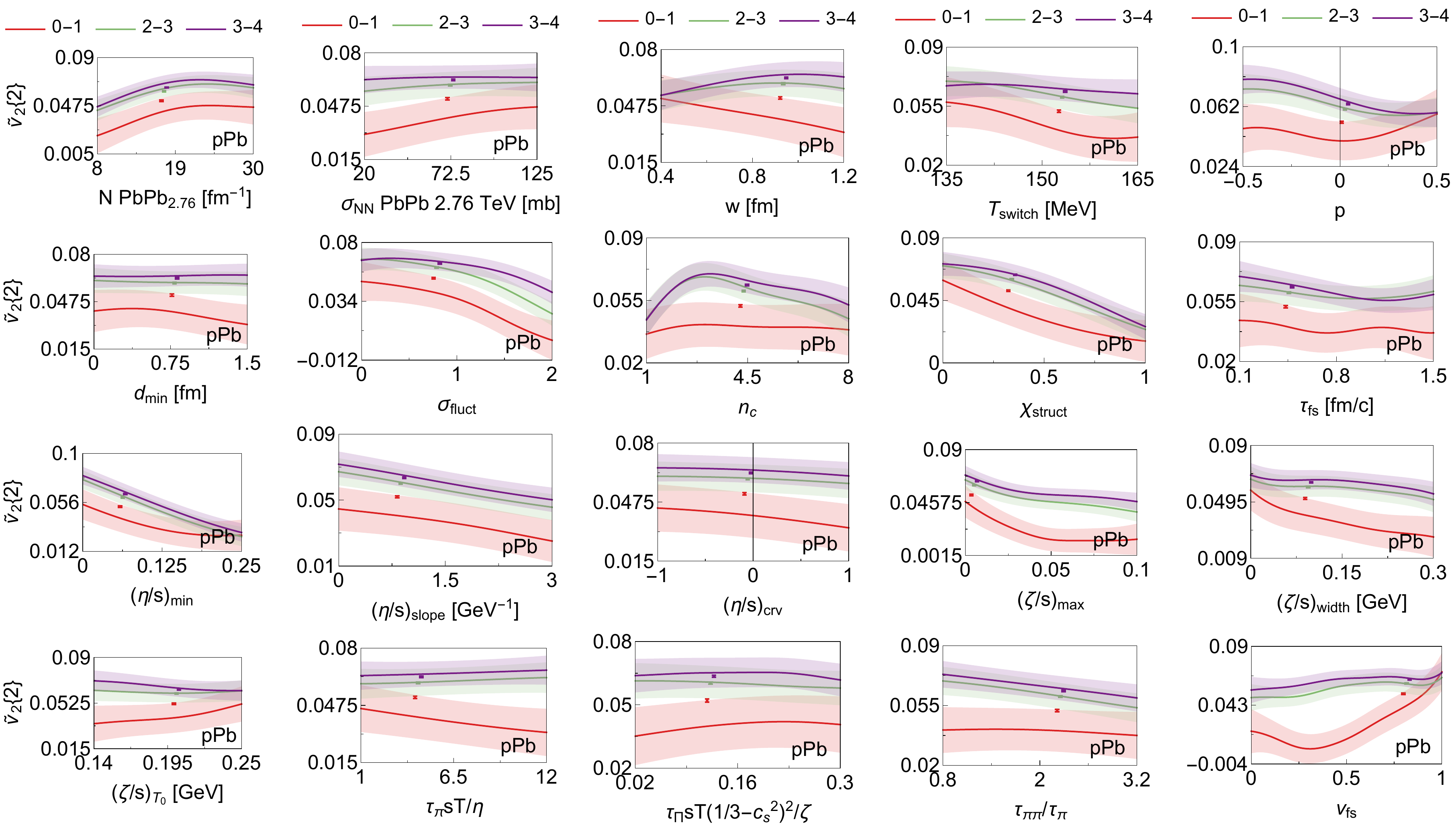}
\caption{\label{fig:emuv2ppb}Similar to \fig{fig:emuv2} and \fig{fig:vnppb} we show the $\tilde v_2\{2\}$ for several multiplicity classes. What is clear that there is a much more refined dependence on all parameters, as opposed to the PbPb case, where the viscosity is mostly dominant. For these small systems also the subnuclear structure, parametrized by $\chi_{\rm struct}$ and $n_c$ is more pronounced.}
\end{figure*}

\begin{figure*}[p]
\includegraphics[width=0.99\textwidth]{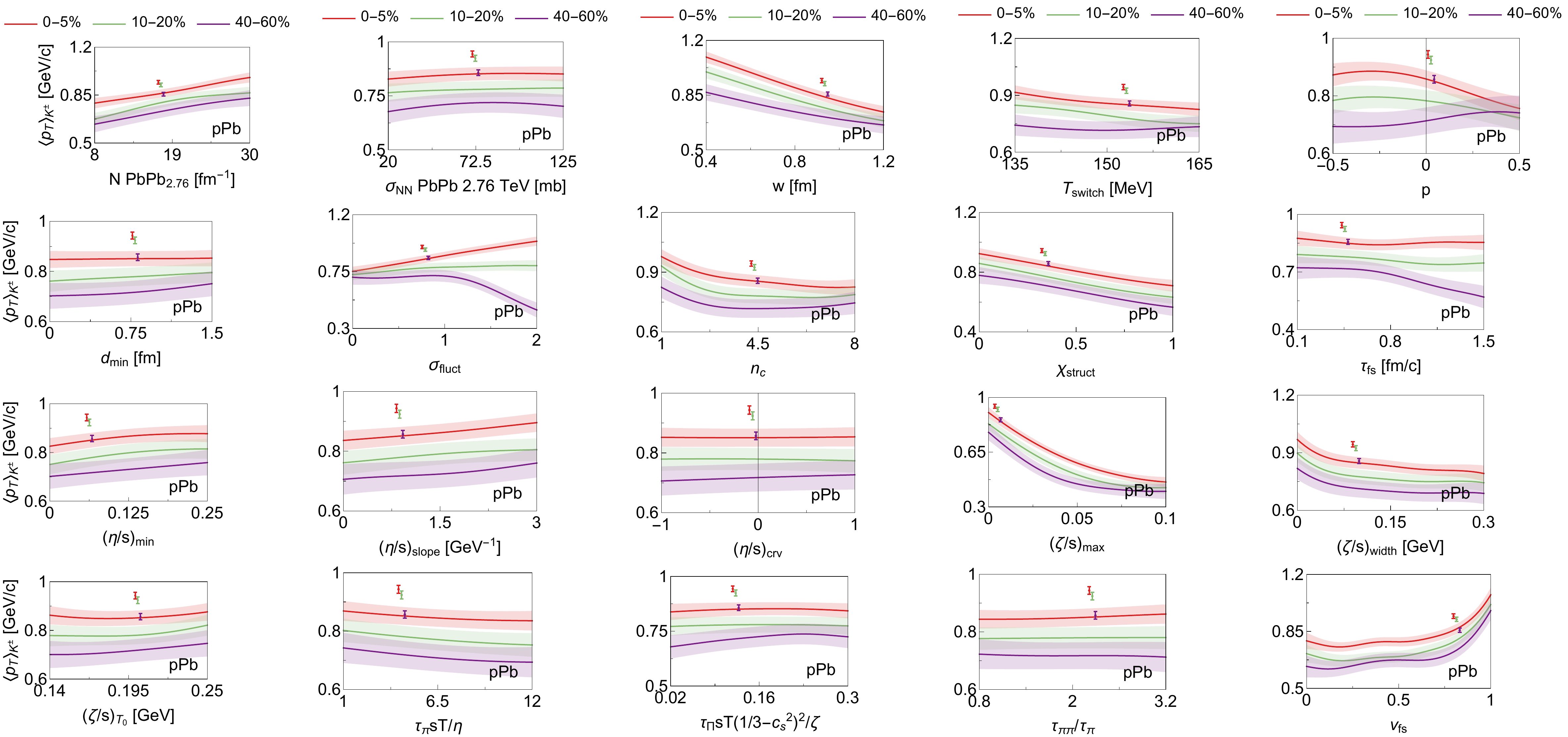}
\caption{\label{fig:emusmeanptppb}%
Similar to \fig{fig:emuv2} we show the mean transverse momentum for kaons for three centrality classes as a function of our model parameters. Data points \cite{Abelev:2013haa} are shown at the MAP values and as discussed in Section \ref{sec:mapspectra} this particular data point does not match well. Note, however, that for this case a different and not excluded value of the width $w$ or free streaming velocity $v_{\rm fs}$ could lead to a better fit.}
\end{figure*}

Another potential improvement lies in the particlization procedure.
Here, thermal particle production is modified by the viscous corrections to ensure continuity of the stress-energy tensor.
However, the precise manner in which this should be done is unknown.
Several different prescriptions exist, and the choice of prescription is known to affect the posterior distributions \cite{Everett:2020yty}\@.
In the future, it would be interesting to improve this part of the simulation, either by data-driven methods such as those presented in \cite{Everett:2020yty}, or by microscopic insights guiding towards the correct prescription.

Currently the difficult emulation of $p$Pb prevents the $p$Pb data from providing a precise constraint in the Bayesian analysis.
Our high statistics MAP results for $\tilde v_4\{2\}$ (see \fig{fig:vnppb}) do not agree well with ATLAS data \cite{Aaboud:2017acw}\@.
This is an indication that perhaps tighter constraints can be obtained by improving the emulation of $p$Pb further.
Furthermore, we observe a deficit of kaon transverse momentum in $p$Pb (see \fig{fig:mapmeanpt})\@.
One hint for resolving this deficit is the strong dependence on $v_{\rm fs}$ in \fig{fig:emusmeanptppb}. Higher values of $v_{\rm fs}$ would increase the mean kaon $p_T$, while these values are not excluded from the Bayesian analysis. 
It would be interesting to see whether an improved emulation and global fit will be able to resolve this discrepancy, or whether the discrepancy will require a non-hydrodynamic explanation.

Lastly, we wish to go beyond the more phenomenological model for the initial stage as presented here. Both the \trento{} initial energy deposition as well as the free streaming phase are not motivated by microscopic models, even though they may be wide enough in scope to describe some of these models. Perhaps most pressing is the fact that the switch from free streaming to hydrodynamics is not smooth, as shown in \ref{fig:taufscheck}, whereas this transition is smooth in holographic \cite{vanderSchee:2012qj,vanderSchee:2013pia} or weakly coupled models \cite{Kurkela:2015qoa, Kurkela:2018vqr}\@.

{\bf Acknowledgements -} We are grateful to Jonah Bernhard and Scott Moreland for making their codes public together with an excellent documentation. We thank Harri Niemi for correspondence. We thank Steffen Bass, Aleksas Mazeliauskas, Ben Meiring and Urs Wiedemann for discussions. 
GN is supported by the U.S. Department of Energy, Office of Science, Office of Nuclear Physics under grant Contract Number DE-SC0011090.

\appendix

\section{Parameter dependence of $p$Pb observables}\label{appendix}
We show the emulator results for $p$Pb collisions in \fig{fig:emuv2ppb} and \ref{fig:emusmeanptppb} for both anisotropic flow coefficients as well as the mean transverse momentum for kaons. The latter is of special interest, as for peripheral $p$Pb collisions there is currently a clear tension between the MAP result and the experimental data (see \fig{fig:mapmeanpt}).

\newpage

\bibliographystyle{apsrev4-1}
\bibliography{main, manual}%

\end{document}